\documentclass[twocolumn,english,aps,prb,nofootinbib,superscriptaddress]{revtex4-1}
\usepackage{CJK}
\usepackage{amsfonts}
\usepackage{amsmath}
\usepackage{ dsfont }
\usepackage{array,multirow}
\usepackage{longtable}
\usepackage{hyperref}
\usepackage{amssymb}
\usepackage{xcolor}
\usepackage{xspace}
\usepackage{ragged2e}
\usepackage{relsize}
\usepackage[bottom]{footmisc}

\usepackage{tikz}
\usetikzlibrary{calc}
\newcommand{\nocontentsline}[3]{}
\newcommand{\tocless}[2]{\bgroup\let\addcontentsline=\nocontentsline#1{#2}\egroup}

\newcommand{\be}{\begin{equation}}
\newcommand{\ee}{\end{equation}}
\newcommand{\bea}{\begin{equation} \begin{aligned}}
\newcommand{\eea}{\end{aligned} \end{equation} }
\newcommand{\bi}{\begin{itemize}}
\newcommand{\ei}{\end{itemize}}

\newcommand{\la}{\lambda}
\renewcommand{\be}{\beta}
\newcommand{\al}{\alpha}
\newcommand{\bpm}{\begin{pmatrix}}
\newcommand{\epm}{\end{pmatrix}}
\newcommand{\eps}{\epsilon}

\renewcommand{\th}{\theta}
\newcommand{\lp}{\left(}
\newcommand{\rp}{\right)}

\newcommand{\del}{\partial}

\newcommand{\Tr}{\text{Tr} \ }

\newcommand{\mbf}[1]{\mathbf{#1}}

\newcommand{\up}{\uparrow}
\newcommand{\dw}{\downarrow}

\usepackage{color}
\usepackage{graphicx}
\usepackage{verbatim}
\usepackage{amsmath}
\usepackage{amssymb}
\usepackage{wasysym}
\usepackage[caption=false]{subfig}
\usepackage{url}
\usepackage{bbold}
\usepackage{slashed}
\usepackage{epstopdf}
\usepackage{braket}
\usepackage{float}
\usepackage[percent]{overpic}

\DeclareRobustCommand{\App}[1]{App.~\ref{#1}}

\DeclareRobustCommand{\Tab}[1]{Table~\ref{#1}}

\DeclareRobustCommand{\Fig}[1]{Fig.~\ref{#1}}

\DeclareRobustCommand{\Eq}[1]{Eq.~(\ref{#1})}
\DeclareRobustCommand{\Eqs}[2]{Eqs.~(\ref{#1}) and (\ref{#2})}
\DeclareRobustCommand{\Ref}[1]{Ref.~\onlinecite{#1}}
\DeclareRobustCommand{\Refs}[2]{Refs.~\onlinecite{#1, #2}}
\DeclareMathAlphabet\mathbfcal{OMS}{cmsy}{b}{n}

\RequirePackage[normalem]{ulem} 
\RequirePackage{color}\definecolor{RED}{rgb}{1,0,0}\definecolor{BLUE}{rgb}{0,0,1} 

\begin{document}

\title{Superfluid Weight Bounds from Symmetry and Quantum Geometry in Flat Bands}

\author{Jonah Herzog-Arbeitman}
\affiliation{Department of Physics, Princeton University, Princeton, NJ 08544}

\author{Valerio Peri}
\affiliation{Institute for Theoretical Physics, ETH Zurich, 8093 Z\"urich, Switzerland}

\author{Frank Schindler}
\affiliation{Princeton Center for Theoretical Science, Princeton University, Princeton, NJ 08544, USA}

\author{Sebastian D. Huber}
\affiliation{Institute for Theoretical Physics, ETH Zurich, 8093 Z\"urich, Switzerland}

\author{B. Andrei Bernevig}
\affiliation{Department of Physics, Princeton University, Princeton, NJ 08544}
\affiliation{Donostia International Physics Center, P. Manuel de Lardizabal 4, 20018
Donostia-San Sebastian, Spain}
\affiliation{IKERBASQUE, Basque Foundation for Science, Bilbao, Spain}

\date{\today}

\begin{abstract}
Flat-band superconductivity has theoretically demonstrated the importance of band topology to correlated phases. In two dimensions, the superfluid weight, which determines the critical temperature through the Berezinksii-Kosterlitz-Thouless criteria, is bounded by the Fubini-Study metric at zero temperature. We show this bound is nonzero within flat bands whose Wannier centers are obstructed from the atoms --- even when they have identically zero Berry curvature. Next, we derive general lower bounds for the superfluid weight in terms of momentum space irreps in all 2D space groups, extending the reach of topological quantum chemistry to superconducting states. We find that the bounds can be naturally expressed using the formalism of real space invariants (RSIs) that highlight the separation between electronic and atomic degrees of freedom. Finally, using exact Monte Carlo simulations on a model with perfectly flat bands and strictly local obstructed Wannier functions, we find that an attractive Hubbard interaction results in superconductivity as predicted by the RSI bound beyond mean-field. Hence, obstructed bands are distinguished from trivial bands in the presence of interactions by the nonzero lower bound imposed on their superfluid weight. 
\end{abstract}

\maketitle

\emph{Introduction}. In a topological insulator, the ground state Wannier functions face an obstruction to exponential localization \cite{Brouder_2007,Thouless_1984,2018CMaPh.tmp....8M,2012RvMP...84.1419M,2020arXiv200604890C}. This real-space picture connects bulk topological invariants computed from the bands in momentum space to the local chemistry of electronic states. Topological states can be either stable or fragile, and are classified by their symmetry properties in momentum space \cite{2017Natur.547..298B,2019arXiv190503262S,2017NatCo...8...50P,PhysRevX.7.041069}. Fragile states can be trivialized by mixing with non-topological bands \cite{2018PhRvL.121l6402P,Bradlyn_2019,2018arXiv180409719B}, while stable states cannot. Moreover, stable topological phases are distinguished by their gapless edge states \cite{2018PhRvX...8c1070K,2010RvMP...82.3045H}, while fragile phases have anomalous boundary signatures exposed by twisted boundary conditions \cite{2020Sci...367..794S,Peri:2020}, magnetic flux \cite{PhysRevLett.125.236804}, or  defects \cite{PhysRevX.9.031003}. Although our understanding of non-interacting topological bands is nearly exhaustive, \cite{2021arXiv210509954V,2020PhRvB.102c5110E,2021arXiv210700647C,2020arXiv201000598E}, this is not the case for interactions within topological bands.

Superconductivity in topological bands \cite{2015NatCo...6.8944P,2021arXiv211100807T} is of interest since its discovery within the fragile flat bands \cite{2018arXiv180710676S,2021PhRvL.126b7002P,2021npjQM...6...82L,2020arXiv200911872S,2020PhRvB.101f0505J,PhysRevB.98.085435,2021arXiv210709090C,2021arXiv210805373C,2020PhRvL.124x7001S,PhysRevB.99.155415} of twisted bilayer graphene \cite{Cao2018UnconventionalSI,2011PNAS..10812233B,2020arXiv201215126S,2019Natur.572..101X,2018Natur.556...80C,5c5f4e1795c94598bec84a15e32355a1,2021arXiv211213401T}. Discovered in \Ref{2016PhRvL.117d5303J}, superconducting order in flat bands, specifically the superfluid weight, originates from quantum geometry characterized by the Fubini-Study metric and has gathered much excitement
\cite{PhysRevX.9.031049,2018PhRvB..98m4513T,2021arXiv210509322H,PhysRevLett.123.237002,2018PhRvB..98v0511T,PhysRevB.102.184504,2021PNAS..11806744V}. The quantum metric, though distinct from the band topology, is bounded by the Chern or (Euler) winding numbers \cite{2015NatCo...6.8944P,PhysRevLett.124.167002,PhysRevB.94.245149}. These early results suggest that topological quantum chemistry could provide general lower bounds, making contact with materials databases \cite{2021arXiv210605272C,2021arXiv210600709W,2020Natur.586..702X}. Our work affirms this suggestion, yielding nonzero bounds in phases without winding numbers. Our bounds are given by another quantized number, the real space invariant (RSI) \cite{2020Sci...367..794S}, which uses symmetries to characterize topological, obstructed Wannier centers (OWCs), and trivial bands.

From a materials perspective, although topological bands are abundant within real crystals, a significant portion are topologically trivial at the Fermi level. Topologically trivial bands with space group symmetries have a finer classification which divides them into trivial atomic bands, where electrons are exponentially localized at the atomic sites, and bands with OWCs which, while exponentially localized, are necessarily centered off the atoms \cite{2021arXiv210610276X,2021arXiv210605287R,2021arXiv210713556S}. Remarkably, we show that, like topological bands, OWCs have a nonzero, lower-bounded Fubini-Study metric even without Berry curvature \cite{PhysRevB.94.134423,liang2017band}. When flat OWC bands are partially filled under attractive interactions, they possess a superconducting instability.

The zero-temperature superfluid weight $[D_s]_{ij}$ of an isolated flat band within BCS theory \cite{liang2017band} is (\App{app:MFSF})
\bea
\label{eq:dsBCS}
[D_s]_{ij} &= 2|\Delta| \sqrt{\nu(1-\nu)} \int \frac{d^2k}{(2\pi)^2} g_{ij}(\mbf{k}) \\
\eea
where $\Delta$ is the superconducting gap, $\nu$ is the filling fraction of the flat bands, $\mbf{k}$ is a momentum in the Brillouin zone (BZ) with area $(2\pi)^2/\Omega_c$ ($\Omega_c$ is the unit cell area), and $g_{ij}$ is the Fubini-Study quantum metric. A nonzero superfluid weight implies a finite critical superconducting temperature \cite{Kosterlitz1973OrderingMA,PhysRevLett.39.1201} and a supercurrent $\mbf{J} = -4D_s \mbf{A}$, where $\mbf{A}$ is the vector potential in the London gauge. In a Hamiltonian with $N_{\mathrm{orb}}$ orbitals and $N_{\mathrm{occ}}$ occupied bands, we define the (abelian) quantum geometric tensor \cite{2011EPJB...79..121R}
\bea
\Tr \mathcal{G}_{ij} &= \Tr P \del_i P \del_j P = g_{ij} + \frac{i}{2} f_{ij},
\eea
where $P(\mbf{k})$ is the $N_{\mathrm{orb}}\times N_{\mathrm{orb}}$ gauge-invariant projection matrix onto the occupied bands, $\del_i$ is a momentum-space derivative, and the trace is over the matrix indices. The abelian Berry curvature  $f_{ij} = - f_{ji}$ is well studied while the positive semi-definite quantum metric $g_{ij} = g_{ji}$ is an object of more recent interest \cite{2021arXiv211014658M,2021arXiv210802216S,2021arXiv210414257J,2010arXiv1012.1337C,2021arXiv210600800M,2021arXiv210811478R,2011EPJB...79..121R,PhysRevB.102.165118,2020arXiv201007751V,2021arXiv210507491W,PhysRevB.87.245103,PhysRevA.103.053311,2015PhRvL.115u6806S,2017PhRvB..96f4511L,ma2021topology,PhysRevB.104.064306,2021arXiv210301241A,2014PhRvL.112p6601G,orenstein2021topology,kumar2020topological,PhysRevLett.65.1697}. With spatial rotation symmetry, $[D_s]_{ij}$ is determined by the trace of $g_{ij}$ \cite{footnote1}
\bea
\label{eq:trg}
G = \frac{1}{2} \int\frac{d^2k}{(2\pi)^2} \Tr \pmb{\nabla} P \cdot  \pmb{\nabla} P  \geq 0,
\eea
which is coordinate invariant, dimensionless, and importantly is quadratic in $P(\mbf{k})$  (\App{app:qgm}). We give an efficient numerical discretization formula in \App{app:SFWformula}.

\emph{Flat Band Model.}
\begin{figure}
 \centering
\includegraphics[width=6.cm]{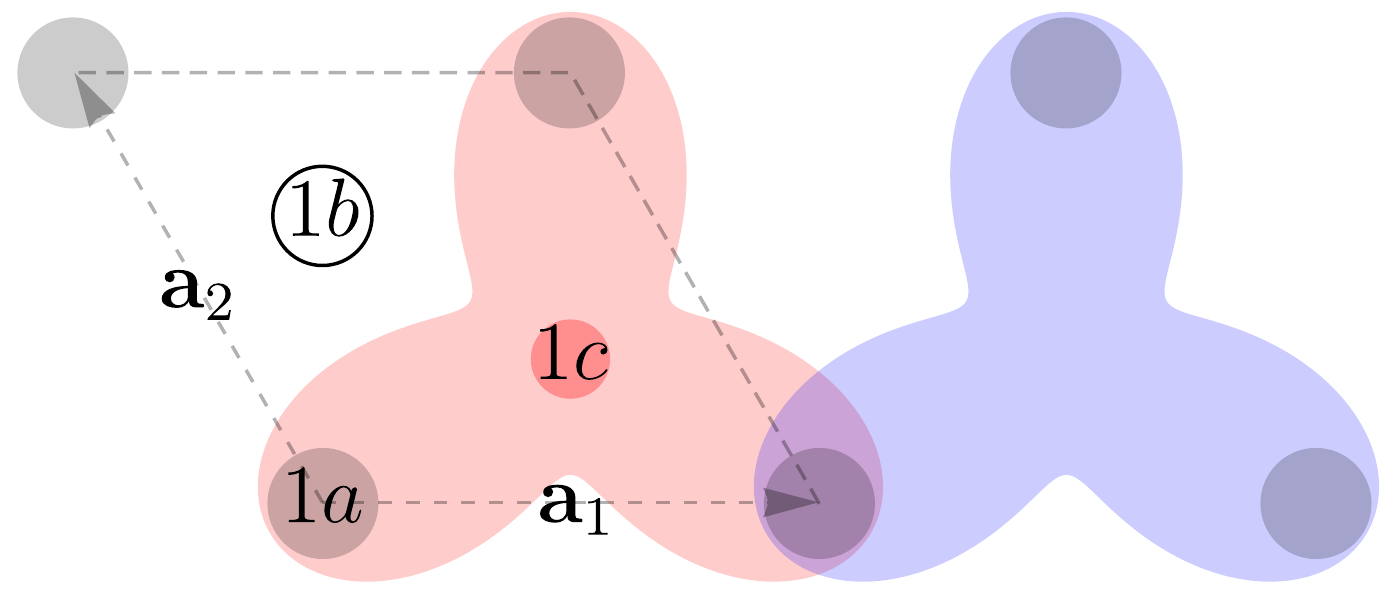}
\caption{Wannier basis at 1c $= \frac{2}{3} \mbf{a}_1 + \frac{1}{3} \mbf{a}_2$. The Wannier state $\ket{A_{1c}}$ (red) centered at 1c is supported only on the neighboring atomic sites (grey) with $A, {}^1E$, and ${}^2E$ orbitals. Only one site overlaps with neighboring Wannier states (blue).}
\label{fig:C3wannier}
\end{figure}
We begin by constructing an OWC model in the space group $p3$ generated by spinless $C_3$ symmetry and translations along the lattice vectors $\mbf{a}_1=(1,0), \mbf{a}_2 = C_3 \mbf{a}_1$. At the origin (the 1a position), we place electrons in the $A, {}^1E, {}^2E$ irreps. These orbitals induce band representations \cite{2018PhRvB..97c5139C,PhysRevB.23.2824} with irreps defined by
\bea
A_{1a} \uparrow p3 &= \Gamma_1 + K_1 + K'_1, \quad C_3 = +1 \\
{}^1E_{1a} \uparrow p3 &= \Gamma_2 + K_2 + K'_3, \quad C_3 = e^{-\frac{2\pi i}{3}} \\
{}^2E_{1a} \uparrow p3 &= \Gamma_3 + K_3 + K'_2, \quad C_3 = e^{\frac{2\pi i}{3}} \\
\eea
where $\Gamma = (0,0)^\mathrm{T}, K = \frac{2\pi}{3}(\mbf{b}_1+\mbf{b}_2), K' = -\frac{2\pi}{3}(\mbf{b}_1+\mbf{b}_2)$ are the high symmetry points, and $\mbf{a}_i \cdot \mbf{b}_j = \delta_{ij}$. The full group theory data can be found on the Bilbao Crystallographic server \cite{Aroyo:firstpaper,Aroyo:xo5013,PhysRevE.96.023310}. To construct a flat band OWC from these orbitals, we will use a Wannier basis centered at the 1c position \emph{off} the atomic sites at 1a as in \Fig{fig:C3wannier}. We form the Wannier states
\bea
\label{eq:c3wan}
\ket{\mbf{R},A_{1c}} = \frac{1}{3} T_{\mbf{R}} \sum_{j=0}^2 \tilde{C}_3^j (\ket{0,A} + \ket{0,{}^1E} + \ket{0,{}^2E}),
\eea
where $T_\mbf{R}$ is the translation operator by $\mbf{R}$, $\tilde{C}_3$ is the rotation operator about the 1c position, and $\ket{0,\rho}$ are the $\rho$ orbitals in unit cell 0. Taking $\tilde{C}_3 \to e^{\mp \frac{2\pi i}{3}} \tilde{C}_3$ in \Eq{eq:c3wan} yields ${}^1E$ and ${}^2E$ states at 1c. It is easy to check that the states $\ket{\mbf{R},A_{1c}}$ are orthonormal: $\braket{\mbf{R},A_{1c}|\mbf{R}',A_{1c}}$ is nonzero only if $\mbf{R}$ and $\mbf{R}'$ are nearest neighbors, and in this case the only overlap is on a single site which vanishes due to $\tilde{C}_3$ eigenvalues of the orbitals. Fourier transforming \Eq{eq:c3wan} to obtain the eigenstate $\ket{\mbf{k},A_{1c}}$ yields the eigenvector $U_\al(\mbf{k}) = \braket{\mbf{k},\al|\mbf{k},A_{1c}}$, $\al = A, {}^1E, {}^2E$:
\bea
U(\mbf{k}) &= \frac{1}{3}\!\bpm 1 \\ 1 \\ 1\epm \! + \frac{1}{3} \!\bpm 1 \\ e^{\frac{4\pi i}{3}} \\ e^{ \frac{2\pi i}{3}} \epm \! e^{i \mbf{k} \cdot (\mbf{a}_1 + \mbf{a}_2)} \!+ \frac{1}{3} \!\bpm 1 \\ e^{\frac{2\pi i}{3}} \\ e^{\frac{4\pi i}{3}} \epm \! e^{i\mbf{k} \cdot  \mbf{a}_2},
\eea
and the local momentum-space Hamiltonian
\begin{equation} \label{eq: flatblandbloch}
h(\mbf{k}) = - \lvert t\rvert U(\mbf{k}) U^\dag(\mbf{k}) \equiv - \lvert t\rvert P(\mbf{k})
\end{equation}
which has three exactly flat bands: the $A_{1c}$ band at energy $-\lvert t\rvert$ and the degenerate ${}^1E_{1c}$ and ${}^2E_{1c}$ bands at zero energy. At filling $1/3$, $h(\mbf{k})$ has the band representation
\bea
\Gamma_1 + K_3 + K'_3 &= A_{1c} \uparrow p3,
\eea
confirming our construction in real space. We also calculate the Berry connection in crystalline coordinates,
\bea
\label{eq:berryAi}
A_i(\mbf{k}) = U^\dag(\mbf{k}) i \del_i U(\mbf{k}) &= \lp -1/3, -2/3 \rp_i,
\eea
which is the expectation value of the lattice position operator in the occupied bands. Noting that the lattice position operator is only defined mod 1, \Eq{eq:berryAi} confirms that the states are located at the 1c position. Because $A_i(\mbf{k})$ is independent of $\mbf{k}$ (up to a gauge choice), the Wilson loop bands are perfectly flat \cite{Alexandradinata:2012sp} and the Berry curvature is identically zero. Topologically, the model is therefore trivial. However, we calculate the quantum metric in cartesian coordinates ($a$ is the lattice constant):
\bea
\label{eq:gmunuthird}
g_{ij}(\mbf{k}) = \frac{1}{2} \Tr \del_i P(\mbf{k}) \del_j P(\mbf{k}) = a^2 \delta_{ij}/6 , 
\eea
so the mean-field superfluid weight in \Eq{eq:dsBCS} is nonzero despite the model being topologically trivial and having compact Wannier functions (zero correlation length) \cite{2021arXiv210713556S}.

It is natural to ask what indices describe these compact OWC phases. By definition, they are induced by off-site atomic orbitals, so topological quantum chemistry can identify them because their symmetry data does not match any of the orbitals present in the lattice. This is different than the stable and fragile indices which are independent of the basis orbitals.

 Wilson loops can also identify OWCs. A useful reference is the SSH chain\cite{PhysRevLett.42.1698}  where an eigenvalue of $\pi$ of the Wilson loop operator identifies the off-site states \cite{2018arXiv181003484N,2020JAP...128v1102S}. Lastly, OWCs can be most naturally defined using the RSI formalism developed in \Ref{2020Sci...367..794S}. RSIs are local quantum numbers which are well-defined in fragile and OWC phases where they supply lower bounds on the number of states at the high symmetry Wyckoff positions. By definition, RSIs are invariant under symmetry-preserving adiabatic deformations. Since the Wannier states in OWCs cannot be moved to atomic sites without closing a gap, they are characterized by a nonzero RSI off an atomic site. In space group $p3$, the RSIs at a $C_3$-symmetric Wyckoff positions are
\bea
\label{eq:c3rsis}
\delta_1 = m({}^1E) - m(A), \quad \delta_2 = m({}^2E) - m(A)
\eea
and $m(\rho)$ is the number of $\rho$ irreps. The RSIs can be conveniently calculated from the momentum space symmetry data \cite{2020Sci...367..794S}. In our model, the only nonzero RSIs are off the atomic sites at the 1c position, $(\delta_{1c,1}, \delta_{1c,2}) = (-1,-1)$. We will now show that, in generality, these offsite RSIs are responsible for a bounded superfluid weight.

\emph{Lower Bounds in Real Space.}
We derive general lower bounds for the superfluid weight in terms of RSIs and orbital positions to show that nonzero superfluid weight is a generic feature of partially filled OWC bands, as has been shown for Chern insulators and Euler insulators \cite{2015NatCo...6.8944P,PhysRevLett.124.167002}. Our bounds also apply to all 2D topological bands, including fragile bands (\App{app:tables}). Here, for simplicity, we prove a lower bound for our model with $C_3$ and all orbitals at the 1a position. Our starting point is a real space expression for $P(\mbf{k})$, the projector onto the occupied $A_{1c}$ band. If all orbitals are at the 1a position, then $h(\mbf{k})$ and $P(\mbf{k})$ are periodic under $\mbf{k} \to \mbf{k} + 2\pi \mbf{b}_i$. Thus there is a Fourier representation
\bea
\label{eq:Fourierharmonics_main}
P(\mbf{k}) &= \sum_\mbf{R} e^{-i \mbf{R} \cdot \mbf{k}} p(\mbf{R}), \ p(\mbf{R}) = \int \frac{dk^1dk^2}{(2\pi)^2} e^{i \mbf{R} \cdot \mbf{k}} P(\mbf{k}),
\eea
defined in terms of the harmonics $p(\mbf{R})$, which are $N_{\mathrm{orb}} \times N_{\mathrm{orb}}$ matrices, the lattice vectors $\mbf{R}$, and the dimensionless crystal momenta $k^i$. Note that $P(\mbf{k})$ is the momentum space Green's function \cite{andreibook}, so $p(\mbf{R})$ is the real space correlation function. The harmonics obey a normalization condition
\bea
\label{eq:normalization}
\sum_\mbf{R} ||p(\mbf{R})||^2 = \int \frac{dk^1dk^2}{(2\pi)^2} \Tr P(\mbf{k}) = N_{\mathrm{occ}} = 1,
\eea
where $||A||^2 = \Tr A^\dag A$ is the squared Frobenius norm (\App{app:realspaceproj}). Rewriting \Eq{eq:trg} in real space, we find
\bea
\label{eq:trgRS}
G = \frac{1}{2} \int \frac{d^2k}{(2\pi)^2} \Tr \pmb{\nabla} P \cdot  \pmb{\nabla} P  = \frac{1}{2\Omega_c} \sum_{\mbf{R}} |\mbf{R}|^2  ||p(\mbf{R})||^2 \ .
\eea
We will use symmetry eigenvalues to show that $||p(\mbf{a}_1)||$ and symmetry-related terms are bounded below. This immediately gives a bound for $G$ because $G \geq \frac{1}{2\Omega_c} |\mbf{a}_1|^2 ||p(\mbf{a}_1)||^2$ since all terms in \Eq{eq:trgRS} are positive semi-definite. Indeed, all are \emph{positive} definite except for the zero mode $p(0)$, the constant mode of $P(\mbf{k})$.

\begin{figure}
 \centering
 \includegraphics[height=3cm]{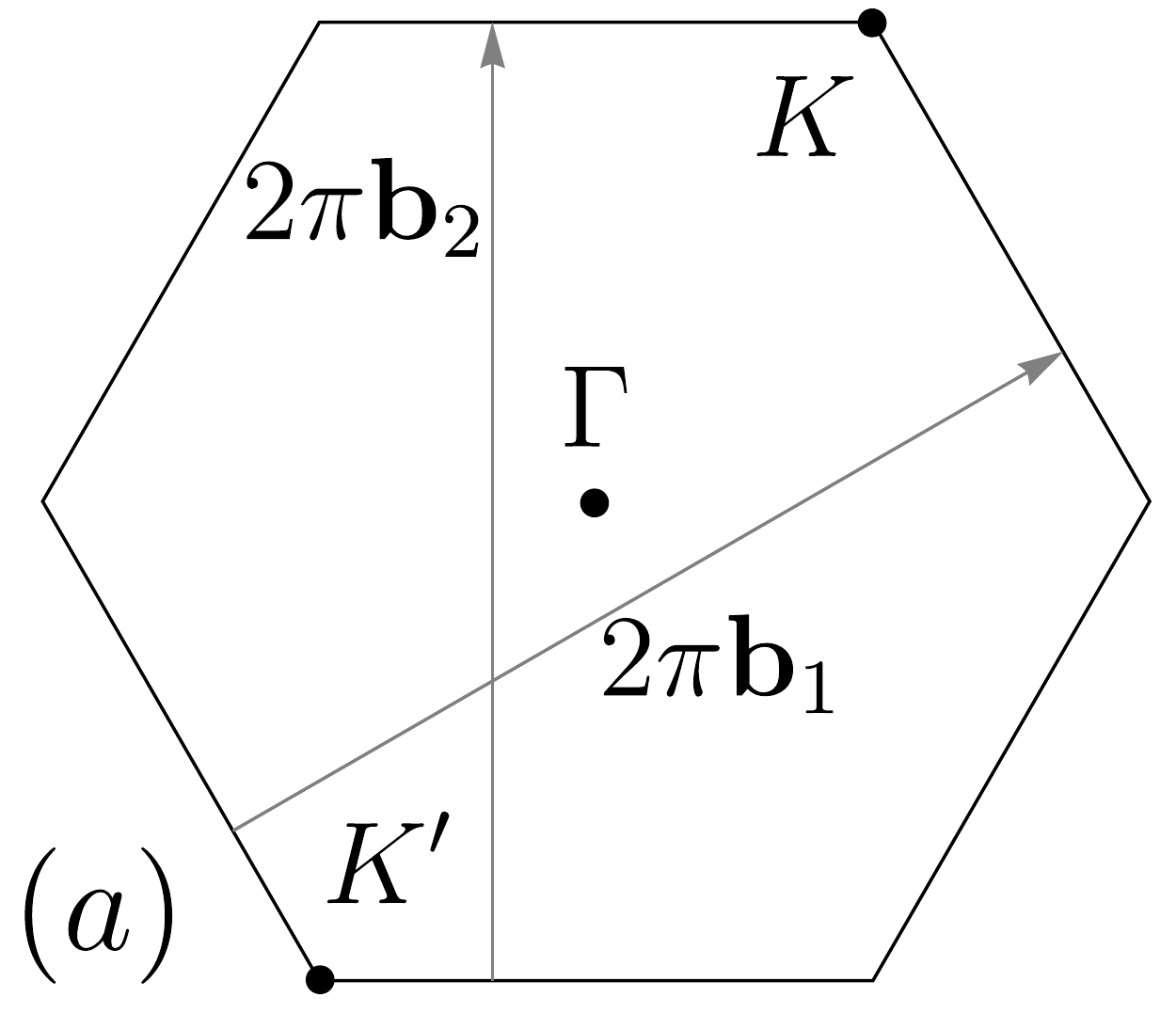} \qquad \includegraphics[height=3cm]{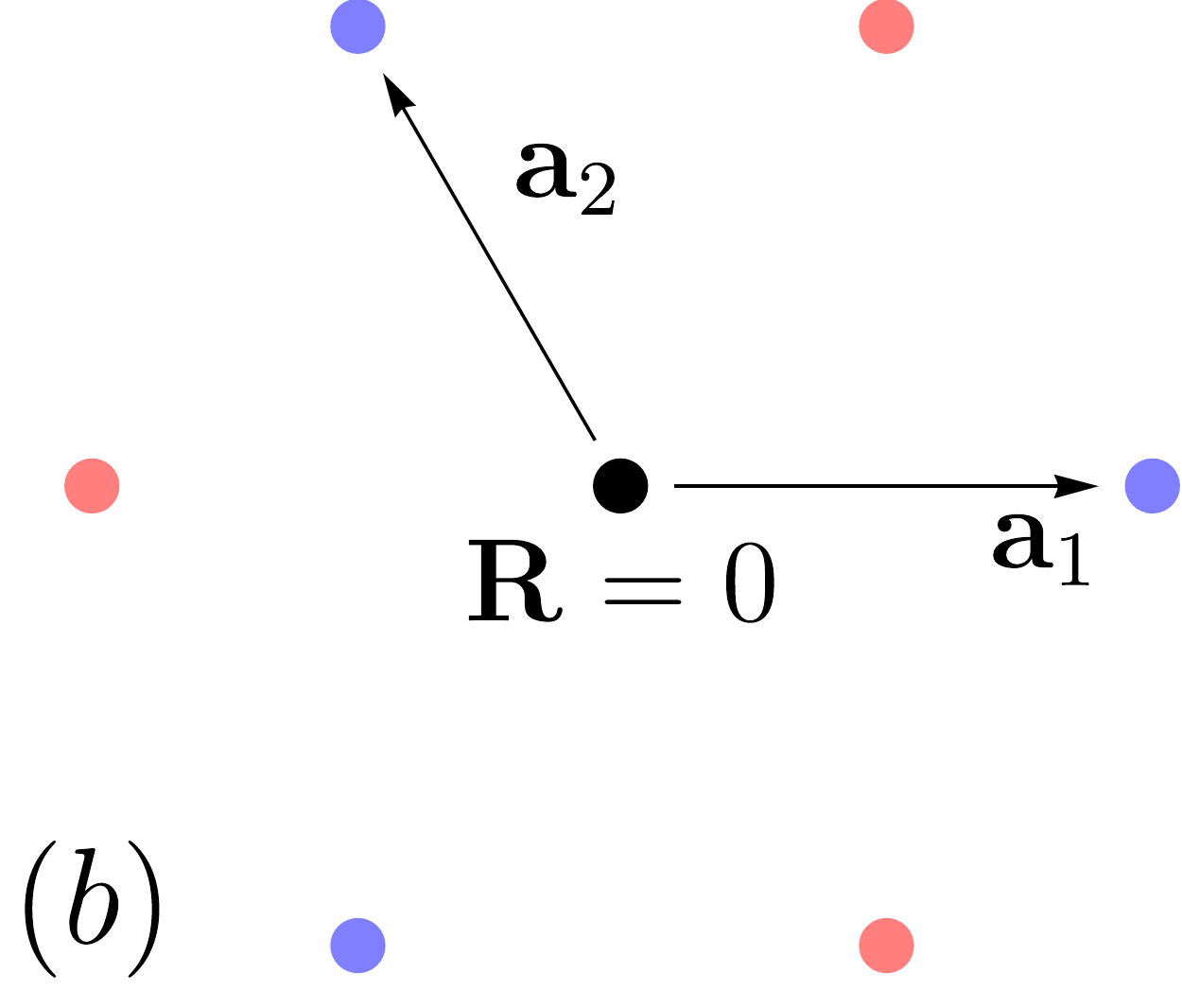}
\caption{$C_3$ Bounds. $(a)$ We label the $C_3$-invariant points $\Gamma$, $K$, and $K'$ in the BZ. $(b)$ By taking linear combinations of $P(\mbf{k})$, we find lower bounds for the harmonics at $|\mbf{R}| = 1$, shown in red and blue. Higher harmonics are not shown.
\label{fig:inv_main}
}
\end{figure}

The momentum space irreps consist of the $C_3$ eigenvalues in the occupied bands at $\Gamma, K,K'$ (see \Fig{fig:inv_main}a). The irrep multiplicities $m(\rho)$ obey (\App{eq:TQC})
\bea
\label{eq:irrepdiffC3}
m(\Gamma_1) + e^{\frac{2\pi i}{3}} m(\Gamma_2)+ e^{-\frac{2\pi i}{3}} m(\Gamma_3) &= \Tr D[C_3] P(\Gamma),
\eea
and similarly for $K$ and $K'$. Here $D[C_3] = \text{diag}(1, e^{-\frac{2\pi i}{3}},e^{\frac{2\pi i}{3}})$ is the representation matrix of $C_3$ on the orbitals. Thus the irrep multiplicities give information about $P(\mbf{k})$. Writing out and summing \Eq{eq:Fourierharmonics_main} at each high-symmetry momentum gives
\bea
\label{eq:TRIMpoints}
P(\Gamma) + e^{\frac{2\pi i}{3}} P(K) + e^{\frac{4\pi i}{3}} P(K')&= 3\sum_{n=1}^3  p(-C_3^n \mbf{a}_1) + \dots,
\eea
where the dots represent higher harmonics $p(\mbf{R})$ for $|\mbf{R}| > |\mbf{a}_i|$ (see \Fig{fig:inv_main}b). Crucially, the roots of unity cancel $p(0)$, so only harmonics at $\mbf{R}\neq0$ appear in \Eq{eq:TRIMpoints}. We now bound \Eq{eq:TRIMpoints} on both sides. To manipulate the momentum space side, we use an elementary inequality\cite{wolkowicz1980bounds}
\bea
||A||^2 \geq |\Tr S A|^2 / \text{Rk}(A) \quad \forall S \text{ unitary},
\eea
proven in \Ref{SM}. Choosing $A = P(\Gamma) + e^{\frac{2\pi i}{3}} P(K) + e^{-\frac{2\pi i}{3}} P(K')$, we see $1/\text{Rk}(A) \geq 1/3$ because $A$ is a $3\times 3$ matrix, and with $S = D[C_3]$, we find with \Eq{eq:irrepdiffC3}:
\bea
|\Tr S A|^2  &= \frac{9}{4} (m(K_3)+m(K'_3)-m(\Gamma_2) - m(\Gamma_3))^2 \  + \\
\frac{3}{4} (m(&\Gamma_2)\! - \!m(\Gamma_3) \!+\! m(K_1)\! - m(K_2)\! - \!m(K'_1) \!+\! m(K'_2))^2 \\
&= 9 (\delta_{1c,1}^2 - \delta_{1c,1} \delta_{1c,2} + \delta_{1c,2}^2),
\eea
where we first used \Eq{eq:irrepdiffC3} to write the trace in terms of momentum space irreps, and then used the tables in \Ref{2020Sci...367..794S} to rewrite them in terms of the RSIs in \Eq{eq:c3rsis}.

Taking the Frobenius norm of \Eq{eq:TRIMpoints} and applying the triangle inequality to the real space side gives
\bea
\label{eq:pa1inequality}
||A|| \leq 3 (||p(\mbf{a}_1)|| \!+\!||p(C_3\mbf{a}_1)|| \!+\!||p(C_3^2\mbf{a}_1)||  \!+ \dots),
\eea
where the dots are higher harmonics and we used $||p(\mbf{R})|| = ||p(-\mbf{R})||$ which follows from \Eq{eq:Fourierharmonics_main}. We check explicitly that \Eq{eq:pa1inequality} is not a tight inequality (by a factor of 3) and prevents our bound from being saturated by this model. All other inequalities are tight.

We next use $C_3$ symmetry which ensures $||p(\mbf{R})|| = ||p(C_3\mbf{R})||$(\App{eq:TQC}). Because we have a lower bound for $||A||^2$, \Eq{eq:pa1inequality} proves that $||p(\mbf{R})|| \neq 0$ for some $\mbf{R} \neq 0$. We now employ an optimization argument:
\bea
\label{eq:opt}
\frac{1}{2\Omega_c}\sum_{\mbf{R}} |\mbf{R}|^2 ||p(\mbf{R})||^2 \geq \min_{\psi_\mbf{R}} \frac{1}{2\Omega_c} \sum_{\mbf{R}} |\mbf{R}|^2 |\psi_\mbf{R}|^2,
\eea
where the minimization is taken over all $\psi_\mbf{R} \in \mathbb{R}$ obeying
\bea
\label{eq:psicon}
\sum_\mbf{R} |\psi_\mbf{R}|^2 = 1, \ |\psi_\mbf{R}| = |\psi_{-\mbf{R}}|, \  ||A|| = 9 |\psi_{\mbf{a}_1}| + \dots,
\eea
and the dots denote terms depending on $\psi_{|\mbf{R}| > |\mbf{a}_1|}$. As such, the space of admissible $|\psi_\mbf{R}|$ described by \Eq{eq:psicon} includes the choice where $|\psi_\mbf{R}| = ||p(\mbf{R})||$. By keeping only the constraints in \Eq{eq:psicon}, and not the restriction that $\psi_\mbf{R}$ be the Fourier transform of a projection matrix, we can perform the minimization in \Eq{eq:opt} directly. 

A lemma we prove in \Ref{SM} shows that the minimum occurs when $||A|| = 9 |\psi_{\mbf{a}_1}|$, i.e., when \Eq{eq:pa1inequality} is saturated with the lowest harmonics possible. This is expected because higher harmonics have larger $|\mbf{R}|^2$ weights. Adding up the contributions from the inner six harmonics $||p(\mbf{R})||^2$ in \Fig{fig:inv_main}, we find (\App{app:SFWbounds})
\bea
\label{eq:boundrsis}
\frac{1}{2\Omega_c} \sum_{\mbf{R}}|\mbf{R}|^2 ||p(\mbf{R})||^2 \geq \frac{a^2}{9\Omega_c} (\delta_{1c,1}^2 - \delta_{1c,1} \delta_{1c,2} + \delta_{1c,2}^2) \ .
\eea
Plugging in $\delta_{1c,1} = \delta_{1c,2} = -1$ from \Eq{eq:c3rsis}, we obtain $G \geq a^2/9\Omega_c = 2/9\sqrt{3}$, a factor of 3 below the exact calculation in \Eq{eq:gmunuthird}. The RSIs in \Eq{eq:boundrsis} show that states off the atomic positions (1a in this case), which define OWCs, enforce a nonzero superfluid weight. We obtain bounds for all 2D space groups in \Ref{SM}.

\emph{Hubbard Model.}
We have shown that single-particle OWCs and fragile states have a nonzero superfluid weight at $T=0$. However, $[D_s]_{ij}$ in \Eq{eq:dsBCS} is obtained from mean-field BCS theory, which may seem unsuitable to treat flat band systems lacking a well-defined Fermi surface. We resort to exact numerical simulations to check its validity at finite temperature. 

Using the Hamiltonian $h(\mbf{k})$ defined in Eq.~\ref{eq: flatblandbloch}, we form a spinful Hamiltonian with $h_\uparrow(\mbf{k})=h(\mbf{k})$ and $h_\downarrow(\mbf{k})= \mathcal{T} h_\uparrow(\mbf{k}) \mathcal{T}^{-1} = h^*(-\mbf{k})$ which preserves time-reversal $\mathcal{T}$. Here $\uparrow,\downarrow$ label the spins. Including an attractive Hubbard term with strength $|U|$, the full Hamiltonian is
\bea
\label{eq:attractivehubbard}
H &= -|t| \sum_{\mbf{R},\sigma} w^\dag_{\mbf{R}\sigma}w_{\mbf{R}\sigma} -\lvert U \rvert\sum_{\mbf{R}\alpha} c^\dagger_{\mbf{R}\alpha\uparrow}c^\dagger_{\mbf{R}\alpha\downarrow}c^{\phantom{\dagger}}_{\mbf{R}\alpha\downarrow}c^{\phantom{\dagger}}_{\mbf{R}\alpha\uparrow},
\eea
where $w^\dag_{\mbf{R}\uparrow}$ creates the Wannier state in \Eq{eq:c3wan}, $w^\dag_{\mbf{R}\downarrow} = \mathcal{T} w^\dag_{\mbf{R},\uparrow} \mathcal{T}^{-1}$, $c^\dag_{\mbf{R}\alpha\sigma}$ is the creation operator in unit cell $\mbf{R}$, orbital $\alpha$, and spin $\sigma=\{\uparrow,\downarrow\}$. The attractive Hubbard model does not suffer from the fermionic sign problem, and lends itself to auxiliary-field quantum Monte Carlo methods \cite{Blankenbecler:1981,Bercx:2017}. We perform finite-temperature simulations in the grand canonical ensemble and tune the chemical potential $\mu(T)$ to half fill the $A_{1c}$ band.  We consider a range of Hubbard interactions $\lvert U \rvert$ smaller than the single-particle gap $|t|$ above the $A_{1c}$ band: $\lvert U \rvert = 3,4,5$, with $\lvert t\rvert=6$. These parameters set us away from the isolated flat band regime $\lvert U \rvert \ll \lvert t \rvert$. We focus on a system with $6\times 6$ unit cells and periodic boundary conditions.

\begin{figure}
 \centering
\includegraphics{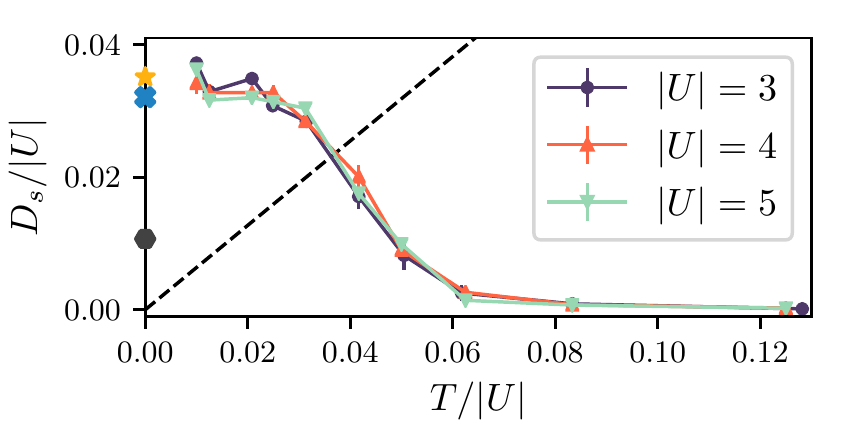}
\caption{Monte Carlo. The superfluid weight $D_s$ as a function of temperature $T$ is computed from Monte Carlo simulations on $H$ in \Eq{eq:attractivehubbard} \cite{Scalapino:1992,Scalapino:1993}. We consider $\lvert U \rvert = 3,4,5$ with $\lvert t \rvert = 6$ in a system with $6\times 6$ unit cells. The crossing of $D_s$ with the dashed line $2T/\pi$ indicates the Berezinskii-Kosterlitz-Thouless superconducting transition. The yellow star/ blue cross indicate the mean-field $D_s(T=0)$ obtained from a multi-band/ isolated flat band calculation (\Eq{eq:gmunuthird}). The gray hexagon shows the RSI bound on $D_s(T=0)$.}
\label{fig:MonteCarlo}
\end{figure}

We can directly extract the finite-temperature superfluid weight $D_s(T)$ from the Monte Carlo results (\App{app:SFMC}). The transition temperature $T_{c}$ is determined by the Nelson-Kosterlitz criterion \cite{PhysRevLett.39.1201}: $T_{c}=\pi D_s^-/2$, where $D_s^-$ is the superfluid weight at the critical temperature approached from below. In Fig.~\ref{fig:MonteCarlo}, we plot $D_s(T)$ for different $\lvert U \rvert$ as a function of $T/\lvert U \rvert$, finding the curves collapse on top of each other. This confirms $T_c\propto \lvert U\rvert$ \cite{Hofmann:2019}. Our results prove that a coherent superconductor emerges upon inclusion of an attractive Hubbard interaction in the OWC flat bands, as in topological bands \cite{Hofmann:2019,2021PhRvL.126b7002P}. \Ref{2021PhRvL.126b7002P} discusses the contrasting case of trivial atomic bands. 

We can compare the results of our Monte Carlo simulations to the zero temperature predictions of BCS theory. In particular, we recall in \Ref{SM} that the BCS wavefunction is an exact zero-temperature ground state of the attractive Hubbard model projected into the flat bands \cite{PhysRevB.94.245149}, as follows from the equal weight of the flat band's Wannier function over all orbitals in the unit cell \cite{2016PhRvL.117d5303J}. The blue cross in \Fig{fig:MonteCarlo} shows the result of the analytical mean-field calculation after projection into the flat band. Alternatively, we solve the multi-band mean-field theory numerically in \Ref{SM}. The result is shown with the yellow star in \Fig{fig:MonteCarlo}. The agreement between our finite-temperature Monte Carlo simulations and the zero temperature mean-field calculations justify the use of the BCS result in \Eq{eq:dsBCS}, showing that our lower bounds successfully describe the many-body physics.

\emph{Discussion.}
We have shown that the RSIs characterizing the quantum geometry have a profound influence on the interacting groundstate when the flat bands are partially filled. Our lower bound for the superfluid weight is nontrivial in OWCs where the Wannier charge centers are obstructed from the atoms. Our bounds are not saturated by the Hamiltonian in \Eq{eq: flatblandbloch}, but we hope that future work can improve these bounds to be tight. Our RSI bounds also apply to OWCs with corner states, as well as stable and fragile topological phases \cite{2017PhRvB..96x5115B,2017Sci...357...61B}. Conceptually, the gauge-invariant expression \Eq{eq:trgRS} in terms of the correlation function shows that long-ranged Wannier functions are \emph{not} essential to the lower bound. Any Wannier function which is supported over multiple unit cells \cite{2021arXiv210713556S,2018PhRvB..98m4513T}, as can be enforced by symmetry in a OWC state, produces a quantized RSI lower bound. Our derivation is general for arbitrary bands and arbitrary symmetries. Although we studied the problem in 2D, our method is generalizable to 3D where flat band OWCs have been exhaustively identified \cite{2021arXiv210605287R}.  

\acknowledgements

J.H-A. thanks Zhi-Da Song and Dylan King for useful discussions, and gratefully acknowledges insight P\"aivi T\"orm\"a. J.H-A. is supported by a Marshall Scholarship funded by the Marshall Aid Commemoration Commission. V.P., and S.D.H. acknowledge support from the Swiss National Science Foundation, the NCCR QSIT, the Swiss National Supercomputing Centre (CSCS) under project ID eth5b, and the European Research Council under the Grant Agreement No. 771503 (TopMechMat). F.S. was supported by a fellowship at the Princeton Center for Theoretical Science. B.A.B thanks funding from the European Research Council under the Grant Agreement no. 101020833 (SuperFlat). B.A.B. was also supported by the U.S. Department of Energy (Grant No. DE-SC0016239) and was partially supported by the National Science Foundation (EAGER Grant No. DMR 1643312), a Simons Investigator grant (No. 404513), the Office of Naval Research (ONR Grant No. N00014-20-1-2303), the Packard Foundation, the Schmidt Fund for Innovative Research, the BSF Israel US foundation (Grant No. 2018226), the Gordon and Betty Moore Foundation through Grant No. GBMF8685 towards the Princeton theory program, a Guggenheim Fellowship from the John Simon Guggenheim Memorial Foundation, and the NSF-MRSEC (Grant No. DMR2011750). The auxiliary-field QMC simulations were carried out with the ALF package available at \href{https://git.physik.uni-wuerzburg.de/ALF/ALF}{https://git.physik.uni-wuerzburg.de/ALF/ALF}.

\let\oldaddcontentsline\addcontentsline
\renewcommand{\addcontentsline}[3]{}
\bibliography{finalbib}

\begin{thebibliography}{102}
\bibitem[foo()]{footnote1}
Note that the off-diagonal $g_{12}$ term do not appear in the trace, but this
  term is only relevant in highly anisotropic systems.
\end{thebibliography}
\bibliographystyle{aipnum4-1}
\bibliographystyle{unsrtnat}
\let\addcontentsline\oldaddcontentsline

\setcitestyle{numbers,square}

\onecolumngrid
\appendix

\tableofcontents

\section{Symmetries and Topological Quantum Chemistry}
\label{eq:TQC}

In this Appendix, we set up the gauge-invariant formalism for obtaining the single-particle symmetry data, defined by the irreps of the occupied bands in momentum space. \App{app:symreal} discusses the electron orbitals in real space which are used to define the space group symmetries and Hamiltonian. In \App{app:symmom}, we work in momentum space to describe the occupied bands and their symmetry properties. \App{app:bandrep} discusses the symmetry data, paying special attention to formulas invariant under the eigenvector gauge freedom.

\subsection{Symmetries in Real Space}
\label{app:symreal}

We review the symmetries of a tight-binding model Hamiltonian from a real space perspective. The canonical electron operators $c^\dag_{\mbf{R}, \al}$ are indexed by their unit cell $\mbf{R} = R_1 \mbf{a}_1 + R_2 \mbf{a}_2, R_i \in \mathds{Z}$ and their orbital index $\al = 1, \dots, N_{\mathrm{orb}}$. The area of the unit cell is $\Omega_c = \mbf{a}_1 \times \mbf{a}_2$. We denote the position of the orbitals with the unit cell as $\mbf{R} + \mbf{r}_\al$. We can think of the orbitals $c^\dag_{\mbf{R},\al}$ as being the atomic orbitals, so $\mbf{r}_\al$ are the positions of the atoms of the crystal. The orbitals $c^\dag_{\mbf{R},\al}$ define the tight-binding Hilbert space and carry information about the positions of the underlying orbitals.

We consider 2D systems with a space group $G$ generated by lattice translations $T_i$, $n$-fold rotations $C_n$, mirrors $M$, and possibly time-reversal symmetry $\mathcal{T}$ which is an anti-unitary operator. In real space, the symmetry operators are defined by their representation on the orbitals $c^\dag_{\mbf{R},\al}$. For a symmetry $g\in G$, we write
\bea
\label{eq:defineDg}
g c^\dag_{\mbf{R},\al} g^\dag = \sum_\be c^\dag_{g(\mbf{R}+ \mbf{r}_\al) - \mbf{r}_\be,\be} D_{\be \al}[g]
\eea
where $D[g]$ is the $N_{\mathrm{orb}} \times N_{\mathrm{orb}}$ representation matrix of $g$ on the orbitals. If $g$ is a symmetry of the lattice, then $D_{\be \al}[g]$ is nonzero only when $g\mbf{r}_\al - \mbf{r}_\be$ is a lattice vector, i.e. $g\mbf{r}_\al - \mbf{r}_\be \mod \mbf{a}_i = 0$. The high-symmetry Wyckoff positions can be found for all space groups (and thus their 2D subgroups) on the \href{https://www.cryst.ehu.es/}{Bilbao Crystallographic Server}.

The non-interacting Hamiltonian of the crystal can be written in terms of the hopping matrix $t_{\al \be}(\mbf{R})$ via
\bea
H = \sum_{\mbf{R},\mbf{R}',\al \be} t_{\al \be}(\mbf{R}-\mbf{R}') c^\dag_{\mbf{R},\al} c_{\mbf{R}',\be} \\
\eea
which manifestly respects translation symmetry. The symmetries $g \in G$ commute with $H$, and hence
\bea
g H g^\dag &= \sum_{\mbf{R},\mbf{R}',\al \be} D_{\al' \al}[g]  t_{\al \be}(\mbf{R}-\mbf{R}') D^\dag_{\be \be'}[g] c^\dag_{g(\mbf{R}+\mbf{r}_\al)-\mbf{r}_{\al'},\al'} c_{g(\mbf{R}'+\mbf{r}_\be)-\mbf{r}_{\be'},\be'} \\
&= \sum_{\mbf{S},\mbf{S}',\al \be} D_{\al' \al}[g]  t_{\al \be}((g^{-1}(\mbf{S}+\mbf{r}_{\al'}) - \mbf{r}_\al)- (g^{-1}(\mbf{S}'+\mbf{r}_{\be'}) - \mbf{r}_\be) ) D^\dag_{\be \be'}[g] c^\dag_{\mbf{S},\al'} c_{\mbf{S}',\be'} \\
\eea
and thus we require
\bea
t_{\al' \be'}(\mbf{R}-\mbf{R}') &= \sum_{\al \be} D_{\al' \al}[g] t_{\al \be}(g^{-1}(\mbf{R}+\mbf{r}_{\al'}) - \mbf{r}_\al - (g^{-1}(\mbf{R}'+\mbf{r}_{\be'}) - \mbf{r}_\be) ) D^\dag_{\be \be'}[g] \ .
\eea
In order to simplify the expressions, we sum over repeated indices from here on.

\subsection{Symmetries in Momentum Space}
\label{app:symmom}

To diagonalize the Hamiltonian, we define momentum space electron operators which are eigenstates of the translation operators. On periodic boundary conditions with $L_1L_2$ total sites, we define
\bea
c^\dag_{\mbf{k},\al} = \frac{1}{\sqrt{L_1L_2}} \sum_{\mbf{R}} e^{- i \mbf{k} \cdot (\mbf{R}+\mbf{r}_\al)} c^\dag_{\mbf{R},\al}
\eea
which is a Fourier transform over the lattice, and $\mbf{k} = k_1 \mbf{b}_1 + k_2 \mbf{b}_2, k_i \in \frac{2\pi}{L_i} \mathds{Z}_{L_i}$.  Note that in this convention, the momentum operators obey
\bea
c^\dag_{\mbf{k}+2\pi \mbf{b}_i,\al} &= e^{- 2\pi i \mbf{b}_i \cdot \mbf{r}_\al} c^\dag_{\mbf{k},\al} \equiv c^\dag_{\mbf{k},\be} V_{\be \al}[2\pi\mbf{b}_i], \qquad V_{\be \al}[2\pi\mbf{b}_i] = \delta_{\be \al} e^{- 2\pi i \mbf{b}_i \cdot \mbf{r}_\al} \ .
\eea
where we defined the embedding matrix $V[2 \pi \mbf{b}_i]$. The embedding matrix will play a very important role in this work because it encodes the position of the orbitals in the unit cell and hence is important for defining obstructed atomic insulators (OWCs).

We also need to compute the action of $g\in G$ on the momentum operators:
\bea
g c^\dag_{\mbf{k},\al} g^\dag &= \frac{1}{\sqrt{N}} \sum_{\mbf{R}} e^{- i \mbf{k} \cdot (\mbf{R} + \mbf{r}_\al)} c^\dag_{g(\mbf{R}+ \mbf{r}_\al) - \mbf{r}_\be,\be} D_{\be \al}[g] \\
&=  \frac{1}{\sqrt{N}} \sum_{\mbf{R}} e^{- i \mbf{k} \cdot g^{-1}(\mbf{R}+ \mbf{r}_\be) } c^\dag_{\mbf{R},\be} D_{\be \al}[g] \\
&= c^\dag_{g\mbf{k},\be}  D_{\be \al}[g] \ . \\
\eea
We now write the Hamiltonian in momentum space. Using the translation invariance of $t_{\al \be}(\mbf{R}-\mbf{R}')$, we find
\bea
H &= \frac{1}{N} \sum_{\mbf{k},\mbf{k}'} \sum_{\mbf{R},\mbf{R}',\al \be} e^{- i \mbf{k} \cdot (\mbf{R} + \mbf{r}_\al)+ i \mbf{k}' \cdot (\mbf{R}'+\mbf{r}_\be)} t_{\al \be}(\mbf{R}-\mbf{R}') c^\dag_{\mbf{k},\al} c_{\mbf{k}',\be} \\
&= \sum_{\mbf{k},\al \be} \lp \sum_{\mbf{d}} e^{- i \mbf{k} \cdot (\mbf{d} + \mbf{r}_\al - \mbf{r}_\be)} t_{\al \be}(\mbf{d}) \rp c^\dag_{\mbf{k},\al} c_{\mbf{k},\be} \\
&= \sum_{\mbf{k}} c^\dag_{\mbf{k},\al} h_{\al \be}(\mbf{k}) c_{\mbf{k},\be}  \\
\eea
where the $\al,\be$ sums are implicit in the last line. We call $h(\mbf{k})$ the single-particle or first-quantized Hamiltonian, which is the Fourier transform of the hopping matrix $t_{\al \be}(\mbf{d})$.

Note that $h(\mbf{k})$ obeys the embedding relations
\bea
h_{\al \be}(\mbf{k}+2\pi \mbf{b}_i) &= \sum_{\mbf{d}} e^{- i (\mbf{k}+2\pi \mbf{b}_i) \cdot (\mbf{d} + \mbf{r}_\al - \mbf{r}_\be)} t_{\al \be}(\mbf{d}) \\
&= e^{- 2\pi i \mbf{b}_i \cdot ( \mbf{r}_\al - \mbf{r}_\be)}  \sum_{\mbf{d}} e^{- i \mbf{k} \cdot (\mbf{d} + \mbf{r}_\al - \mbf{r}_\be)} t_{\al \be}(\mbf{d}) \\
&=  e^{- 2\pi i \mbf{b}_i \cdot \mbf{r}_\al} h_{\al \be}(\mbf{k})  e^{2\pi i \mbf{b}_i \cdot \mbf{r}_\be}  \\
&= \left[ V[2\pi\mbf{b}_i] h(\mbf{k}) V^\dag[2\pi\mbf{b}_i] \right]_{\al \be}
\eea
which can be written simply in matrix notation as $h(\mbf{k} + 2\pi \mbf{b}_i) = V[2\pi\mbf{b}_i] h(\mbf{k}) V^\dag[2\pi\mbf{b}_i]$. We write $V[\mbf{G}]$ with brackets to emphasize that the embedding matrix is defined with $\mbf{G}$ a reciprocal lattice vector.

We now compute the action of the symmetries on the single-particle Hamiltonian. We find
\bea
g H g^\dag &= \sum_{\mbf{k}} c^\dag_{g\mbf{k},{\al'}}  D_{\al' \al}[g]  h_{\al \be}(\mbf{k})  D^\dag_{\be \be'}[g]  c_{g\mbf{k},\be'}  \\
&= \sum_{\mbf{k}} c^\dag_{\mbf{k},{\al'}} D_{\al' \al}[g] h_{\al \be}(g^{-1} \mbf{k})  D^\dag_{\be \be'}[g]  c_{\mbf{k},\be'}  \\
\eea
and thus we derive $D[g] h(g^{-1} \mbf{k}) D^\dag[g] = h(\mbf{k})$. At high symmetry momenta $\mbf{K}$ where $g \mbf{K} = \mbf{K} \mod 2\pi \mbf{b}_i$, we find that there are symmetries which act locally on $\mbf{K}$. To give explicit expressions, we define $\mbf{G} = g\mbf{K} - \mbf{K}$ and find
\bea
D[g] h(\mbf{K}) D^\dag[g] &= h(g\mbf{K}) = h(\mbf{K}+ \mbf{G}) = V[\mbf{G}] h(\mbf{K}) V[\mbf{G}]^\dag
\eea
so the unitary matrix $V[\mbf{G}]^\dag D[g]$ commutes with $h(\mbf{K})$. Thus $h(\mbf{K})$ has a nontrivial symmetry group which is conventionally called the little group $G_{\mbf{K}}$. We need to prove that $V[\mbf{G}]^\dag D[g]$ is a good representation of $g$, i.e. that it obeys the group multiplication. To do so, we need the following identity
\bea
\label{eq:DgVgproof}
D[g] V[ \mbf{G}] D[g]^\dag &=  V[g \mbf{G}]  \qquad \text{for } \mbf{G} \text{ a reciprocal lattice vector.}
\eea
This is proven by direct calculation (indices unsummed):
\bea
\label{eq:DgVg}
\null [D[g] V[\mbf{G}]]_{\al \be} &=  D_{\al \be}[g] e^{- i \mbf{G} \cdot \mbf{r}_{\be}} \\
\null [V[g\mbf{G}] D[g] ]_{\al \be} &=  e^{- i g\mbf{G} \cdot \mbf{r}_{\al}} D_{\al \be}[g] \ .  \\
\eea
Recall that $D_{\al \be}[g]$ is only nonzero when $g \mbf{r}_\be - \mbf{r}_{\al} = \mbf{R}$ where $\mbf{R}$ is a lattice vector. Using $\mbf{G} \cdot \mbf{R} = 0 \mod 2\pi$ because $\mbf{G}$ is a reciprocal lattice vector, we find $e^{- i \mbf{G} \cdot \mbf{r}_{\be}} = e^{- i \mbf{G} \cdot g^{-1} \mbf{r}_{\al}}  = e^{- ig\mbf{G} \cdot \mbf{r}_{\al}} $
and hence the first and second lines of \Eq{eq:DgVg} are equal. This proves \Eq{eq:DgVgproof}. Now we can show that the matrices $V[\mbf{G}]^\dag D[g]$ do indeed form a representation of the little group at $\mbf{K}$. For $\mbf{G} = g\mbf{K} - \mbf{K}$ and $\mbf{G}' = g'\mbf{K} - \mbf{K}$, we check
\bea
\label{eq:orbitalrep}
\lp V[\mbf{G}]^\dag D[g] \rp\lp V[\mbf{G}']^\dag D[g'] \rp &=  V[\mbf{G}]^\dag V[g\mbf{G}']^\dag D[g]  D[g'] \\
&=  V[\mbf{G} +g\mbf{G}']^\dag D[g g'] \\
&=  V[ g\mbf{K} - \mbf{K} + g (g'\mbf{K} - \mbf{K}) ']^\dag D[g g'] \\
&=  V[gg' \mbf{K} - \mbf{K}]^\dag D[g g'] \\
\eea
which satisfies the group multiplication so, $V[\mbf{G}]^\dag D[g]$ is a good representation of the little group. Thus we have shown that the high-symmetry points on the BZ have nontrivial symmetries with $N_{\mathrm{orb}} \times N_{\mathrm{orb}}$ representation matrices.

\subsection{Symmetry data}
\label{app:bandrep}

Having discussed the symmetry group of the momentum space Hamiltonian, we now focus on the representations at the high symmetry points of a general set of gapped bands. To do so, we first need to discuss the band structure.

Because $h(\mbf{k})$ is a Hermitian matrix, it has a spectral decomposition
\bea
h_{\al \be}(\mbf{k}) &= \sum_n U_{\al n}(\mbf{k}) E_n(\mbf{k}) U^*_{\be,n}(\mbf{k})
\eea
where $U_n(\mbf{k})$ is a column eigenvector of $h(\mbf{k})$ with eigenvalue $E_n(\mbf{k})$. The spectrum $E_n(\mbf{k})$ is referred to as the band structure. By defining the energy eigenstates through a unitary transition $\gamma^\dag_{\mbf{k},n} = c^\dag_{\mbf{k},\al} U_{\al,n}(\mbf{k})$, the Hamiltonian can be put into diagonal form
\bea
H &= \sum_{\mbf{k},n} E_n(\mbf{k}) \gamma^\dag_{\mbf{k},n} \gamma_{\mbf{k},n} \\
\eea
We say that the Hamiltonian has a gap at filling $N_{\mathrm{occ}}$ if $E_{N_{\mathrm{occ}}+1}(\mbf{k}) - E_{N_{\mathrm{occ}}}(\mbf{k}) > 0$ for all $\mbf{k}$. States in the occupied bands (which are uniquely defined due to the gap) are all filled in the many-body ground state of the Hamiltonian. We now show that the occupied bands furnish representations of the little group as well \cite{2018PhRvB..97c5139C,PhysRevB.23.2824}.

For ease of notation, we define $U(\mbf{k}) = [U_1(\mbf{k}), \dots, U_{N_{\mathrm{occ}}}(\mbf{k})]$ as the $N_{\mathrm{orb}} \times N_{\mathrm{occ}}$ matrix of occupied eigenvectors. Because of the orthonormality of the eigenvectors, $U(\mbf{k})$ satisfies the important properties
\bea
\label{eq:Uprop}
U^\dag(\mbf{k}) U(\mbf{k}) = \mathbb{1}_{N_{\mathrm{occ}} \times N_{\mathrm{occ}}}, \qquad U(\mbf{k}) U^\dag(\mbf{k}) \equiv P(\mbf{k})
\eea
where $P(\mbf{k})$ is a Hermitian projector onto the occupied bands. It satisfies $P(\mbf{k})^2 = P(\mbf{k})$ and $P(\mbf{k}) U(\mbf{k}) = U(\mbf{k})$ as is direct to check from \Eq{eq:Uprop}.

Let us study the action of the symmetries at $\mbf{K}$ on the occupied bands. Because $V[g\mbf{K}-\mbf{K}]^\dag D[g]$ commutes with $h(\mbf{K})$ for all $g \in G_{\mbf{K}}$, $V[g\mbf{K}-\mbf{K}]^\dag D[g] U_n(\mbf{k})$ is an eigenstate of $h(\mbf{K})$ with energy $E_n(\mbf{k})$. If we take $n$ to be in the occupied bands, then $V[g\mbf{K}-\mbf{K}]^\dag D[g] U_n(\mbf{k})$ can be expanded using the basis of occupied states at $\mbf{K}$, so
\bea
V[\mbf{G}]^\dag D[g] U(\mbf{K}) = U(\mbf{K}) \mathcal{D}_{\mbf{K}}[g]
\eea
where $\mathcal{D}_{\mbf{K}}[g]$ is an $N_{\mathrm{occ}} \times N_{\mathrm{occ}}$ matrix. At generic $\mbf{k}$ points, $\mathcal{D}_{\mbf{K}}[g]$ is called the sewing matrix. We use the notation $\mathcal{D}$ at the high-symmetry points to emphasize that it is a representation (proved momentarily).

An expression for $\mathcal{D}_{\mbf{K}}[g]$ is obtained by left-multiplying $U^\dag(\mbf{k})$:
\bea
\label{eq:Drepmat}
\mathcal{D}_{\mbf{K}}[g] &= U^\dag(\mbf{K}) V[\mbf{G}]^\dag D[g] U(\mbf{K})  \ .
\eea
We now check that $\mathcal{D}_{\mbf{K}}[g] $ is unitary:
\bea
\mathcal{D}_{\mbf{K}}[g]^\dag \mathcal{D}_{\mbf{K}}[g] &= U^\dag(\mbf{K}) (V[\mbf{G}]^\dag D[g])^\dag U(\mbf{K}) U^\dag(\mbf{K}) V[\mbf{G}]^\dag D[g] U(\mbf{K})  \\
&= U^\dag(\mbf{K}) (V[\mbf{G}]^\dag D[g])^\dag  V[\mbf{G}]^\dag D[g]U(\mbf{K})  \\
&= U^\dag(\mbf{K}) U(\mbf{K})  \\
&= \mathbb{1}_{N_{\mathrm{occ}} \times N_{\mathrm{occ}}} \ .
\eea
The crucial step $U(\mbf{K})U^\dag(\mbf{K}) V[\mbf{G}]^\dag D[g] U(\mbf{K})  = P(\mbf{K}) V[\mbf{G}]^\dag D[g] U(\mbf{K}) =  V[\mbf{G}]^\dag D[g] U(\mbf{K})$ in the second equality follows because $V[\mbf{G}]^\dag D[g] U(\mbf{K})$ is in the occupied subspace --- note that $V[\mbf{G}]^\dag D[g]$ commutes with $h(\mbf{K})$. Thus the projector $P(\mbf{K})$ acts on $V[\mbf{G}]^\dag D[g] U(\mbf{K})$ as the identity. The requirement of a gap is essential here for the occupied subspace to be well-defined. In fact with this requirement, $ \mathcal{D}_{\mbf{K}}[g] $ forms a representation of the little group $G_{\mbf{K}}$. This follows from a very similar calculation
\bea
\mathcal{D}_{\mbf{K}}[g] \mathcal{D}_{\mbf{K}}[g'] &= U^\dag(\mbf{K}) V[\mbf{G}]^\dag D[g] U(\mbf{K}) U^\dag(\mbf{K}) V[\mbf{G}']^\dag D[g'] U(\mbf{K})  \\
&= U^\dag(\mbf{K}) V[\mbf{G}]^\dag D[g] V[\mbf{G}']^\dag D[g'] U(\mbf{K})  \\
&= U^\dag(\mbf{K}) V[\mbf{G}'']^\dag D[gg'] U(\mbf{K}), \qquad \mbf{G}'' = gg' \mbf{K} - \mbf{K}  \\
&= \mathcal{D}_{\mbf{K}}[gg']
\eea
where we used \Eq{eq:orbitalrep} in the third line. As is suggested by \Eq{eq:Drepmat}, $\mathcal{D}_{\mbf{K}}[g]$ is simply the representation of the orbitals projected into the flat bands. As long as there is a gap, this projection still results in a well-defined representation. Note that $\mathcal{D}_{\mbf{K}}[g]$ is a $N_{\mathrm{occ}}$-dimensional representation and hence is decomposable into irreps. The multiplicity of the $\chi$ irrep is
\bea
\label{eq:BRtrace}
m(\chi) = \frac{1}{|G_{\mbf{K}}|} \sum_{g\in G_\mbf{K}} \chi^*[g] \Tr \mathcal{D}_{\mbf{K}}[g]
\eea
where $|G_{\mbf{K}}|$ is the number of elements of $G_{\mbf{K}}$. A list of irreps and their characters $\chi[g]$ in all the 2D little groups may be found at the \href{https://www.cryst.ehu.es/rep/point.html}{Bilbao crystallographic server}. The symmetry data vector $B$ is simply a list of the irrep multiplicities at all high symmetry points in the BZ \cite{2020arXiv200604890C,2017Natur.547..298B}. $B$ is an invariant of the occupied subspace: the irrep multiplicities cannot be changed unless a gap is closed. As such, some (but not all) topological invariants can be computed from the symmetry data alone to diagnose stable and fragile topological phases \cite{2020PhRvB.102c5110E}.

We conclude this section with a discussion of the gauge freedom of the occupied eigenvectors. Recall that an individual eigenvector is only well-defined up to an overall phase factor. Thus there is the gauge freedom $U_n(\mbf{k}) \to U_n(\mbf{k}) e^{i \la_n(\mbf{k})} $ at all $\mbf{k}$. If there are degeneracies, then the eigenvectors are only defined up to a unitary transformation that mixes the degenerate bands. In the flat band limit where all occupied bands are degenerate everywhere, the occupied eigenvectors are only defined up to the gauge transformation $U(\mbf{k}) \to U(\mbf{k}) \mathcal{W}(\mbf{k})$ where $\mathcal{W}(\mbf{k})$ is an arbitrary $N_{\mathrm{occ}} \times N_{\mathrm{occ}}$ unitary matrix. (This general case includes the non-degenerate case. If there are no degeneracies then $\mathcal{W}(\mbf{k})$ is a diagonal matrix of phases.)

It is important to understand the effect of the gauge freedom on the expressions derived in this section. Notably, the little group representation $\mathcal{D}_{\mbf{K}}[g]$ is not gauge-invariant, as can be seen from \Eq{eq:Drepmat}. Under $U(\mbf{k}) \to U(\mbf{k}) \mathcal{W}(\mbf{k})$, we find $\mathcal{D}_{\mbf{K}}[g] \to \mathcal{W}(\mbf{k})^\dag \mathcal{D}_{\mbf{K}}[g] \mathcal{W}(\mbf{k})$. This is expected because the explicit matrices of a representation are not unique. However their character, or equivalently the irrep multiplicities \emph{are} unique because \Eq{eq:BRtrace} is defined in terms of $\Tr \mathcal{D}_{\mbf{K}}[g]$ which is gauge-invariant.

Another gauge-invariant object is the projector matrix which plays a central role in \App{app:qgm}. We see that
\bea
P(\mbf{k}) = U(\mbf{k}) U^\dag(\mbf{k}) \to U(\mbf{k}) \mathcal{W}(\mbf{k}) \mathcal{W}(\mbf{k})^\dag U^\dag(\mbf{k}) = P(\mbf{k})
\eea
is invariant under the gauge freedom. All quantities written in terms of $P(\mbf{k})$ can be evaluated numerically without needing to choose a smooth gauge for the eigenvectors. As such, it is useful to have an expression for the irrep multiplicities directly in terms of the projector. This is easily derived from \Eq{eq:BRtrace} with the cyclicity of the trace:
\bea
m(\chi) &= \frac{1}{|G_{\mbf{K}}|} \sum_{g \in G_\mbf{K}} \chi^*[g] \Tr V[\mbf{G}]^\dag D[g] U(\mbf{K}) U^\dag(\mbf{K}) = \frac{1}{|G_{\mbf{K}}|} \sum_{g \in G_\mbf{K}} \chi^*[g] \Tr V[\mbf{G}]^\dag D[g]  P(\mbf{K}) \ . \\
\eea
Using the character orthogonality theorems, we can invert the expression to find that the trace of $\Tr V[\mbf{G}]^\dag D[g]  P(\mbf{K})$ is determined by the characters of the representation:
\bea
\label{eq:trPtoirrep}
\sum_\chi m(\chi) \chi[g] &= \frac{N_\chi}{|G_{\mbf{K}}|} \Tr V[\mbf{G}]^\dag D[g]  P(\mbf{K}), \qquad \mbf{G} = g \mbf{K}-\mbf{K}
\eea
where $N_\chi$ is the number of irreps (equivalently, the number of conjugacy classes). \Eq{eq:trPtoirrep} will be directly applicable to the lower bounds in \App{app:SFWbounds}. We prove two simple properties of the projectors $P(\mbf{k})$ which will be useful there as well. First we study the behavior of $P(\mbf{k})$ under the spatial symmetries. Because $h(\mbf{k}) = D^\dag[g] h(g\mbf{k}) D[g]$, the eigenvectors at symmetry related points $g\mbf{k}$ and $\mbf{k}$ are related by $U(g^{-1} \mbf{k}) = D^\dag_{\mbf{k}}[g] U(\mbf{k}) \mathcal{B}(\mbf{k}) $ where $\mathcal{B}(\mbf{k})$ is a unitary sewing matrix \cite{2017PhRvB..96x5115B}. At the high-symmetry points, the sewing matrix is the representation matrix in \Eq{eq:Drepmat}. The projector transforms simply as
\bea
P(g^{-1} \mbf{k}) = U(g^{-1} \mbf{k}) U^\dag(g^{-1} \mbf{k}) = D^\dag[g] U(\mbf{k}) \mathcal{B}(\mbf{k}) \mathcal{B}^\dag(\mbf{k}) U^\dag(\mbf{k}) D[g]  = D^\dag[g] P(\mbf{k}) D[g]
\eea
so $P(g^{-1} \mbf{k})$ and $P(\mbf{k})$ are unitarily related. In the same manner, we check the periodicity of $P(\mbf{k})$ on the BZ:
\bea
\label{eq:Vperiod}
P(\mbf{k} + 2\pi \mbf{b}_i) = V[2\pi \mbf{b}_i] P(\mbf{k}) V^\dag[2\pi \mbf{b}_i]
\eea
which shows that $P(\mbf{k})$ is only periodic up to a unitary transform when the orbitals of the model are at different positions.

\section{Gauge-Invariant Quantum Geometry}
\label{app:qgm}

In this Appendix, we discuss the quantum geometry of the occupied eigenstates. In \App{app:QGT}, we set our index conventions and introduce the quantum geometric tensor as a natural gauge-invariant object in which the non-abelian Berry curvature and non-abelian Fubini-Study metric are contained. \App{app:projectors} gives simple expressions for the abelian Berry curvature and abelian Fubini-Study metric in terms of the projector matrices, making them manifestly gauge invariant. \App{app:SFWformula} derives point-split formula for the abelian Fubini-Study metric which is suitable for numerical implementation, in analogy of the projector formula for Wilson loops.

\subsection{Quantum Geometric Tensor}
\label{app:QGT}
We now take the lattice period $L_i \to \infty$ so that we work on infinite boundary conditions and the momentum $k_i \in (-\pi, \pi]$ is a continuous variable in the Brillouin zone (BZ). The BZ is a smooth manifold, and in 2D is a torus. As in \App{app:bandrep}, we define $U(\mbf{k})$ as the $N_{\mathrm{orb}} \times N_{\mathrm{occ}}$ matrix whose columns are the eigenvectors of the occupied bands. Because of the gauge freedom, $U(\mbf{k})$ is only defined up to $U(\mbf{k}) \to U(\mbf{k}) \mathcal{W}(\mbf{k})$ where $\mathcal{W}(\mbf{k})$ is an arbitrary $N_{\mathrm{occ}} \times N_{\mathrm{occ}}$ unitary matrix in the most generic degenerate flat band case. The projector $P(\mbf{k}) = U(\mbf{k})U^\dag(\mbf{k})$ is gauge invariant. Going forward, we suppress the $\mbf{k}$ dependence in our notation. We let $\del_\mu$ denote the momentum derivative $\frac{\del}{\del k^\mu}$ where $\mu$ is a spatial index on the BZ. The BZ has a spatial metric $\eta_{\mu \nu}$ inherited from real space. In general coordinate systems, $|\mbf{k}|^2 = \eta_{\mu \nu} k^\mu k^\nu$ and we sum over repeated indices. On the BZ, the natural coordinate system is crystalline coordinates. Denoting the real space lattice vectors by $\mbf{a}^i = a_\mu^i \hat{x}^\mu$, the BZ is parameterized by $\mbf{k} = k^i \mbf{b}_i = k^i b^\mu_i \hat{x}_\mu$ where the reciprocal lattice vectors satisfy $\mbf{a}^i \cdot \mbf{b}_j = \delta^i_j$. In indices, we have the identities $a^i_\mu b_j^\mu = \delta^i_j$ and $a^i_\mu b_i^\nu = \delta^\nu_\mu$. The metric $\eta_{ij}$ in crystalline coordinates is derived from $|\mbf{k}|^2 =(\delta_{\mu \nu} b^\mu_i b^\nu_j) k^i k^j $, so $\eta_{ij} = \mbf{b}_i \cdot \mbf{b}_j$. We define the inverse metric $\eta^{ij}$ by $\eta^{ij} \eta_{jk} = \delta^i_k$. 
For concreteness, the two most often used coordinate systems are
\bea
\mbf{b}_1 = (1,0), \mbf{b}_2 = C_4 \mbf{b}_1, & \quad \eta_{ij} = \bpm 1 & 0 \\ 0 & 1 \\ \epm \\
\mbf{b}_1 = (1,0), \mbf{b}_2 = C_3 \mbf{b}_1 , & \quad \eta_{ij} = \bpm 1 & -\frac{1}{2} \\ -\frac{1}{2} & 1 \\ \epm \\
\eea
for $C_2,C_4$ -symmetric lattices and $C_3,C_6$-symmetric lattices respectively. Here we set the reciprocal lattice constants to one.

The projector $P(\mbf{k})$ is a simple gauge-invariant object on the BZ. It has a natural geometric interpretation since it describes the $N_{\mathrm{occ}}$-dimensional occupied subspace within the $N_{\mathrm{orb}}$ dimensional Hilbert space. We will now derive another gauge-invariant object, the abelian quantum geometric tensor, with two derivatives. In 2D, a two derivative tensor is fundamental because it is dimensionless when integrated on the BZ. Indeed, we will see that the Berry curvature and quantum metric appear in this construction.

Our starting point is to use $P(\mbf{k})$ to define a covariant derivative $(1-U(\mbf{k}) U^\dag(\mbf{k})) \del_i U(\mbf{k})$. (Here covariant refers to the eigenvector gauge freedom, not to the simple transformation of the spatial index $i$.) Under a gauge transformation of the $N_{occ}$ eigenvectors $U \to U \mathcal{W}$, we find
\bea
\label{eq:covder}
(1-U U^\dag) \del_i U \to (1-U U^\dag) \del_i (U \mathcal{W}) &=  (1-U U^\dag) (\del_i U \mathcal{W} + U  \del_i \mathcal{W}) \\
&= \big( (1-U U^\dag) \del_i U \big)\, \mathcal{W} \\
\eea
where in the last line we used the projector property $(1-U U^\dag) U = U (1- U^\dag U) = 0$. Note that the naive partial derivative $\del_i U$ does not transform covariantly. Now we can define two-derivative a gauge-covariant $N_{\mathrm{occ}} \times N_{\mathrm{occ}}$ matrix called the quantum geometric tensor:
\bea
\label{eq:QGTu}
\mathcal{G}_{ij}(\mbf{k}) &= \del_i U^\dag (1 - U U^\dag) \del_j U
\eea
which can be written in terms of the covariant derivatives ($\mbf{k}$ dependence suppressed) as $\mathcal{G}_{ij}  = \big( (1-U U^\dag) \del_j U \big)^\dag (1-U U^\dag) \del_i U$ because $(1-U U^\dag)^2 = (1-U U^\dag)$. Hence $\mathcal{G}_{ij}\to \mathcal{W}^\dag \mathcal{G}_{ij} \mathcal{W}$ under gauge transformations, and we find that the \emph{abelian} quantum geometric tensor defined by
\bea
\Tr \mathcal{G}_{ij}(\mbf{k}) &\to \Tr \mathcal{W}^\dag(\mbf{k}) \mathcal{G}_{ij}(\mbf{k}) \mathcal{W}(\mbf{k}) = \Tr \mathcal{G}_{ij}(\mbf{k})
\eea
is gauge invariant. In the next section, we give an expression for $\Tr \mathcal{G}_{ij}(\mbf{k})$ in terms of gauge-invariant projectors.

\subsection{Berry Curvature and Quantum Metric in terms of Projectors}
\label{app:projectors}

It is desirable to have expressions for the abelian quantum geometric tensor (the trace of \Eq{eq:QGTu}) in terms of projector matrix which is the natural gauge-invariant object. We need the simple identity $(\del_i U^\dag) U = - U^\dag \del_i U$ which follows from $\del_i (U^\dag U) = \del_i \mathbb{1} = 0$. With this, a direct calculation establishes that
\bea
\del_i P &= \del_i U \, U^\dag + U \del_i U^\dag \\
(\del_i P) U&= \del_i U  + U (\del_i U^\dag) U = \del_i U  - U U^\dag \del_i U = (1- U U^\dag) \del_i U \\
\eea
which is the covariant derivative in \Eq{eq:covder}. From \Eq{eq:QGTu} it follows that $\mathcal{G}_{ij} = U^\dag \del_i P  \del_j P \, U $ and thus
\bea
\Tr \mathcal{G}_{ij} &= \Tr U^\dag \del_i P  \del_j P \, U =  \Tr U  U^\dag \del_i P  \del_j P = \Tr P \del_i P  \del_j P \
\eea
from the cyclicity of the trace. This expression for $\Tr \mathcal{G}_{ij}$ is manifestly gauge invariant. Note that $\Tr \mathcal{G}^\dag_{ij} =  \Tr \del_j P  \del_i P  P = \Tr \mathcal{G}_{ji}$ so the symmetric part of $\Tr \mathcal{G}_{ij}$ is real and the anti-symmetric part is imaginary. We now separate $\Tr G_{ij}$ into symmetric and anti-symmetric parts:
\bea
\Tr \mathcal{G}_{ij} &= g_{ij} - \frac{i}{2} f_{ij}, \qquad g_{ij} =  \frac{1}{2} \Tr P \{\del_i P, \del_j P \}, \quad f_{ij} = i \Tr P [\del_i P, \del_j P] \\
\eea
where $g_{ij}$ is the quantum metric and $f_{ij}$ is the Berry curvature. The choice of the explicit $1/2$ in $g_{ij}$ is conventional. Because $f_{ij}$ is anti-symmetric, it can be integrated as a differential form without a metric \cite{Brouder_2007}:
\bea
\label{eq:Chernclass}
 \int_{\mathrm{BZ}} \frac{1}{2} f_{ij} \ dk^i \wedge dk^j &= \int_{\mathrm{BZ}} f_{12} \, dk^1dk^2 = 2\pi C
\eea
where $C \in \mathds{Z}$ is the Chern number. We now turn out attention to the quantum metric. There is slightly simpler expression using the identities $\{P,\del_i P\} = \del_i P$ (derived from $P^2 = P$) and $\Tr A \{B,C\} = \Tr B \{A,C\}$, from which we find
\bea
g_{ij} &= \frac{1}{2} \Tr P \{\del_i P, \del_j P \} = \frac{1}{2} \Tr \del_i P \del_j P \ .
\eea
Strikingly, this expression shows that the quantum metric $g_{ij}$ is simply the induced metric (or pullback) from the Euclidean space of $N_{\mathrm{orb}} \times N_{\mathrm{orb}}$ matrices. To see this, recall that the induced metric is usually written $g_{ij}(x) = \del_i X^\mu \del_j X^\nu \tilde{g}_{\mu \nu}$ where $X$ is an embedding map from a manifold parameterized by $x$ to a different manifold parameterized by $x$ with metric $\tilde{g}$. The metric $\tilde{g}$ defines a natural ``induced metric" $g_{ij}(x)$ through the embedding. Here we identify $P_{\al \be}(\mbf{k})$ as the embedding, and $\tilde{g}_{\al \be, \sigma \rho} = \delta_{\al \rho} \delta_{\be \sigma}$ as the metric corresponding to the Frobenius norm.

We now obtain a scalar (coordinate-independent quantity) by contracting the quantum metric with the inverse coordinate metric on the BZ: $\eta^{ij} g_{ij} = g^i_i$. Integrating over the BZ in analogy to the Berry curvature in \Eq{eq:Chernclass}, we define the dimensionless number
\bea
\label{eq:Gtwoforms}
G &= \int \frac{d^2 k}{(2\pi)^2} \sqrt{\det \eta} \ g^i_i = \int \frac{dk^1 dk^2}{(2\pi)^2\Omega_c} \frac{1}{2} \eta^{ij} \Tr \del_i P \del_j P \ .
\eea
The first expression in \Eq{eq:Gtwoforms} is a coordinate-invariant expression, noting that $\int d^2k \sqrt{ \det \eta} = (2\pi)^2 |\mbf{b}_1 \times \mbf{b}_2| = (2\pi)^2/\Omega_c$ is the invariant area of the BZ, and the second writes the integral in terms of the dimensionless lattice momenta $k^i \in (-\pi,\pi)$. Briefly, we remark that the quantum geometric tensor is positive semi-definite giving lower bound on $g^i_i$ in terms of $|f_{12}|$ \cite{2015NatCo...6.8944P}. However, these bounds are not of interest in this work because fragile and OWC phases can have vanishing Berry curvature. In fact, the quantum metric itself is positive semi-definite. This is easily seen by checking
\bea
v^i g_{ij}  v^j &= \frac{1}{2} \Tr v^i \del_i P \del_j P v^j = \frac{1}{2} || v^i \del_iP||^2 \geq 0  \quad \forall v_i
\eea
where $v_i$ is a real 2D vector and $||A||^2 = \Tr A^\dag A$ is the Frobenius norm.
In any state where $P(\mbf{k})$ is not constant in $\mbf{k}$, $\del_i P$ will make a nonzero contribution to $g_{ij}$. We will show that the symmetry data of a fragile or OAL phase requires the projector to change between different high symmetry points in the BZ, which we exploit to give lower bounds in \App{app:SFWbounds}. To give an example, we consider the 1D dimerized chain with two atoms per unit cell at locations $\delta_1, \delta_2$ with the real space Hamiltonian
\bea
\label{eq:Hdimer}
H &= \sum_R t c^\dag_{R,2} c_{R,1} + t' c^\dag_{R+1,1} c_{R,2} + h.c. \ . \\
\eea
We keep the orbital locations $\delta_1,\delta_2$ general so it is clear that the choice of unit cell does not matter. Fourier transforming, the momentum space Hamiltonian is
\bea
h(k) &= \bpm 0 & t' e^{- i k(1+\delta_1-\delta_2)} \\
t e^{- i k(\delta_2-\delta_1)} & \epm + h.c. = \bpm 0 & * \\ e^{- i k(\delta_1-\delta_2)} (t + e^{i k} t') & 0 \epm
\eea
which has a nontrivial embedding matrix $V(G) = \text{diag}(e^{- i G \delta_1},e^{- i G \delta_2})$. The energies are $\pm|t + e^{i k} t'|$ with eigenvectors
\bea
\label{eq:thprime}
U_\pm(k) &= \bpm e^{-i k \delta_1} & \\ &  e^{-i k \delta_2} \epm  \bpm \pm1 \\ e^{i \th(k)} \epm / \sqrt{2}, \qquad e^{i \th(k)} = \frac{t+t'e^{i k}}{|t+t'e^{i k}|} \ .
\eea
It is now direct to compute the occupied band projector $P(k) = U_-(k)U_-^\dag(k)$ and the superfluid weight:
\bea
\label{eq:Gdimer}
G =  \int \frac{dk}{2\pi} \frac{1}{2} \Tr \del_k P(k)\del_k P(k) = \int \frac{dk}{2\pi} \frac{1}{2} ||\del_k P(k)||^2 &= \frac{1}{4} \int \frac{dk}{2\pi} \big( \th'(k) + \delta_1-\delta_2 \big)^2 \ . \\
\eea
Note that in 1D, the superfluid weight (or Drude weight) has dimensions of length, but we set the lattice constant to one for ease. To see how the obstructed phase has nonzero $G$ (while the trivial phase has $G=0$), it is sufficient to go to the inversion symmetric case where $\delta_1 = -\delta, \delta_2 =  + \delta$. We can compute the  Berry connection
\bea
\label{eq:A}
A(k) &= -i U^\dag \del_k U = \frac{1}{2} ( \th'(k) - \delta_1 - \delta_2) = \frac{1}{2} \th'(k)
\eea
which is the projected position operator into the occupied band at momentum $k$. (Note that $\th(k)$ has a branch cut that corresponds to the ambiguity of the position operator mod 1). We can interpret $A(k)$ as the Wannier center of the electron at a given momentum. Rewriting \Eq{eq:Gdimer}, we find
\bea
G &= \int \frac{dk}{2\pi} \big( A(k) - \delta \big)^2 \\
\eea
which shows plainly that $G$ is minimized when $A(k)$ (the Wannier center) is as close to $\delta$ (the orbital location) as possible. If $A(k) \neq \delta$, then the model is in an obstructed phase and $G$ is nonzero. As a simple check, we set $t'=0$ where we find $\th'(k) = A(k) = 0$ from \Eq{eq:thprime}. Then $G = \delta^2$ which is only nonzero when $\delta = 0$, \emph{exactly} when the Wannier center and the orbital are in the same location. Otherwise, the phase is obstructed and $G\neq 0$.

For completeness, we also include an exact expression for all $t,t' >0, |\delta| \leq 1/2$:
\bea
G &= \int \frac{dk}{2\pi} \frac{1}{2} ||\del_k P(k)||^2 =  \int \frac{dk}{2\pi} \lp \frac{t'^2 + t t' \cos k- 2(t^2 + t'^2 + 2 t t' \cos k) \delta}{2(t^2 + t'^2 + 2 tt' \cos k)} \rp^2 \\
&= \frac{1}{16} \lp (1 - 4 \delta)^2 + (8\delta-1) \text{sign}(t-t') + \frac{2t'^2}{(t+t')|t-t'|} \rp \ .
\eea
The discontinuities occur at the gap closing $t = t'$.

\subsection{Projector Formula for Numerical purposes}
\label{app:SFWformula}

In this section, we derive a convenient numerical formula for $G$ defined in \Eq{eq:Gtwoforms}. It has roughly the same computational complexity as a Wilson loop, which requires the projectors $P(\mbf{k})$ on a fine mesh over the BZ \cite{Alexandradinata:2012sp}.

We have derived the expression
\bea
\label{eq:delP}
G &= \int \frac{d^2 k}{(2\pi)^2} \sqrt{\det \eta} \ g^i_i = \int \frac{dk^1 dk^2}{(2\pi)^2\Omega_c} \frac{1}{2} \delta^{\mu \nu} \Tr \del_\mu P \del_\nu P
\eea
where in the last line we used crystalline coordinates as a chart on the BZ, but we evaluated $g^\mu_\mu$ in cartesian coordinates where $\eta^{\mu\nu} = \delta^{\mu \nu}$ is diagonal, so there are only two terms in the $\mu \nu$ sum. Recall that the derivatives are numerically stable because $P(\mbf{k})$ is gauge invariant. Thus we can derive a simple point-split expression:
\bea
\frac{1}{2} \Tr (\del_{k_x} P)^2 &= \frac{1}{2\eps^2} \Tr \lp P(\mbf{k} + \eps \hat{x}) - P(\mbf{k}) \rp^2 \\
&= \frac{1}{2\eps^2} \Tr \lp P(\mbf{k} + \eps \hat{x})^2 + P(\mbf{k})^2 - \{P(\mbf{k} + \eps \hat{x}) ,P(\mbf{k})  \}\rp \\
&= \frac{1}{2\eps^2} \Tr \lp P(\mbf{k} + \eps \hat{x}) + P(\mbf{k}) - \{P(\mbf{k} + \eps \hat{x}) ,P(\mbf{k})  \}\rp \\
&= \frac{1}{2\eps^2} \lp 2N_{\mathrm{occ}} - 2\Tr P(\mbf{k} + \eps \hat{x}) P(\mbf{k}) \rp \\
&= \frac{1}{\eps^2} \lp N_{\mathrm{occ}} - \Tr P(\mbf{k} + \eps \hat{x}) P(\mbf{k}) \rp \\
\eea
where we used $P(\mbf{k})^2 = P(\mbf{k})$ and $\Tr P(\mbf{k}) = N_{\mathrm{occ}}$. If we choose an $L \times L$ grid for $\mbf{k}$ on the BZ, then we find
\bea
G &= \frac{1}{\Omega_c L^2} \sum_{\mbf{k}} \frac{1}{\eps^2} \lp N_{\mathrm{occ}} - \Tr P(\mbf{k} + \eps \hat{x}) P(\mbf{k}) +N_{\mathrm{occ}} - \Tr P(\mbf{k} + \eps \hat{y}) P(\mbf{k})  \rp  \\
&= \frac{2N_{\mathrm{occ}} }{\eps^2 \Omega_c} - \frac{1}{\eps^2 \Omega_c L^2} \sum_{\mbf{k}} \Tr P(\mbf{k}) \lp P(\mbf{k} + \eps \hat{x})  + P(\mbf{k} + \eps \hat{y}) \rp \\
\eea
which we find numerically converges very rapidly. In particular, if one choose $\eps = 1/L$, then $G$ can be computed from only $L^2$ evaluations of $P(\mbf{k})$ on the BZ mesh.

\section{Superfluid Weight from auxiliary-field Monte Carlo simulations}
\label{app:SFMC}

To compute $D_s(T)$, we introduce an external electromagnetic field via its electromagnetic potential $\mathbf{A}$ and Peierls substitution:
	$t_{\mbf{R}\alpha \mbf{R}'\beta}\rightarrow t_{\mbf{R}\alpha \mbf{R}'\beta}\,\exp[i\mathbf{A}\cdot(\mbf{R}+\mbf{r}_\alpha-\mbf{R}'-\mbf{r}_\beta)]=t_{\mbf{R}\alpha \mbf{R}'\beta}(\mathbf{A})$. Here, $\mbf{R}=R_1\mbf{a}_1+R_2\mbf{a}_2$ with $R_1,R_2\in\mathbb{Z}$ is the position of the unit cell, $\mbf{r}_\alpha$ is the position of the orbital $\alpha$ inside the unit cell \cite{PhysRevLett.125.236804}. We can then expand $H(\mathbf{A})$ up to second order in $\mathbf{A}$ (here $ij$ are cartesian indices):
\begin{equation}
H(\mathbf{A})=H+ j^p_i A_i+\frac{1}{2}T_{ij}A_i A_j,
\end{equation}
where $j^p_i$ is the paramagnetic current operator and $T_{ij}A_j$ is the diamagnetic current operator. These operators are defined as $j^p_i=\sum_\mbf{R}j^p_i(\mbf{R})$ with
\begin{equation}
	\label{eq:defj}
	j^p_i(\mbf{R})=\sum_{\alpha \mbf{R}'\beta\sigma}\frac{\partial t_{\mbf{R}\alpha \mbf{R}'\beta}(\mathbf{A})}{\partial A_i}c^\dagger_{\mbf{R}\alpha\sigma}c^{\phantom{\dagger}}_{\mbf{R}'\beta\sigma},
\end{equation}
and
\begin{equation}
	T_{ij}=\sum_{\mbf{R}\alpha \mbf{R}'\beta\sigma}\frac{\partial^2 t_{\mbf{R}\alpha \mbf{R}'\beta}(\mathbf{A})}{\partial A_i\partial A_j}c^\dagger_{\mbf{R}\alpha\sigma}c^{\phantom{\dagger}}_{\mbf{R}'\beta\sigma}.
\end{equation}
The superfluid weight characterizes the zero-frequency, long-wavelength response to the external field, $J_i=-4[D_{s}]_{ij}A_j$. It is given by (see \Refs{Scalapino:1992}{Scalapino:1993})
\begin{equation}
	\label{eq:boundWeight}
	[D_s]_{ij}=\frac{1}{4\Omega_c}\left[\langle T_{ij} \rangle-\Lambda_{ij}(k_\parallel=0,k_\perp\rightarrow 0,i\omega_m=0)\right],
\end{equation}
where $k_{\parallel (\perp)}$ is the momentum component parallel (perpendicular) to $\mathbf{A}$, and $\langle\cdot\rangle$ represents the expectation value over the many-body ground state at temperature $T$. Here, $\Lambda_{ij}(\mathbf{k},\omega)$ is the paramagnetic current susceptibility in momentum and frequency space:
\begin{equation}
\Lambda_{ij}(\mathbf{k},i\omega_m)= \int_0^\beta \text{d}\tau e^{i\omega_m \tau}\left\langle \left[j^p_i(\mathbf{k},\tau),j^p_j(-\mathbf{k},0) \right] \right\rangle,
\end{equation}
with $\omega_m=2\pi mT$, $m\in \mathbb Z$, and $\beta=1/k_{B}T$ the inverse temperature. 
Note that with the factor $1/4$ in \Eq{eq:boundWeight}, the BKT transition occurs at $T_{c}=\pi D_s^-/2$, where $D_s^-$ is the superfluid weight at the critical temperature approached from below \cite{PhysRevLett.39.1201}.
In the simulations, we consider $D_s=[D_s]_{yy} = \frac{1}{2} \text{tr} D_s$ and a gauge potential $\mathbf{A}=A\mathbf{\hat{y}}$.

\section{Superfluid Weight from Mean-Field Theory}
\label{app:MFSF}
We now want to obtain the superfluid weight in the mean-field regime. As a first step, we carefully derive the Bogoliubov-de Gennes (BdG) Hamiltonian obtained from mean-field decoupling of an attractive interaction in the pair channel.

We neglect FFLO order and consider exclusively pairing between electrons with opposite spin and momentum $\mbf{k}$ and $-\mbf{k}$. We further consider an interaction diagonal in orbital space: 
\begin{equation} H=\sum_{\mbf{k}\alpha\beta\sigma}[h_\sigma(\mathbf{k})-\mu\mathbb{1}]_{\alpha\beta}c^\dagger_{\mbf{k}\alpha\sigma}c^{\phantom{\dagger}}_{\mbf{k}\beta\sigma}-\lvert U \rvert \sum_{\mbf{k}\alpha}c^\dagger_{\mbf{k}\alpha\up} c^\dagger_{-\mbf{k}\alpha\dw}c^{\phantom{\dagger}}_{-\mbf{k}\alpha\dw}c^{\phantom{\dagger}}_{\mbf{k}\alpha\up} \end{equation}
and perform the mean-field decoupling:
\begin{equation}
    \Delta_{\alpha}(\mbf{k})=-\lvert U \rvert\langle c_{-\mbf{k}\alpha\dw}c_{\mbf{k}\alpha\up}\rangle.
\end{equation}
The BdG Hamiltonian is then:
\begin{equation}
    \label{eq:BdGH}
    \begin{split}
    H_{\text{BdG}} & =  \sum_{\mbf{k}\alpha\beta}[h_\up(\mathbf{k})-\mu\delta_{\alpha\beta}]_{\alpha\beta}c^\dagger_{\mbf{k}\alpha\up}c^{\phantom{\dagger}}_{\mbf{k}\beta\up} -\sum_{\mbf{k}\alpha\beta}[h_\dw(-\mathbf{k})-\mu\delta_{\alpha\beta}]_{\alpha\beta}c^{\phantom{\dagger}}_{-\mbf{k}\beta\dw}c^\dagger_{-\mbf{k}\alpha\dw}+\sum_{\mbf{k}\alpha\beta}[h_\dw(\mathbf{k})-\mu\delta_{\alpha\beta}]_{\alpha\beta}\delta_{\alpha\beta}\\ &+\sum_{\mbf{k}\alpha}\left(\Delta_\alpha(\mathbf{k})c^\dagger_{\mbf{k}\alpha\up}c^\dagger_{-\mbf{k}\alpha\dw}+\Delta^*_\alpha(\mathbf{k})c^{\phantom{\dagger}}_{-\mbf{k}\alpha\dw}c^{\phantom{\dagger}}_{\mbf{k}\alpha\up}\right)\\
     & =  \sum_{\mathbf{k}}\Psi^\dagger_\mathbf{k}
     \begin{pmatrix}
         h_\up(\mathbf{k})-\mu& \Delta(\mathbf{k})\\
         \Delta^\dagger(\mathbf{k})& -[h_\dw(-\mathbf{k})]^T+\mu \\
     \end{pmatrix} \Psi^{\phantom{\dagger}}_\mathbf{k}+\sum_\mbf{k}\text{Tr}(E_\mbf{k}-\mu)\\
     & = \sum_{\mathbf{k}}\Psi^\dagger_\mathbf{k}
     \begin{pmatrix}
         h_\up(\mathbf{k})-\mu& \Delta(\mathbf{k})\\
         \Delta^\dagger(\mathbf{k})& -h_\up(\mathbf{k})+\mu \\
     \end{pmatrix} \Psi^{\phantom{\dagger}}_\mathbf{k}+\sum_\mbf{k}\text{Tr}(E_\mbf{k}-\mu)\\
	 & = \sum_{\mathbf{k}}\Psi^\dagger_\mathbf{k}
     \mathcal{H}_\mbf{k} \Psi^{\phantom{\dagger}}_\mathbf{k}+\sum_\mbf{k}\text{Tr}(E_\mbf{k}-\mu).
\end{split}
\end{equation}
In the last line we used $h_\up(\mathbf{k}) = h^*_\dw(-\mathbf{k})$, due to time-reversal symmetry, and introduced the Nambu spinor $\Psi_\mathbf{k}=(c^{\phantom{\dagger}}_{\mbf{k}1\up},c^{\phantom{\dagger}}_{\mbf{k}2\up},...,c^\dagger_{-\mbf{k}1\dw},c^\dagger_{-\mbf{k}2\dw},...)^T$ which has $2N_{\mathrm{orb}}$ components. $E_\mbf{k}$ is a diagonal matrix, whose eigenvalues are the single-particle energies $E_n(\mbf{k})$: $E_\mbf{k}=U^\dagger(\mbf{k})h_\up(\mathbf{k})U(\mbf{k})$, with $U(\mbf{k})$ the matrix that diagonalizes the free fermion-Hamiltonian. In our model, we have $E_\mbf{k}=E^{(\uparrow)}_\mbf{k}=E^{(\downarrow)}_\mbf{-k}$. 

Here, we assumed $\Delta(\mathbf{k})=\text{diag}(\Delta_{1}(\mbf{k}),\dots, \Delta_{N_{\mathrm{orb}}}(\mbf{k}))$. The zero-temperature gap at the mean-field level can be obtained self-consistently:
\begin{equation}
    \Delta_{\alpha}=\frac{1}{N_c}\sum_\mbf{k}\Delta_{\alpha}(\mbf{k})=-\frac{\lvert U \rvert}{N_c}\sum_{n\mathbf{k}}\theta[\epsilon_n(\mathbf{k})]\bra{n\mathbf{k}}\partial_{\Delta_\alpha}H_{\text{BdG}}(\mathbf{k})\ket{n\mathbf{k}},
\end{equation}
where $N_c$ is the number of unit cell. We restrict to a spatially uniform ansatz for $\Delta_\alpha$, $n$ is the Bogoliubov band label, and $H_{\text{BdG}}(\mathbf{k})\ket{n\mathbf{k}}=\epsilon_n(\mathbf{k})\ket{n\mathbf{k}}$. $\theta[\epsilon_n(\mathbf{k})]$ is the zero-temperature Fermi-Dirac distribution where only negative Bogoliubov eigenenergies are occupied. Together with the gap equation, the chemical potential $\mu$ is set to satisfy the number equation:
\begin{equation}
N_{\mathrm{occ}}=\sum_{n\mathbf{k}} \theta[\epsilon_n(\mathbf{k})].
\end{equation}

\subsection{Uniform pairing and BCS ground state}
\label{app:uniformPairing}
As a first step we justify why the BdG Hamiltonian might be a good starting point for the zero temperature superfluid weight in our flat-band model.

Consider an isolated flat band with Wannier functions $W_{\alpha\sigma}(\mbf{R}-\mbf{R}')$, where $\alpha$ indicates the orbital, $\sigma$ the spin, and $\mbf{R}$, $\mbf{R}'$ are the position of the unit-cells $\mbf{R}=R_1\mbf{a}_1+R_2\mbf{a}_2$ and $\mbf{R}'=R'_1\mbf{a}_1+R'_2\mbf{a}_2$, with $R_1,R_2,R'_1,R'_2 \in \mathbb{Z}$.
Due to time-reversal symmetry, we have $[W_{\alpha\uparrow}(\mbf{R}-\mbf{R}')]^*=W_{\alpha\downarrow}(\mbf{R}-\mbf{R}')$ and in the following we will drop the spin index and refer to $W_{\alpha}(\mbf{R}-\mbf{R}')=W_{\alpha\uparrow}(\mbf{R}-\mbf{R}')$.
Note that these Wannier functions are eigenstates of the flat-band single-particle Hamiltonian \cite{PhysRevB.94.245149}.

Given the Wannier functions, we can check whether they have the same weight on all orbitals:
\begin{equation}
	n_{\phi\alpha}=\sum_{\mbf{R}}\lvert W_\alpha(\mbf{R}) \rvert^2=\sum_{\mbf{R}}\pi_{\alpha\alpha}(\mathbf{0}),
	\end{equation}
where we introduced the real-space projector onto the flat band:
\begin{equation}
	\pi_{\alpha\beta}(\mbf{R}-\mbf{R}')=\sum_\mbf{R''}W_\alpha(\mbf{R}-\mbf{R''})W^*_\beta(\mbf{R}'-\mbf{R''}).
	\end{equation}
If $n_{\phi\alpha}=n_\phi$ for all $\alpha$ over which the Wannier function has nonzero weight, we say that the band satisfies the uniform pairing condition \cite{PhysRevB.94.245149}.

The uniform pairing condition is of prominent importance once we consider an attractive local interaction:
\begin{equation}
	H_{\text{int}}=-\lvert U \rvert \sum_{\mbf{R}\alpha} c^\dagger_{\mbf{R}\alpha\uparrow}c^\dagger_{\mbf{R}\alpha\downarrow}c^{\phantom{\dagger}}_{\mbf{R}\alpha\downarrow}c^{\phantom{\dagger}}_{\mbf{R}\alpha\uparrow}.
	\end{equation}
Let us define the field operator projected into the flat band:
\begin{equation}
	\bar{c}_{\mbf{R}\alpha\sigma}=\sum_{\mbf{R}'}W_{\alpha\sigma}(\mbf{R}-\mbf{R}')d_{\mbf{R}'\sigma},
	\end{equation}
where we introduced the annihilation operator of the Wannier orbital of the flat band:
	\begin{equation}
		d_{\mbf{R}\sigma}=\sum_{\mbf{R}',\beta}W^*_{\beta,\sigma}(\mbf{R}'-\mbf{R})c_{\mbf{R}'\beta\sigma}.
		\end{equation}
With the help of the projection operator, we can define the projected field operators in terms of the original ones as:
\begin{equation}
	\bar{c}_{\mbf{R}\alpha\uparrow}=\sum_{\mbf{R}'\beta}\pi_{\alpha \beta}(\mbf{R}-\mbf{R}')c_{\mbf{R}'\beta\uparrow}.
	\end{equation}
Note that $c_{\mbf{R}\alpha\sigma}$, $c^\dagger_{\mbf{R}\alpha\sigma}$ and $d_{\mbf{R}\sigma}$, $d^\dagger_{\mbf{R}\sigma}$ satisfy the standard fermionic anti-commutation relations. The projected operators, on the other hand, obey \cite{PhysRevB.94.245149}:
\begin{equation}
	\{\bar{c}_{\mbf{R}\alpha\uparrow},\bar{c}^\dagger_{\mbf{R}'\beta\uparrow}\}=\pi_{\alpha\beta}(\mbf{R}-\mbf{R}'),\quad \{\bar{c}_{\mbf{R}\alpha\downarrow},\bar{c}^\dagger_{\mbf{R}'\beta\downarrow}\}=\pi^*_{\alpha\beta}(\mbf{R}-\mbf{R}'),
	 \end{equation}
with the other anti-commutators trivial, and we define $\bar{n}_{\mbf{R}\alpha\sigma}=\bar{c}^\dagger_{\mbf{R}\alpha\sigma}\bar{c}_{\mbf{R}\alpha\sigma}$ as the projected number operator.

We now consider the interacting Hamiltonian projected onto the flat band:
\begin{equation}
    \label{eq:projectedHubbard}
	\bar{H}_\text{int}=-\lvert U\rvert\sum_{\mbf{R}\alpha}\bar{n}_{\mbf{R}\alpha\uparrow}\bar{n}_{\mbf{R}\alpha\downarrow},
	\end{equation}
and we show that, when the uniform pairing condition is satisfied, the BCS wave function is an exact ground state at zero temperature for the Hamiltonian projected into the flat bands \cite{PhysRevB.94.245149}.

The BCS wave function in real space is given by:
\begin{equation}
	\ket{\Psi}=u^{L^2}\exp\left(\frac{v}{u}b^\dagger_0\right)\ket{0}=\prod_\mbf{R}\left(u+v d^\dagger_{\mbf{R}\uparrow}d^\dagger_{\mbf{R}\downarrow}\right)\ket{0},
	\end{equation}
	where $u=\sqrt{1-\nu}$ and $v=\sqrt{\nu}$ with $\nu$ is the filling of the flat band \cite{2015NatCo...6.8944P} and
	\begin{equation}
		b^\dagger_0=\sum_{\mbf{R}'}d^\dagger_{\mbf{R}'\uparrow} d^\dagger_{\mbf{R}'\downarrow}.
		\end{equation}

The operator $b^\dagger_0$ commutes with the projected spin operator $\bar{S}^z_{\mbf{R}\alpha}=(\bar{n}_{\mbf{R}\alpha\uparrow}-\bar{n}_{\mbf{R}\alpha\downarrow})/2$ 
\cite{PhysRevB.94.245149}. Hence, the BCS wavefunction is a zero energy eigenstate of the positive semidefinite Hamiltonian:
\begin{equation}
	\bar{H}'_{\text{int}}=\frac{\lvert U \rvert}{2}\sum_{\mbf{R}\alpha} \left(\bar{n}_{\mbf{R}\alpha\uparrow}-\bar{n}_{\mbf{R}\alpha\downarrow}\right)^2.
\end{equation}

We further have that \cite{PhysRevB.94.245149}
\begin{equation}
    \bar{H}'_{\text{int}} = \frac{\lvert U \rvert}{2}\sum_{\mbf{R}\alpha}\pi_{\alpha\alpha}(\mathbf{0})\left(\bar{n}_{\mbf{R}\alpha\uparrow}+\bar{n}_{\mbf{R}\alpha\downarrow}\right)-\lvert U \rvert\sum_{\mbf{R}\alpha}\bar{n}_{\mbf{R}\alpha\uparrow}\bar{n}_{\mbf{R}\alpha\downarrow}.
\end{equation}
The first term is generically an orbital dependent potential, whereas the second term is exactly \Eq{eq:projectedHubbard}. If the uniform pairing condition is satisfied, the first term becomes a simple energy shift proportional to the number of particles in the flat bands, i.e., a chemical potential.
Therefore, if the uniform pairing condition is satisfied, the BCS wave function is an exact zero-temperature ground state of the attractive Hubbard attractive model projected into the flat band with energy:
\begin{equation}
	\label{eq:BCSEnergy}
	\frac{\epsilon_\text{BCS}}{L^2}=\left( 2E_0-n_\phi \lvert U  \rvert\right) \nu,
	\end{equation}
where $E_0$ is the flat band energy.

\subsection{Superfluid weight from the multi-band mean-field}
After justifying the use of the mean-field theory at zero temperature, here we numerically compute the superfluid weight of the multi-band BdG Hamiltonian from the response function of the current operators to an external gauge potential $\mathbf{A}$, as described in \App{app:SFMC} \cite{2021PNAS..11806744V}. The gauge potential couples exclusively to the kinetic part of the Hamiltonian:
\begin{equation}
	\label{eq:trtr}
	\begin{split}
    H_{\text{BdG}} & = \sum_{\mathbf{k}}\Psi^\dagger_\mathbf{k}
    \begin{pmatrix}
        h_\up(\mathbf{k}-\mathbf{A})-\mu& \Delta^{\phantom{\dagger}}(\mathbf{k})\\
        \Delta^\dagger(\mathbf{k})& -h_\up(\mathbf{k}+\mathbf{A})+\mu \\
    \end{pmatrix} \Psi^{\phantom{\dagger}}_\mathbf{k}+\sum_\mbf{k}\text{Tr}(E_{\mbf{k}-\mbf{A}}-\mu)\\ & = \sum_{\mathbf{k}}\Psi^\dagger_\mathbf{k}
    \mathcal{H}_\mbf{k}(\mbf{A}) \Psi^{\phantom{\dagger}}_\mathbf{k}+\sum_\mbf{k}\text{Tr}(E_{\mbf{k}-\mbf{A}}-\mu).
	\end{split}
\end{equation}
We compute $D_s(T=0)$ from the Kubo formula of \Eq{eq:boundWeight}. Here, the diamagnetic part is
\begin{equation}
   \label{eq:tmf} [T]_{ij}=\sum_{\mbf{k}n}\theta[\epsilon_n(\mathbf{k})]\bra{n\mathbf{k}}\partial_{k_i}\partial_{k_j}H_{\text{BdG}}(\mathbf{k},\Delta=0)\ket{n\mathbf{k}}
\end{equation}
and the paramagnetic current susceptibility is
\begin{equation}
	\label{eq:lambdamf}
    [\Lambda]_{ij}=\sum_{\mbf{k}nm}\frac{\theta[\epsilon_n(\mathbf{k})]-\theta[\epsilon_m(\mathbf{k})]}{\epsilon_n(\mathbf{k})-\epsilon_m(\mathbf{k})}
    \bra{n\mathbf{k}}\partial_{k_i}H_{\text{BdG}}(\mathbf{k},\Delta=0)\gamma^z \ket{m\mathbf{k}}
    \bra{m\mathbf{k}}\gamma^z \partial_{k_j}H_{\text{BdG}}(\mathbf{k}, \Delta=0)\ket{n\mathbf{k}},
\end{equation}
where the pre-factor is zero when $\epsilon_n(\mathbf{k})=\epsilon_m(\mathbf{k})$, $H_{\text{BdG}}(\mathbf{k},\Delta=0)$ indicates the Hamiltonian of \Eq{eq:trtr} with $\Delta(\mathbf{k})=0$, $\gamma^z=\begin{pmatrix}1&0\\0&-1\end{pmatrix}\otimes \mathbb{1}$ accounts for the opposite charge of holes and particles, and $i,j$ are spatial indices.

As before, in our simulations we only compute $D_s=[D_s]_{yy}$.
Namely, the $C_3$ symmetry of our model imposes $[C_3^{-1} D_s C_3]_{ij} = [D_s]_{ij}$ which sets $[D_s]_{xx}=[D_s]_{yy}$ and  $[D_s]_{xy}=0$ because $[D_s]_{ij}$ is symmetric. We can then consider a gauge potential $\mathbf{A}=A\mathbf{\hat{y}}$ and identify $\frac{1}{2}[D_{s}]^i_i$=$D_s$.

We self-consistently solve the BdG Hamiltonian of \Eq{eq:trtr} and numerically compute the superfluid weight for $\lvert U \rvert=4$ and $\lvert t\rvert=6$ from \Eqs{eq:tmf}{eq:lambdamf}. Note that we are not in the limit $\lvert U \rvert \ll \lvert t \rvert$, where a projection on the flat band would be rigorously justified. The yellow star in Fig. 3 of the Main Text shows the result of this numerical multi-band mean-field calculation.

\subsection{Superfluid weight upon projection on isolated flat band}
\label{app:MFsingle}
To make analytical progress, we consider the problem projected into the flat band \cite{2015NatCo...6.8944P,PhysRevLett.124.167002}.
As a first step, we diagonalize the BdG Hamiltonian in the presence of a gauge potential $\mbf{A}$:
\begin{equation}
\epsilon_\mbf{k}(\mbf{A})=W_\mbf{k}^\dagger(\mbf{A})\mathcal{H}_{\mbf{k}}(\mbf{A}) W_\mbf{k}(\mbf{A}),
\end{equation}
where $\epsilon_\mbf{k}(\mbf{A})$ is a diagonal matrix whose eigenvalues $\epsilon_{n\mathbf{k}}(\mathbf{A})$ are the Bogoliubov bands.
We then compute the free energy in the presence of an external gauge potential $\Omega(\mathbf{A})$:
\begin{equation}
    \begin{split}
    \Omega(\mathbf{A}) & =\sum_{\mathbf{k},\epsilon_{n\mathbf{k}}(\mathbf{A})<0} \epsilon_{n\mathbf{k}}(\mathbf{A}) + \sum_\mbf{k}\text{Tr}\left(E_{\mbf{k}-\mbf{A}}-\mu\right)\\
    & = \frac{1}{2}\sum_{\mathbf{k}}\left(\sum_{\epsilon_{n\mathbf{k}}(\mathbf{A})<0} \epsilon_{n\mathbf{k}}(\mathbf{A})+\sum_{\epsilon_{n\mathbf{k}}(\mathbf{A})>0} \epsilon_{n\mathbf{k}}(\mathbf{A})+\sum_{\epsilon_{n\mathbf{k}}(\mathbf{A})<0} \epsilon_{n\mathbf{k}}(\mathbf{A})-\sum_{\epsilon_{n\mathbf{k}}(\mathbf{A})>0} \epsilon_{n\mathbf{k}}(\mathbf{A}) \right)\\ & \quad+ \sum_\mbf{k}\text{Tr}\left(E_{\mbf{k}-\mbf{A}}-\mu\right) \\
    & = \frac{1}{2}\sum_{n\mathbf{k}}\epsilon_{n\mathbf{k}}(\mathbf{A})-\frac{1}{2}\sum_{n\mathbf{k}}\lvert \epsilon_{n\mathbf{k}}(\mathbf{A})\rvert + \sum_\mbf{k}\text{Tr}\left(E_{\mbf{k}-\mbf{A}}-\mu\right)\\
    & =  \frac{1}{2}\sum_\mbf{k}\text{Tr}\left(\mathcal{H}_{\mathbf{k}}(\mathbf{A})\right)-\frac{1}{2}\sum_{n\mathbf{k}}\lvert \epsilon_{n\mathbf{k}}(\mathbf{A})\rvert+\sum_\mbf{k}\text{Tr}\left(E_{\mbf{k}-\mbf{A}}-\mu\right)\\
	& = \frac{1}{2}\left(\sum_\mbf{k}\text{Tr}\left(E_{\mbf{k}-\mbf{A}}-\mu\right)-\sum_\mbf{k}\text{Tr}\left(E_{\mbf{k}+\mbf{A}}-\mu\right)\right) -\frac{1}{2}\sum_{n\mathbf{k}}\lvert \epsilon_{n\mathbf{k}}(\mathbf{A})\rvert+ \sum_\mbf{k}\text{Tr}\left(E_{\mbf{k}-\mbf{A}}-\mu\right)\\
    & = -\frac{1}{2}\sum_{n\mathbf{k}}\lvert \epsilon_{n\mathbf{k}}(\mathbf{A})\rvert+ \sum_\mbf{k}\text{Tr}\left(E_{\mbf{k}-\mbf{A}}-\mu\right).
\end{split}
\end{equation}
For exactly flat bands, the term $\sum_\mbf{k}\Tr{\left(E_{\mbf{k}-\mbf{A}}-\mu\right)}$ does not depend on momentum and hence it does not contribute to the superfluid weight. 

Now consider a change of basis that diagonalizes the kinetic part of the BdG Hamiltonian. The off-diagonal block is then given by
\begin{equation}
    \mathcal{D}_\mathbf{k}(\mathbf{A})=U^\dagger(\mathbf{k}-\mathbf{A})\Delta(\mathbf{k}) U(\mathbf{k}+\mathbf{A})=\Delta U^\dagger(\mathbf{k}-\mathbf{A})U(\mathbf{k}+\mathbf{A}),
\end{equation}
where the unitary matrix $U(\mathbf{k})$ diagonalizes the kinetic term at momentum $\mathbf{k}$ and we assumed $\Delta_{\alpha}(\mathbf{k})=\Delta$.

If the ratio $\lvert U \rvert / \delta$ is sufficiently small, with $\delta$ the energetic gap between the flat band of interest and the other bands, we can solve the Hamiltonian by projecting onto the flat bands at energy $\epsilon_0$. Take $\tilde{U}(\mathbf{k})$ to be the projection of $U(\mathbf{k})$ onto the flat band, we have \cite{PhysRevLett.124.167002}
\begin{equation}
    \tilde{\mathcal{H}}_{\mathbf{k}}(\mathbf{A})=
    \begin{pmatrix}
        E_0-\mu & \tilde{\mathcal{D}}_\mathbf{k}(\mathbf{A}) \\
        \tilde{\mathcal{D}}^\dagger_\mathbf{k}(\mathbf{A}) & -E_0+\mu\\
    \end{pmatrix}
\end{equation}
with
\begin{equation}
    \tilde{\mathcal{D}}_\mathbf{k}(\mathbf{A}) = \Delta \tilde{U}^\dagger(\mathbf{k}-\mathbf{A})\tilde{U}(\mathbf{k}+\mathbf{A}).
\end{equation}

The eigenvalues of the projected BdG Hamiltonian can be obtained by taking its square \cite{PhysRevLett.124.167002}:
\begin{equation}
\tilde{\mathcal{H}}^2_{\mathbf{k}}(\mathbf{A})=
\begin{pmatrix}
    (E_0-\mu)^2+\tilde{\mathcal{D}}^{\phantom{\dagger}}_\mathbf{k}(\mathbf{A})\tilde{\mathcal{D}}^{\dagger}_\mathbf{k}(\mathbf{A})& 0 \\
    0 & (E_0-\mu)^2+\tilde{\mathcal{D}}^{\dagger}_\mathbf{k}(\mathbf{A})\tilde{\mathcal{D}}^{\phantom{\dagger}}_\mathbf{k}(\mathbf{A}) \\
\end{pmatrix}.
\end{equation}
Denote by $\lambda_{n\mathbf{k}}(\mathbf{A})$ and $\varphi_{n\mathbf{k}}(\mathbf{A})$ the eigenvalues of the matrices $\tilde{\mathcal{D}}^{\phantom{\dagger}}_\mathbf{k}(\mathbf{A})\tilde{\mathcal{D}}^{\dagger}_\mathbf{k}(\mathbf{A})$ and $\tilde{\mathcal{D}}^{\dagger}_\mathbf{k}(\mathbf{A})\tilde{\mathcal{D}}^{\phantom{\dagger}}_\mathbf{k}(\mathbf{A})$, respectively. These matrices are Hermitian and semi-positive definite. Therefore, the eigenvalues of the projected BdG Hamiltonian satisfy $\epsilon^2_{n\mathbf{k}}(\mathbf{A})=(E_0 -\mu)^2+\lambda_{n\mathbf{k}}(\mathbf{A})$ when $1\leq n\leq N_F$ and $\epsilon^2_{n\mathbf{k}}(\mathbf{A})=(E_0 -\mu)^2+\varphi_{n\mathbf{k}}(\mathbf{A})$ when $N_F+1\leq n\leq 2N_F$, where $N_F$ is the number of flat bands in the system.

We readily obtain the free energy as:
\bea
\begin{split}
    \Omega(\mathbf{A})& =-\frac{1}{2}\sum_{n\mathbf{k}}\lvert \epsilon_{n\mathbf{k}}(\mathbf{A})\rvert +E_0-\mu\\ & = -\frac{1}{2}\sum_{n\mathbf{k}} \left(\sqrt{(E_0 -\mu)^2+\lambda_{n\mathbf{k}}(\mathbf{A})}+\sqrt{(E_0 -\mu)^2+\varphi_{n\mathbf{k}}(\mathbf{A})}\right) +E_0-\mu \ . \\
	\end{split}
\eea
Note that the free energy $\Omega$ is extensive because there are $L^2$ terms in the sum when discretized on an $L\mbf{a}_1 \times L\mbf{a}_2$ crystal. To derive the superfluid weight, we take the second derivative \cite{PhysRevLett.124.167002}:
\begin{equation}
	\label{eq:supvp}
    \begin{split}
    [D_s]_{ij}&=\frac{1}{4L^2|\mbf{a}_1\times \mbf{a}_2|}\frac{\partial^2\Omega(\mathbf{A})}{\partial A_i\partial A_j}\Bigg\rvert_{\mathbf{A}=0}\\
    & =-\frac{1}{16L^2|\mbf{a}_1\times \mbf{a}_2|}\sum_{n\mathbf{k}}\Bigg(\frac{\partial_{A_i}\partial_{A_j}\lambda_{n\mathbf{k}}(\mathbf{A})}{\sqrt{(E_0 -\mu)^2+\lambda_{n\mathbf{k}}(0)}}-
    \frac{\partial_{A_i}\lambda_{n\mathbf{k}}(\mathbf{A})\partial_{A_j}\lambda_{n\mathbf{k}}(\mathbf{A})}{2\left[(E_0 -\mu)^2+\lambda_{n\mathbf{k}}(0)\right]^{3/2}}\\
    &\phantom{=}+\frac{\partial_{A_i}\partial_{A_j}\varphi_{n\mathbf{k}}(\mathbf{A})}{\sqrt{(E_0 -\mu)^2+\varphi_{n\mathbf{k}}(0)}}-
    \frac{\partial_{A_i}\varphi_{n\mathbf{k}}(\mathbf{A})\partial_{A_j}\varphi_{n\mathbf{k}}(\mathbf{A})}{2\left[(E_0 -\mu)^2+\varphi_{n\mathbf{k}}(0)\right]^{3/2}}\Bigg)\Bigg\rvert_{\mathbf{A}=0}.
\end{split}
\end{equation}
Note that with the factor $1/4$ in the first line of \Eq{eq:supvp}, we defined the superfluid weight such that the Kosterlitz-Thouless transition occurs at $\frac{D_s(T_c)}{k_BT_c}=\frac{2}{\pi}$.
With our choice of $\Delta(\mathbf{k})$, we have $\lambda_{n\mathbf{k}}(0)=\varphi_{n\mathbf{k}}(0)=\Delta^2$. Therefore
\begin{equation}
    [D_s]_{ij}=- \frac{1}{16}\int_{BZ} \frac{d^2k}{(2\pi)^2}\sum_n\Bigg(\frac{\partial_{A_i}\partial_{A_j}\lambda_{n\mathbf{k}}(\mathbf{A})}{E_B}+\frac{\partial_{A_i}\partial_{A_j}\varphi_{n\mathbf{k}}(\mathbf{A})}{E_B}-
    \frac{\partial_{A_i}\lambda_{n\mathbf{k}}(\mathbf{A})\partial_{A_j}\lambda_{n\mathbf{k}}(\mathbf{A})}{2E_B^{3}}-
    \frac{\partial_{A_i}\varphi_{n\mathbf{k}}(\mathbf{A})\partial_{A_j}\varphi_{n\mathbf{k}}(\mathbf{A})}{2E_B^{3}}\Bigg)\Bigg\rvert_{\mathbf{A}=0},
\end{equation}
with $E_B=\sqrt{(E_0-\mu)^2+\Delta^2}$ and the integral is over the physical BZ with volume $(2\pi)^2 |\mbf{b}_1 \times \mbf{b}_2|$, noting that $|\mbf{b}_1 \times \mbf{b}_2|^{-1}=|\mbf{a}_1 \times \mbf{a}_2|$ is the unit cell area. We further have
\begin{equation}
    \partial_{\mathbf{A}}\left(\tilde{\mathcal{D}}_\mathbf{k}(\mathbf{A})\right)\big\rvert_{\mathbf{A}=0}=\Delta\left(-\partial_\mathbf{k}\tilde{U}^\dagger(\mathbf{k})\tilde{U}(\mathbf{k})+\tilde{U}^\dagger(\mathbf{k})\partial_\mathbf{k}\tilde{U}(\mathbf{k})\right),
\end{equation}
\begin{equation}
    \partial_{\mathbf{A}}\left(\tilde{\mathcal{D}}^\dagger_\mathbf{k}(\mathbf{A})\right)\big\rvert_{\mathbf{A}=0}=\Delta\left(\partial_\mathbf{k}\tilde{U}^\dagger(\mathbf{k})\tilde{U}(\mathbf{k})-\tilde{U}^\dagger(\mathbf{k})\partial_\mathbf{k}\tilde{U}(\mathbf{k})\right).
\end{equation}
Then, the first order derivatives of the products $\tilde{\mathcal{D}}^\dagger_\mathbf{k}(\mathbf{A})\tilde{\mathcal{D}}_\mathbf{k}(\mathbf{A})$ and $\tilde{\mathcal{D}}_\mathbf{k}(\mathbf{A})\tilde{\mathcal{D}}^\dagger_\mathbf{k}(\mathbf{A})$ are zero when evaluated at $\mathbf{A}=0$. By the Hellmann-Feynman theorem we have $\partial_{A_i}\lambda_{n\mathbf{k}}(\mathbf{A})\big\rvert_{\mathbf{A}=0}=\partial_{A_i}\varphi_{n\mathbf{k}}(\mathbf{A})\big\rvert_{\mathbf{A}=0}=0$ and \cite{PhysRevLett.124.167002}
\begin{equation}
    \begin{split}
    [D_s]_{ij}&=-\frac{1}{16E_B}\int_{BZ} \frac{d^2k}{(2\pi)^2} \text{Tr}\left[\partial_{A_i}\partial_{A_j}\left(\tilde{\mathcal{D}}_\mathbf{k}(\mathbf{A})\tilde{\mathcal{D}}^\dagger_\mathbf{k}(\mathbf{A})\right)\Big\rvert_{\mathbf{A}=0}+\partial_{A_i}\partial_{A_j}\left(\tilde{\mathcal{D}}^\dagger_\mathbf{k}(\mathbf{A})\tilde{\mathcal{D}}_\mathbf{k}(\mathbf{A})\right)\Big\rvert_{\mathbf{A}=0}\right]\\
    & =-\frac{1}{16E_B}\int_{BZ} \frac{d^2k}{(2\pi)^2}\partial_{A_i}\partial_{A_j} \left[\text{Tr}\left(\tilde{\mathcal{D}}_\mathbf{k}(\mathbf{A})\tilde{\mathcal{D}}^\dagger_\mathbf{k}(\mathbf{A})+\tilde{\mathcal{D}}^\dagger_\mathbf{k}(\mathbf{A})\tilde{\mathcal{D}}_\mathbf{k}(\mathbf{A})\right)\right]\Bigg\rvert_{\mathbf{A}=0}\\
    & = -\frac{1}{8E_B}\int_{BZ} \frac{d^2k}{(2\pi)^2}\text{Tr}\left[\partial_{A_i}\partial_{A_j}\left(\tilde{\mathcal{D}}_\mathbf{k}(\mathbf{A})\tilde{\mathcal{D}}^\dagger_\mathbf{k}(\mathbf{A})\right)\Big\rvert_{\mathbf{A}=0}\right]\\
    & = \frac{\Delta^2}{\sqrt{(E_0-\mu)^2+\Delta^2}}\int_{BZ} \frac{d^2k}{(2\pi)^2}\text{Tr}\left[\frac{1}{2}\left(\partial_{k_i}\tilde{U}^\dagger_\mathbf{k}\partial_{k_j}\tilde{U}_\mathbf{k}+\partial_{k_j}\tilde{U}^\dagger_\mathbf{k}\partial_{k_i}\tilde{U}_\mathbf{k}\right)+\left(\tilde{U}^\dagger_\mathbf{k}\partial_{k_i}\tilde{U}_\mathbf{k}\tilde{U}^\dagger_\mathbf{k}\partial_{k_j}\tilde{U}_\mathbf{k}\right)\right].
\end{split}
\end{equation}
The integrand is the Fubini-Study metric $g_{ij}(\mbf{k})$. Abbreviating $\del_{k_i}$ as $\del_i$, this is because
\bea
\label{eq:gabeliannote}
g_{ij} &= \frac{1}{2} \Tr \del_i P \del_j P \\
&= \frac{1}{2} \Tr \del_i (\tilde{U} \tilde{U}^\dag)  \del_j (\tilde{U} \tilde{U}^\dag) \\
&= \frac{1}{2} \Tr \Big( \del_i \tilde{U} \del_j \tilde{U}^\dag +  \tilde{U} \del_i \tilde{U}^\dag \del_j \tilde{U} \, \tilde{U}^\dag + \del_i \tilde{U} \,\tilde{U}^\dag \del_j \tilde{U} \,\tilde{U}^\dag + \tilde{U} \del_i \tilde{U}^\dag \tilde{U} \del_j \tilde{U}^\dag\Big) \\
&=  \frac{1}{2}\Tr \Big( \del_i \tilde{U} \del_j \tilde{U}^\dag +\del_i \tilde{U}^\dag \del_j \tilde{U} + \tilde{U}^\dag \del_i \tilde{U} \,\tilde{U}^\dag \del_j \tilde{U} \, + \del_i \tilde{U}^\dag \tilde{U} \del_j \tilde{U}^\dag \tilde{U} \Big) \\
&=  \frac{1}{2}\Tr \Big(\del_i \tilde{U}^\dag  \del_j \tilde{U}  + \del_j \tilde{U}^\dag  \del_i \tilde{U} + 2 \tilde{U}^\dag \del_i \tilde{U} \,\tilde{U}^\dag \del_j \tilde{U}  \Big) \\
\eea
using $\del_i \tilde{U}^\dag \, \tilde{U} = - \tilde{U}^\dag \del_i \tilde{U}$. It is worth noting from \Eq{eq:gabeliannote} that only the diagonal terms of the quantum metric appear in the superfluid weight due to the trace over the occupied bands. 

The $C_3$ symmetry of our model imposes $[C_3^{-1} D_s C_3]_{ij} = [D_s]_{ij}$. We can then consider, as above, a gauge potential $\mathbf{A}=A\mathbf{\hat{y}}$ and identify $\frac{1}{2}[D_{s}]^i_i$=$D_s$.
We have
\begin{equation}
    D_s=\frac{\Delta^2}{2\sqrt{(E_0-\mu)^2+\Delta^2}}\int \frac{d^2k}{(2\pi)^2} g_i^i(\mathbf{k})
\end{equation}

We can further simplify the equation for the superfluid weight by expresing the chemical potential in terms of the flat band filling $\nu$. The filling at momentum $\mathbf{k}$ is given by $\lvert v_{\mathbf{k}} \rvert^2$, with
\begin{equation}
    v_\mathbf{k}^2=\frac{1}{2}\left(1-\frac{E_0-\mu}{\sqrt{(E_0-\mu)^2+\Delta^2}}\right),
\end{equation}
where we explicitly took advantage of the fact that the energy is momentum-independent in a flat band and hence $v_\mathbf{k}$ is also momentum independent. Therefore, $\nu=v_{\mathbf{k}}^2$ and
\begin{equation}
    D_s=\lvert\Delta\rvert\sqrt{\nu(1-\nu)}\int \frac{d^2k}{(2\pi)^2} g_i^i(\mathbf{k})
\end{equation}
which is coordinate invariant and gauge invariant under transformations of the eigenvectors. 


We can also solve the gap equation that relates $\Delta$ to $\lvert U \rvert$. This goal can be achieved analytically in the isolated flat band limit considered here \cite{2015NatCo...6.8944P}.
The gap equation at zero temperature is
\begin{equation}
    \Delta_\alpha=\Delta=-\frac{\lvert U \rvert}{N_c}\sum_{n\mathbf{k}}u_\mathbf{k}v_\mathbf{k} [\tilde{U}_\mathbf{k}\tilde{U}^\dagger_\mathbf{k}]_{\alpha\alpha},
\end{equation}
where $N_c$ is the number of unit cell. In the simple case of an isolated flat band $u_\mathbf{k}=u=\sqrt{1-\nu}$ and $v_\mathbf{k}=v=\sqrt{\nu}$ \cite{2015NatCo...6.8944P}. Moreover, since our model satisfies the uniform pairing condition, we have $[\tilde{U}_\mathbf{k}\tilde{U}^\dagger_\mathbf{k}]_{\alpha\alpha}=n_\phi$, where $n_\phi$ is the inverse of the number of orbitals over which the flat band has nonzero weight. Finally, we obtain a simple expression for the superfluid weight projected into the flat band of an attractive Hubbard model:
\begin{equation}
    D_s = \lvert U\rvert n_\phi \sqrt{\nu(1-\nu)}\int \frac{d^2k}{(2\pi)^2} g_i^i(\mathbf{k}).
\end{equation}
The superfluid weight obtained for the model studied is shown by the blue cross in Fig. 3 of the main text.

\section{Lower Bounds on the Superfluid Weight from Topological Quantum Chemistry}
\label{app:SFWbounds}

In this Appendix, we prove general lower bounds on the trace of the quantum metric which depend on the orbitals of the lattice and the symmetry data. Our strategy is outlined in \App{app:outline}. In \App{app:realspaceproj}, we construct lower bounds in the abelian groups with either rotations or a single mirror when all orbitals are the the center of the unit cell. Then in \App{app:tables}, we use RSI reduction to tabulate bounds in all 2D space groups. \App{app:lemmas} contains two simple proofs, one of an elementary Frobenius norm inequality and the other of the ``concentration lemma" used to perform an optimization.

\subsection{Outline}
\label{app:outline}

The general procedure for obtaining RSI bounds on the superfluid weight follows the same logic as in the Main Text. The essential pieces are (1) a Fourier transform of $P(\mbf{k})$ to real space harmonics $p(\mbf{R})$, (2) taking appropriate linear combinations of $P(\mbf{k})$ to isolate symmetry-related harmonics $p(\mbf{R})$, (3) using a Frobenius norm inequality to bound the projectors by irrep multiplicities, and (4) minimizing $p(\mbf{R})$ subject to the irrep constraints. If all orbitals are in the same location, then these steps follow essentially identically to the Main Text. This is because the embedding matrix $V[\mbf{G}]$ is proportional to the identity if all orbitals are in the same location, which simplifies many formulas. In \App{app:realspaceproj}, we derive explicit bounds assuming that all orbitals are at the same position which we can take to be the 1a position (origin) without loss of generality. We emphasize that arbitrary irreps may appear at this position. The method to obtain these bounds is significantly different than early work which relied on the Chern number and Euler number \cite{2015NatCo...6.8944P,PhysRevLett.124.167002}. Our bounds are nonzero in obstructed atomic insulators which have zero Berry curvature and flat Wilson loop spectra (zero Chern number and Euler number). Our bound is instead formulated in real space and gives lower bounds on the superfluid weight in all symmetry-protected phases: obstructed atomic, fragile topological, and stable topological. Expressions for the RSIs in all space groups can be found in \Ref{2020Sci...367..794S}. Thus our bounds can be calculated from only the symmetry data and the location of the underlying orbitals.

Complications arise when orbitals are in general locations because the embedding matrix is nontrivial. In this case, $P(\mbf{k})$ is only periodic over the BZ up to an embedding matrix $V[\mbf{G}]$, and the momentum space symmetry operators $V^\dag[g\mbf{K}-\mbf{K}] D[g]$ depend on the high-symmetry point $\mbf{K}$. Nevertheless, we obtain bounds in this general case in \App{app:genorb}. The final form of the bounds is the same: nonzero RSIs \emph{off} the atomic sites contribute to the superfluid weight. However, evaluating the bounds requires computing some geometric quantities that depend on the positions of the orbitals.

\subsection{Lower bounds from Rotational Symmetry and Mirror Symmetry}
\label{app:realspaceproj}

We first obtain a real space expression for
\bea
G &= \int \frac{dk^1 dk^2}{(2\pi)^2\Omega_c} \frac{1}{2}\Tr \del^\mu P \del_\mu P
\eea
where $P(\mbf{k})$ is the projector onto the occupied bands (see \App{app:qgm}). Practically, the advantage of the real space decomposition, i.e. Fourier transforming $P(\mbf{k})$, is that each term in the expansion (Eq. 14 of the Main Text) has positive definite coefficients except for a single ``zero mode."  From \Eq{eq:Vperiod}, we know that $P(\mbf{k})=P(\mbf{k}+\mbf{G})$ is periodic over the BZ when there are only 1a orbitals because the embedding matrix satisfyies $V[\mbf{G}] = \mathbb{1}$. Thus $P(\mbf{k})$ has a Fourier decomposition
\bea
\label{eq:Fourierharmonics}
P(\mbf{k}) &= \sum_\mbf{R} e^{-i \mbf{R} \cdot \mbf{k}} p(\mbf{R}), \qquad p(\mbf{R}) = \int \frac{dk^1dk^2}{(2\pi)^2} e^{i \mbf{R} \cdot \mbf{k}} P(\mbf{k})
\eea
defined in terms of the harmonics $p(\mbf{R})$, which are $N_{\mathrm{orb}} \times N_{\mathrm{orb}}$ matrices, and the lattice vectors $\mbf{R}$. Note that we use a continuous BZ, e.g. infinite boundary conditions, in \Eq{eq:Fourierharmonics} because we need momentum space derivatives to define the quantum metric.  However in practice, the integrals can be approximated by sums assuming periodic boundary conditions. Here and for the rest of this section, we work in crystalline coordinates. For some intuition, recall that $P(\mbf{k})$ is the equal-time momentum space Green's function:
\bea
P(\mbf{k}) = \frac{1}{2\pi i} \oint \frac{dz}{h(\mbf{k}) - z \mathbb{1}}
\eea
where the contour surrounds the energies of the occupied bands. Thus the Fourier transform $p(\mbf{R})$ is just the one-body Green's function matrix. We refer to $p(\mbf{R})$ as the harmonics of $P(\mbf{k})$. Because $D[g]P(\mbf{k})D^\dag[g] = P(g\mbf{k})$ and $P^\dag(\mbf{k}) =P(\mbf{k})$, we find that the harmonics obey
\bea
D[g]p(\mbf{R})D^\dag[g] = p(g \mbf{R}), \qquad p^\dag(\mbf{R}) = p(-\mbf{R})
\eea
by plugging into \Eq{eq:Fourierharmonics}. Taking norms, we find $||p(\mbf{R})|| = || p(-\mbf{R})|| = ||p(g\mbf{R})||$. Importantly, the harmonics obey a normalization condition
\bea
\label{eq:normalization}
\sum_\mbf{R} ||p(\mbf{R})||^2 &= \int \frac{dk^1dk^2}{(2\pi)^2} \frac{dk'^1dk'^2}{(2\pi)^2} \sum_{\mbf{R}}e^{i \mbf{R} \cdot (\mbf{k}-\mbf{k}')} \Tr P^\dag(\mbf{k}') P(\mbf{k}) \\
&= \int \frac{dk^1dk^2}{(2\pi)^2} \Tr P(\mbf{k})^2 \\
&= \int \frac{dk^1dk^2}{(2\pi)^2} N_{\mathrm{occ}} \\
&= N_{\mathrm{occ}} \\
\eea
using $\Tr P(\mbf{k})^2 =\Tr  P(\mbf{k}) = N_{\mathrm{occ}}$. We now transform $G$ to real space:
\bea
\label{eq:Grealspace1a}
G &= \int \frac{dk^1 dk^2}{(2\pi)^2\Omega_c} \frac{1}{2} \sum_{\mbf{R},\mbf{R}'} \mbf{R}\cdot \mbf{R}' e^{i \mbf{k}\cdot(\mbf{R}-\mbf{R}')} \Tr p^\dag(\mbf{R}') p(\mbf{R}) =\sum_{\mbf{R}}  \frac{1}{2\Omega_c} |\mbf{R}|^2 \Tr p^\dag(\mbf{R}) p(\mbf{R}) \\
&=\sum_{\mbf{R}}  \frac{1}{2\Omega_c} |\mbf{R}|^2 ||p(\mbf{R})||^2 \ .\\
\eea
We note that $G$ is dimensionless because $\Omega_c = |\mbf{a}_1\times \mbf{a}_2|$ and $|\mbf{R}|^2$ have units of area. \Eq{eq:Grealspace1a} shows that $G$ is entirely determined by the Bravais lattice and the harmonics (or Green's function). We now show that the symmetry data places constraints on $||p(\mbf{R})||$. Our approach is to show that $||p(\mbf{R})||^2 > p$ for some known $p$ and finite $|\mbf{R}|$, in which case $G \geq \frac{1}{2}|\mbf{R}|^2 p$.

\begin{figure}
 \centering
 \includegraphics[height=5cm]{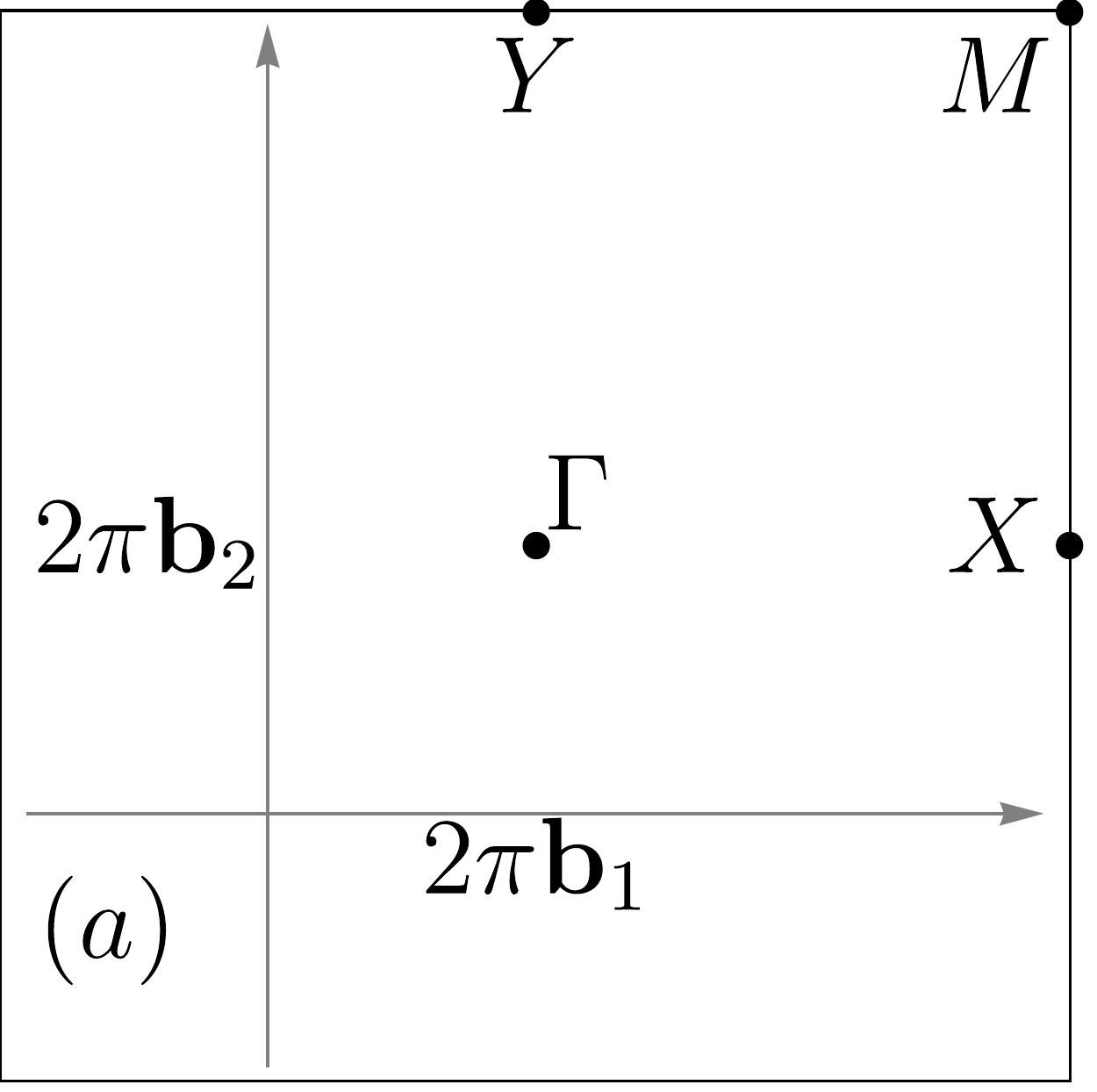} \qquad \includegraphics[height=5cm]{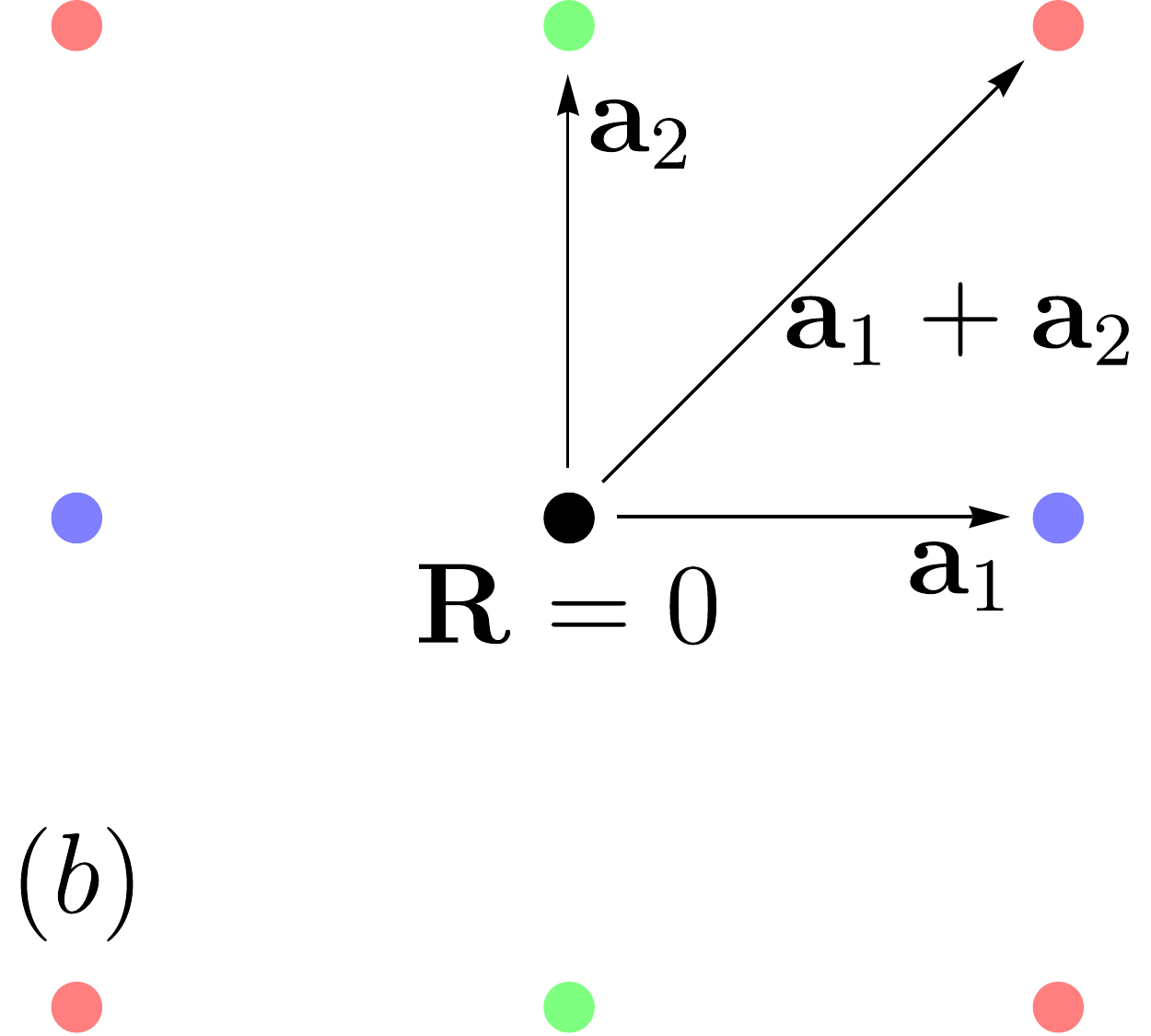}
\caption{$C_2,C_4$ Bounds. $(a)$ We label the $C_2$-invariant points in the BZ. $(b)$ By taking linear combinations of $P(\mbf{K})$, we find lower bounds for the harmonics at $|\mbf{R}| = |\mbf{a}_1|$, shown in red, blue, and green. Higher harmonics are not shown. \label{fig:inv}
\label{fig:appc2}
}
\end{figure}

We first consider the rectangular Bravais lattice first with $C_2$ symmetry. In momentum space, the high-symmetry points are $\Gamma = (0,0), X = \pi \mbf{b}_1, Y = \pi \mbf{b}_2, M = \pi \mbf{b}_1+\pi \mbf{b}_2$ which are all symmetric under $C_2$, and $\Gamma, M$ are symmetric under $C_4$ (see \Fig{fig:appc2}a). At these momenta, $P(\mbf{k})$ is constrained by the symmetry data which we consider known.

By taking linear combinations of projectors at these momenta, we find four linearly independent relations
\bea
\label{eq:P4}
P(\Gamma) + P(X) + P(Y) + P(M) &=  \sum_{\mbf{L} = 2\mathds{Z} \mbf{a}_1+ 2\mathds{Z} \mbf{a}_2} 4p(\mbf{L}) = 4p(0) + \dots \\
P(\Gamma) - P(X) + P(Y) - P(M) &=  \sum_{\mbf{L} = (2\mathds{Z}+1) \mbf{a}_1+ 2\mathds{Z} \mbf{a}_2} 4p(\mbf{L}) = 4p(\mbf{a}_1)+4p(-\mbf{a}_1) + \dots\\
P(\Gamma) + P(X) - P(Y) - P(M) &=  \sum_{\mbf{L} = 2\mathds{Z} \mbf{a}_1+ (2\mathds{Z}+1) \mbf{a}_2} 4p(\mbf{L}) = 4p(\mbf{a}_2)+4p(-\mbf{a}_2) + \dots\\\\
P(\Gamma) - P(X) - P(Y) + P(M) &=  \sum_{\mbf{L} = (2\mathds{Z}+1) \mbf{a}_1+ (2\mathds{Z}+1) \mbf{a}_2} 4p(\mbf{L}) = 4(p(\mbf{a}_1+\mbf{a}_2)+p(\mbf{a}_1-\mbf{a}_2)+p(-\mbf{a}_1+\mbf{a}_2)+p(-\mbf{a}_1-\mbf{a}_2) )+ \dots\\\\
\eea
where we see that the cancelations of the phases have isolated certain sublattices $\mbf{L}$ of the Bravais lattice, illustrated by keeping the first few terms in the series with the dots corresponding to higher harmonics. The smallest $|\mbf{R}|$ terms in each sublattices $\mbf{L}$ are shown in \Fig{fig:appc2}b. Note that the last three lines of \Eq{eq:P4} have canceled the $p(0)$ harmonic, so all terms in the sum have nonzero $|\mbf{R}|^2$. Applying the triangle inequality $||A+B|| \leq ||A|| + ||B||$ to \Eq{eq:P4}, we obtain
\bea
||P(\Gamma) - P(X) + P(Y) - P(M) || &\leq \sum_{\mbf{L} = (2\mathds{Z}+1) \mbf{a}_1+ 2\mathds{Z} \mbf{a}_2} 4||p(\mbf{L})|| \\
||P(\Gamma) + P(X) - P(Y) - P(M)|| &\leq  \sum_{\mbf{L} = 2\mathds{Z} \mbf{a}_1+ (2\mathds{Z}+1) \mbf{a}_2} 4||p(\mbf{L})|| \\
||P(\Gamma) - P(X) - P(Y) + P(M)|| &\leq  \sum_{\mbf{L} = (2\mathds{Z}+1) \mbf{a}_1+ (2\mathds{Z}+1) \mbf{a}_2} 4||p(\mbf{L})|| \\
\eea
which gives lower bounds on the (sum of) nonzero harmonics on the right-hand side. Focusing on the left-hand side, we invoke the inequality $||A||^2 \geq | \Tr S A|^2 / \text{Rk}(A)$ for all unitary $S$ (proven in \App{app:lemmas}). Here $\text{Rk}$ is the matrix rank. Choosing $S = D[C_2]$ where $D[C_2]$ is the (unitary) representation of the $C_2$ operator on the orbitals will allow us to bound e.g. $||P(\Gamma) - P(X) + P(Y) - P(M) ||$ by the symmetry data using the character formulas in \Eq{eq:trPtoirrep}. Comparing with the RSI expressions in \Ref{2020Sci...367..794S}, we compute
\bea
\label{eq:trDc2U}
\Tr D[C_2] (P(\Gamma) - P(X) + P(Y) - P(M) ) &= m(\Gamma_1) - m(\Gamma_2) - m(X_1) + m(X_2)+ m(Y_1) - m(Y_2) - m(M_1) + m(M_2) \\
&= 2 m(\Gamma_1) - 2m(X_1) - 2m(Y_2) + 2m(M_2) \\
&= - 4 \delta_{1b} , \\
\Tr D[C_2] (P(\Gamma) + P(X) - P(Y) - P(M) ) &=  m(\Gamma_1) - m(\Gamma_2)+ m(X_1) - m(X_2)-m(Y_1) + m(Y_2) - m(M_1) + m(M_2) \\
&=  m(\Gamma_1) - m(\Gamma_2)+ m(X_1) - m(X_2)-m(Y_1) + m(Y_2) - m(M_1) + m(M_2) \\
&=  2m(\Gamma_1) + 2m(X_1)+ 2 m(Y_2) - m(M_1) + m(M_2) - 3N_{\mathrm{occ}}\\
&= -4 \delta_{1c} + 3m(M_1) + 3 m(M_2) - 3N_{\mathrm{occ}} \\
&= -4 \delta_{1c} , \\
\Tr D[C_2] (P(\Gamma) - P(X) - P(Y) + P(M) ) &= m(\Gamma_1) - m(\Gamma_2) - m(X_1) + m(X_2) - m(Y_1) + m(Y_2) + m(M_1) - m(M_2) \\
&= 2m(\Gamma_1)  - 2m(X_1)+ 2m(Y_2) - 2m(A_2) \\
&= -4 \delta_{1d}
\eea
where $m(\rho)$ is the multiplicity of the $\rho$ irrep in little group $G_{\mbf{K}}$, and we have made use of the compatibility relations $\Tr P(\mbf{K}) = \sum_{\rho \in G_{\mbf{K}}} m(\rho) \dim \rho = N_{\mathrm{occ}}$. The RSIs $\delta_{1b},\delta_{1c},\delta_{1d} \in \mathds{Z}$ are protected by the $C_2$ point group at the 1b = $\mbf{a}_1/2$, 1c = $\mbf{a}_2/2$, and 1d = $\mbf{a}_1/2+\mbf{a}_2/2$ Wyckoff positions respectively \cite{2020Sci...367..794S}. We also need the basic rank inequalities $\text{Rk} (P_1 + \dots + P_n) \leq \text{Rk } P_1 + \dots + \text{Rk } P_n = n N_{\mathrm{occ}}$ and also $\text{Rk} (P_1 + \dots + P_n) \leq \dim (P_1 + \dots + P_n) = N_{\mathrm{orb}}$ for rank $N_{\mathrm{occ}}$ projectors $P_n$. Hence we have the bounds
\bea
\label{eq:c2Lcon}
\frac{1}{\sqrt{\min \{N_{\mathrm{orb}},4N_{\mathrm{occ}}\}}} |4 \delta_{1b}| &\leq \sum_{\mbf{L} = (2\mathds{Z}+1) \mbf{a}_1+ 2\mathds{Z} \mbf{a}_2} 4||p(\mbf{L})|| = 8 ||p(\mbf{a}_1)|| + \dots  \\
\frac{1}{\sqrt{\min \{N_{\mathrm{orb}},4N_{\mathrm{occ}}\}}} |4 \delta_{1c}| &\leq  \sum_{\mbf{L} = 2\mathds{Z} \mbf{a}_1+ (2\mathds{Z}+1) \mbf{a}_2} 4||p(\mbf{L})|| = 8 ||p(\mbf{a}_2)|| + \dots \\
\frac{1}{\sqrt{\min \{N_{\mathrm{orb}},4N_{\mathrm{occ}}\}}} |4 \delta_{1d}|&\leq  \sum_{\mbf{L} = (2\mathds{Z}+1) \mbf{a}_1+ (2\mathds{Z}+1) \mbf{a}_2} 4||p(\mbf{L})|| = 8 ||p(\mbf{a}_1+\mbf{a}_2)|| + 8 ||p(\mbf{a}_1-\mbf{a}_2)|| + \dots \\
\eea
where we used $||p(\mbf{R})|| = ||p(-\mbf{R})||$ to separate out the lowest harmonics. The constraints in \Eq{eq:c2Lcon} provide lower bounds on $G$ because they show that the sum of the norms of nonzero harmonics is lower bounded, and hence at least one of the harmonics cannot be zero. Intuitively, we expect a lower bound for $G$ to be obtained if the harmonics $||p(\mbf{R})||$ with the \emph{smallest} $|\mbf{R}|^2$ saturate \Eq{eq:c2Lcon}, and all the rest were zero. A simple result called the concentration lemma shows that this is the case. It is proved in \App{app:lemmas} using an optimization argument. Hence a lower bound for $G$ comes from taking saturating \Eq{eq:c2Lcon} on the lowest harmonics (already indicated)
\bea
\label{eq:c2LconB}
\frac{1}{2\sqrt{\min \{N_{\mathrm{orb}},4N_{\mathrm{occ}}\}}} |\delta_{1b}| &= ||p(\mbf{a}_1)|| \\
\frac{1}{2\sqrt{\min \{N_{\mathrm{orb}},4N_{\mathrm{occ}}\}}} |\delta_{1c}| &= ||p(\mbf{a}_2)|| \\
\frac{1}{2\sqrt{\min \{N_{\mathrm{orb}},4N_{\mathrm{occ}}\}}} |\delta_{1d}|&= ||p(\mbf{a}_1+\mbf{a}_2)|| + ||p(\mbf{a}_1-\mbf{a}_2)|| \ . \\
\eea
The first two lines are simple to plug into $G$ for a lower bound. The third line requires one more step because $||p(\mbf{a}_1+\mbf{a}_2)||$ and $||p(\mbf{a}_1-\mbf{a}_2)||$ need not be equal with only $C_2$ symmetry. To deal with this, we use the basic fact that $\min_{x+y=c} 2(a x^2 + b y^2) = \frac{ab}{a+b}c^2$ which can be proven with calculus. With this fact, we use \Eq{eq:Grealspace1a} to obtain the bound
\bea
\label{eq:GboundC2}
G &\geq \frac{1}{2|\mbf{a}_1 \times \mbf{a}_2|} \lp 2 |\mbf{a}_1|^2 \cdot ||p(\mbf{a}_1)||^2 + 2  |\mbf{a}_2|^2 \cdot ||p(\mbf{a}_2)||^2 + 2  |\mbf{a}_1+\mbf{a}_2|^2 \cdot ||p(\mbf{a}_1+\mbf{a}_2)||^2 + 2   |\mbf{a}_1-\mbf{a}_2|^2 \cdot ||p(\mbf{a}_1-\mbf{a}_2)||^2  \rp \\
&\geq \frac{1}{4 \min \{N_{\mathrm{orb}},4N_{\mathrm{occ}}\}} \lp |\mbf{a}_1|^2\delta^2_{1b} +|\mbf{a}_2|^2\delta_{1c}^2+\frac{ |\mbf{a}_1+\mbf{a}_2|^2 |\mbf{a}_1-\mbf{a}_2|^2}{ |\mbf{a}_1+\mbf{a}_2|^2+ |\mbf{a}_1-\mbf{a}_2|^2}\delta_{1d}^2 \rp/|\mbf{a}_1 \times \mbf{a}_2| \\
\eea
where the explicit factors of 2 in the first line of \Eq{eq:GboundC2} indicate the contributions from symmetry-related harmonics, e.g. $||p(\mbf{a}_1)||$ and $||p(-\mbf{a}_1)||$. We see that any nonzero RSI off the 1a position contributes to a positive lower bound for $G$. We emphasize that the RSIs can be computed for any symmetry data using the expressions in \Ref{2020Sci...367..794S}, and thus \Eq{eq:GboundC2} gives a bound for OAL, fragile, and stable phases.

We next consider improvements to the bound when $C_4$ symmetry is present. In this case, the $\mbf{a}_1/2$ and $\mbf{a}_2/2$ positions are related by $C_4$ and are conventionally called the 2c position, and the $\mbf{a}_1/2 + \mbf{a}_2/2$ position, which has a $C_4$ symmetry, is called the 1b position. There are two ways to get bounds. First, the $C_2$ bounds in \Eq{eq:GboundC2} also hold in this case. However, we can also use the $C_4$ eigenvalues at the $\Gamma$ and $M$ points. To cancel the $p(0)$ harmonic, we use
\bea
P(\Gamma) - P(M) &=  \sum_{\mbf{L} = \mathds{Z} (\mbf{a}_1 -\mbf{a}_2)+ (2\mathds{Z}+1) \mbf{a}_2} 2p(\mbf{L}) = 2p(\mbf{a}_1) + 2p(-\mbf{a}_1)+ 2p(\mbf{a}_2)+ 2p(-\mbf{a}_2) + \dots \ .
\eea
Consulting the RSI tables in \Ref{2020Sci...367..794S}, the $C_4$ trace formulas give
\bea
\Tr D[C_4] \lp P(\Gamma) -  P(M) \rp &= m(\Gamma_1) - m(\Gamma_2) + i m(\Gamma_3) - i m(\Gamma_4) - m(M_1) + m(M_2) - i m(M_3) + i m(M_4) \\
&= \big( m(\Gamma_1)-m(\Gamma_2)-m(M_1)+ m(M_2) \big) - i \big(m(M_1) + m(M_2) + 2(M_3) - m(\Gamma_1)- m(\Gamma_2)- 2 m(\Gamma_4) \big) \\
&= -2 \delta_{1b,2} + 2 i (\delta_{1b,1} - \delta_{1b,3}) \\
\eea
where we note there are 3 RSIs $\delta_{1b,i} \in \mathds{Z}$ at the 1b position and 1 RSI $\delta_{2c}$ at the 2c position. Applying the triangle inequality and using $\text{Rk}(P(\Gamma) - P(M)) \leq \min\{N_{orb}, 2N_{occ}\}$ yields
\bea
\label{eq:c4con1}
\frac{1}{\sqrt{\min \{N_{\mathrm{orb}},2N_{\mathrm{occ}}\}}} | -2 \delta_{1b,2} + 2 i (\delta_{1b,1} - \delta_{1b,3})| &\leq \sum_{\mbf{L} = \mathds{Z} (\mbf{a}_1 -\mbf{a}_2)+ (2\mathds{Z}+1) \mbf{a}_2} 2||p(\mbf{L})|| = 8 ||p(\mbf{a}_1)|| + \dots  \\
\eea
using $||p(C_4 \mbf{R})|| = ||p(\mbf{R})||$. We also need to write down the $C_2$ bounds in \Eq{eq:c2LconB} in terms of the RSIs in the $C_4$ case. This follows from irrep reduction: the RSIs in the higher symmetry groups reduce to RSIs of the lower symmetry group \cite{2020Sci...367..794S}. We have
\bea
\delta^{(2)}_{1b} = \delta^{(2)}_{1c}  = \delta^{(4)}_{2c} , \qquad \delta^{(2)}_{1d} = \delta^{(4)}_{1b,1}+ \delta^{(4)}_{1b,3}- \delta^{(4)}_{1b,2}
\eea
where the superscript distinguishes the $C_2$ and $C_4$ RSIs. Reducing the $C_2$ bounds in \Eq{eq:c2LconB}, we find
\bea
\label{eq:c4Lcon2}
\frac{1}{\sqrt{\min \{N_{\mathrm{orb}},4N_{\mathrm{occ}}\}}} |4 \delta_{2c}| &\leq \sum_{\mbf{L} = (2\mathds{Z}+1) \mbf{a}_1+ 2\mathds{Z} \mbf{a}_2} 4||p(\mbf{L})|| = 8 ||p(\mbf{a}_1)|| + \dots  \\
\frac{1}{\sqrt{\min \{N_{\mathrm{orb}},4N_{\mathrm{occ}}\}}} |4 (\delta_{1b,1}+\delta_{1b,3}-\delta_{1b,2})|&\leq  \sum_{\mbf{L} = (2\mathds{Z}+1) \mbf{a}_1+ (2\mathds{Z}+1) \mbf{a}_2} 4||p(\mbf{L})|| = 16 ||p(\mbf{a}_1+\mbf{a}_2)|| + \dots  \\
\eea
using $||p(C_4\mbf{R})|| = ||p(\mbf{R})||$ to simplify the expressions. With the concentration lemma given in \App{app:lemmas}, a lower bound for $G$ comes from saturating the inequalities \Eqs{eq:c4con1}{eq:c4Lcon2} to find
\bea
G &\geq \frac{1}{2} \lp 4 |\mbf{a}_1|^2 \cdot ||p(\mbf{a}_1)||^2 + 4 |\mbf{a}_1+\mbf{a}_2|^2 \cdot ||p(\mbf{a}_1+\mbf{a}_2)||^2 \rp / |\mbf{a}_1\times \mbf{a}_2|\\
&\geq \max \left\{ \frac{\delta_{1b,2}^2 + (\delta_{1b,1}-\delta_{1b,3})^2}{8\min\{N_{\mathrm{orb}},2 N_{\mathrm{occ}}\}} ,  \frac{\delta_{2c}^2}{2\min\{N_{\mathrm{orb}},4 N_{\mathrm{occ}}\}} \right\}+ \frac{(\delta_{1b,1}+\delta_{1b,3}-\delta_{1b,2})^2}{4 \min \{N_{\mathrm{orb}},4N_{\mathrm{occ}}\}} \\
\eea
where the max appears because there are two constraints on $||p(\mbf{a}_1)||$, and we used $\mbf{a}_1 \cdot \mbf{a}_2 = 0$. Again we see that nonzero RSIs off the 1a position give a nontrivial lower bound.

\begin{figure}
 \centering
 \includegraphics[height=5cm]{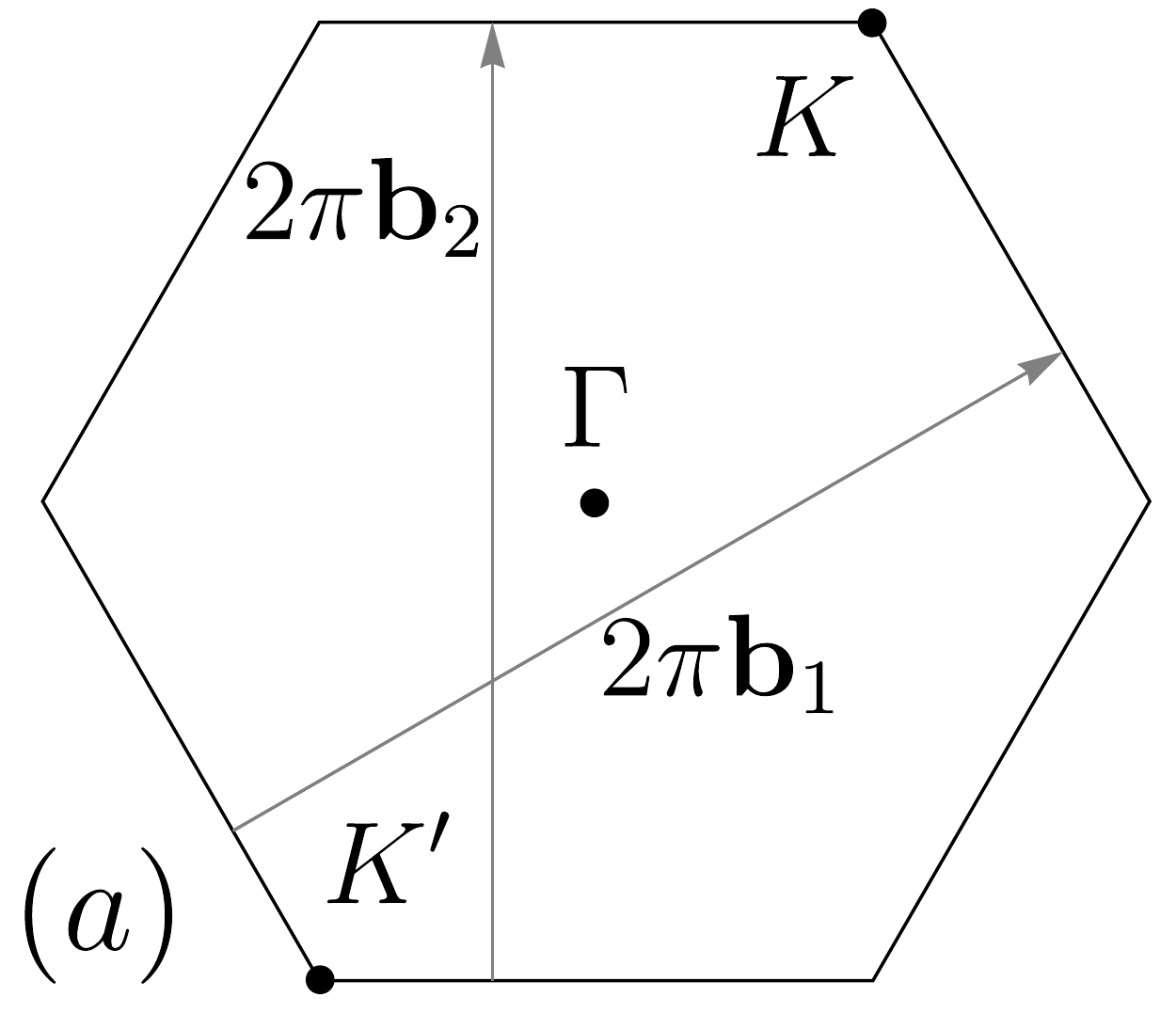} \qquad \includegraphics[height=5cm]{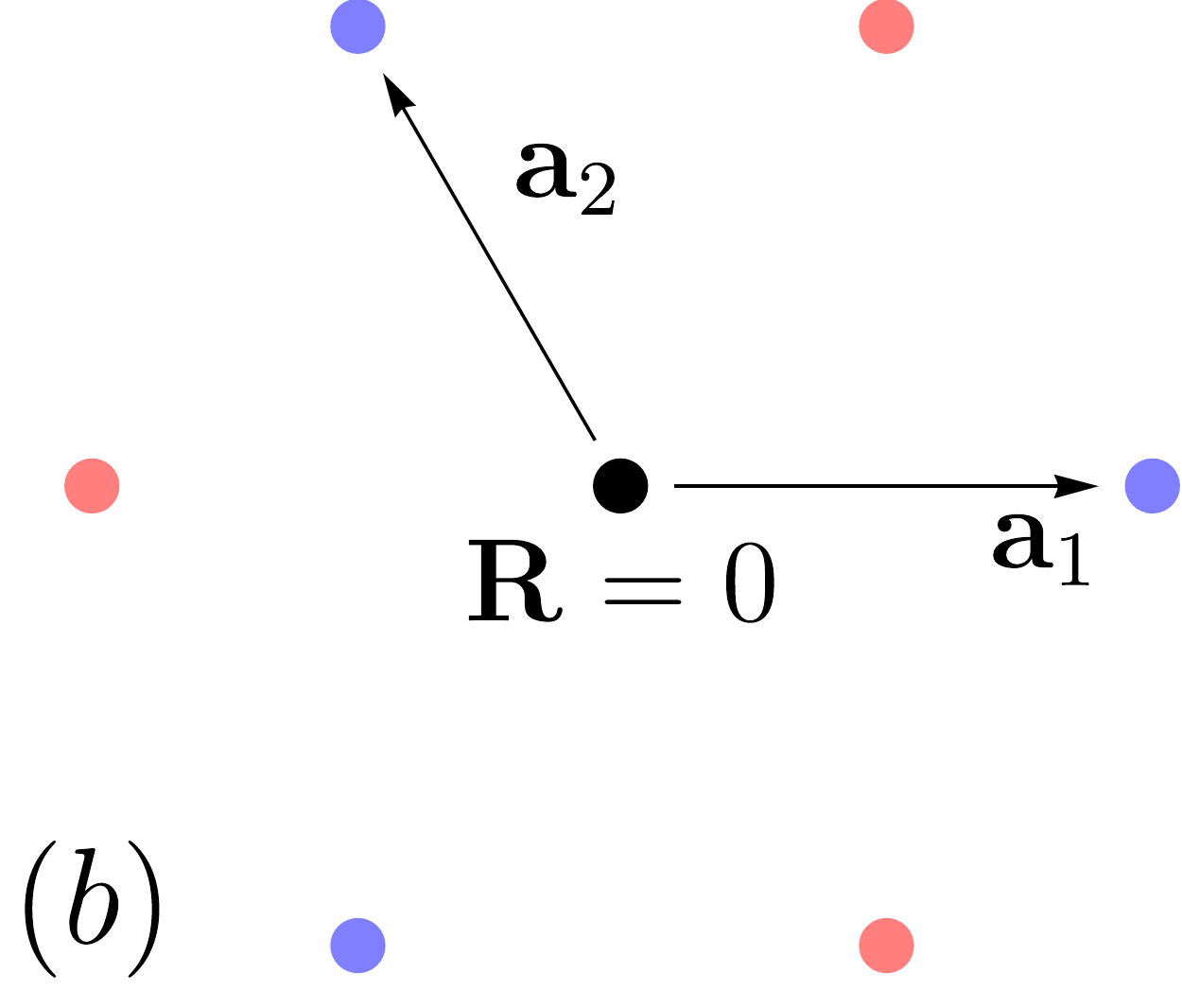}
\caption{$C_3$ Bounds. $(a)$ We label the $C_3$-invariant points in the BZ. $(b)$ By taking linear combinations of $P(\mbf{K})$, we find lower bounds for the harmonics at $|\mbf{R}| = |\mbf{a}_1|$, shown in red and blue. Higher harmonics are not shown.
 \label{fig:c3app}
}
\end{figure}

We now move to the cases of $C_3$ and $C_6$.  In momentum space, the high-symmetry points are $\Gamma = (0,0), K = \frac{2\pi}{3} (\mbf{b}_1+\mbf{b}_2), K' = -\frac{2\pi}{3} (\mbf{b}_1+\mbf{b}_2)$ which are $C_3$ symmetric. $K$ and $K'$ are related by $C_6$, and because $C_6^3 = C_2$, $X,Y$, and $M$ are high-symmetry points and are related by $C_6$ (they are all $M$ points).  At these momenta, $P(\mbf{k})$ is constrained by the symmetry data which we consider known. By taking linear combinations of projectors at these momenta, we find three linearly independent relations (generalizing Eq. 16 of the Main Text)
\bea
\label{eq:P3}
P(\Gamma) + P(K) + P(K') &= \sum_{\mbf{L} = (2\mbf{a}_1+\mbf{a}_2)\mathds{Z}+ (\mbf{a}_1+2\mbf{a}_2)\mathds{Z}} 3 p(\mbf{L}) = 3 p(0) + \dots \\
P(\Gamma) + e^{- \frac{2\pi i}{3}} P(K) + e^{\frac{2\pi i}{3}} P(K') &= \sum_{\mbf{L} = (2\mbf{a}_1+\mbf{a}_2)\mathds{Z}+ (\mbf{a}_1+2\mbf{a}_2)\mathds{Z}+\mbf{a}_1} 3 p(\mbf{L}) = 3 p(\mbf{a}_1) +  3 p(\mbf{a}_2)+ 3 p(-\mbf{a}_1-\mbf{a}_2) + \dots \\
P(\Gamma) + e^{\frac{2\pi i}{3}} P(K) + e^{-\frac{2\pi i}{3}} P(K') &= \sum_{\mbf{L} = (2\mbf{a}_1+\mbf{a}_2)\mathds{Z}+ (\mbf{a}_1+2\mbf{a}_2)\mathds{Z}-\mbf{a}_1} 3 p(\mbf{L}) = 3 p(-\mbf{a}_1) +  3 p(-\mbf{a}_2)+ 3 p(\mbf{a}_1+\mbf{a}_2) + \dots \\
\eea
where we see that the cancelations of the phases have isolated certain sublattices $\mbf{L}$ of the Bravais lattice, illustrated by keeping the first few terms in the series with the dots corresponding to higher harmonics. The smallest $|\mbf{R}|$ terms in each sublattices $\mbf{L}$ are shown in \Fig{fig:appc2}b. Note that the last two lines of \Eq{eq:P3} have canceled the $p(0)$ harmonic, so all terms in the sum have nonzero $|\mbf{R}|^2$. Applying the triangle inequality $||A+B|| \leq ||A|| + ||B||$ to \Eq{eq:P3}, we obtain
\bea
||P(\Gamma) + e^{- \frac{2\pi i}{3}} P(K) + e^{\frac{2\pi i}{3}} P(K')|| &\leq  \sum_{\mbf{L} = (2\mbf{a}_1+\mbf{a}_2)\mathds{Z}+ (\mbf{a}_1+2\mbf{a}_2)\mathds{Z}+\mbf{a}_1} 3 ||p(\mbf{L})|| \\
||P(\Gamma) + e^{\frac{2\pi i}{3}} P(K) + e^{-\frac{2\pi i}{3}} P(K')|| &\leq  \sum_{\mbf{L} = (2\mbf{a}_1+\mbf{a}_2)\mathds{Z}+ (\mbf{a}_1+2\mbf{a}_2)\mathds{Z}-\mbf{a}_1} 3|| p(\mbf{L}) ||
\eea
which gives lower bounds on the (sum of) nonzero harmonics. Focusing on the left-hand side, we invoke the inequality $||A||^2 \geq | \Tr S A|^2 / \text{Rk}(A)$ for all unitary $S$ (proven in \App{app:lemmas}). Choosing $S = D[C_3]$, using the character formulas in \Eq{eq:trPtoirrep}, and comparing with the RSI expressions in \Ref{2020Sci...367..794S}, we compute
\bea
\Tr D[C_3](P(\Gamma) + e^{- \frac{2\pi i}{3}} P(K) + e^{\frac{2\pi i}{3}} P(K')) &= m(\Gamma_1) + e^{\frac{2\pi i}{3}} m(\Gamma_2)+ e^{-\frac{2\pi i}{3}} m(\Gamma_3)  \\
& \qquad + e^{-\frac{2\pi i}{3}} m(K_1) + m(K_2)+ e^{\frac{2\pi i}{3}} m(K_3) \\
& \qquad +  e^{\frac{2\pi i}{3}} m(K'_1) +  m(K'_2)+ e^{-\frac{2\pi i}{3}} m(K'_3)  \\
&= \sqrt{3}(e^{i \frac{5\pi}{6}} m(\Gamma_2)+e^{-i \frac{5\pi}{6}} m(\Gamma_3) - i m(K_1) + e^{-i \frac{\pi}{6}} m(K_2)+ i m(K'_1) + e^{i \frac{\pi}{6}} m(K'_2) )\\
&= 3(e^{\frac{2\pi i}{3}} \delta_{1b,1}+e^{-\frac{2\pi i}{3}} \delta_{1b,2}) \\
\Tr D[C_3](P(\Gamma) + e^{\frac{2\pi i}{3}} P(K) + e^{-\frac{2\pi i}{3}} P(K')) &= m(\Gamma_1) + e^{\frac{2\pi i}{3}} m(\Gamma_2)+ e^{-\frac{2\pi i}{3}} m(\Gamma_3)  \\
& \qquad + e^{\frac{2\pi i}{3}} m(K_1) + e^{-\frac{2\pi i}{3}}m(K_2)+  m(K_3) \\
& \qquad +  e^{-\frac{2\pi i}{3}} m(K'_1) + e^{\frac{2\pi i}{3}} m(K'_2)+ m(K'_3)  \\
&= \sqrt{3}(\sqrt{3}m(\Gamma_1)+e^{i \frac{\pi}{6}} m(\Gamma_2)+e^{-i \frac{\pi}{6}} m(\Gamma_3) \\
& \qquad +e^{i \frac{5\pi}{6}} m(K_1)+e^{-i \frac{5\pi}{6}} m(K_2)+e^{-i \frac{5\pi}{6}} m(K'_1)+e^{i \frac{5\pi}{6}} m(K'_2)) \\
&= 3(e^{\frac{2\pi i}{3}} \delta_{1c,1}+e^{-\frac{2\pi i}{3}} \delta_{1c,2}) \\
\eea
where $m(\rho)$ is the multiplicity of the $\rho$ irrep in little group $G_{\mbf{K}}$, and we have made use of the compatibility relations $\Tr P(\mbf{K}) = \sum_{\rho \in G_{\mbf{K}}} m(\rho) \dim \rho = N_{\mathrm{occ}}$. The RSIs $\delta_{1b,1},\delta_{1b,2},\delta_{1c,1},\delta_{1c,2}\in \mathds{Z}$ are protected by the $C_3$ point group at the 1b = $\frac{1}{3}\mbf{a}_1+\frac{2}{3}\mbf{a}_2$ and 1c = $\frac{2}{3}\mbf{a}_1+\frac{1}{3}\mbf{a}_2$ Wyckoff positions \cite{2020Sci...367..794S}. Using the rank inequalities, we have the bounds
\bea
\label{eq:c3Lcon}
\frac{1}{\sqrt{\min \{N_{\mathrm{orb}},3N_{\mathrm{occ}}\}}} |3(e^{\frac{2\pi i}{3}} \delta_{1b,1}+e^{-\frac{2\pi i}{3}} \delta_{1b,2})|  &\leq  \sum_{\mbf{L} = (2\mbf{a}_1+\mbf{a}_2)\mathds{Z}+ (\mbf{a}_1+2\mbf{a}_2)\mathds{Z}+\mbf{a}_1} 3 ||p(\mbf{L})|| = 9 ||p(\mbf{a}_1)||+\dots \\
\frac{1}{\sqrt{\min \{N_{\mathrm{orb}},3N_{\mathrm{occ}}\}}} |3(e^{\frac{2\pi i}{3}} \delta_{1c,1}+e^{-\frac{2\pi i}{3}} \delta_{1c,2})|  &\leq  \sum_{\mbf{L} = (2\mbf{a}_1+\mbf{a}_2)\mathds{Z}+ (\mbf{a}_1+2\mbf{a}_2)\mathds{Z}-\mbf{a}_1} 3|| p(\mbf{L}) || = 9 ||p(-\mbf{a}_1)|| + \dots
\eea
where we used $||p(\mbf{R})|| = ||p(-\mbf{R})||$ due to $p(-\mbf{R}) = p^\dag(\mbf{R})$ to separate out the lowest harmonics. An interesting feature of the $C_3$ case is that $||p(-\mbf{a}_1)|| = ||p(\mbf{a}_1)||$ because $p(-\mbf{R}) = p^\dag(\mbf{R})$, even though the vectors $\mbf{a}_1$ and $-\mbf{a}_1$ are not related by symmetry. Applying the concentration lemma and saturating the inequalities, we find (generalizing Eq. 22 of the Main Text)
\bea
G \geq \frac{1}{2|\mbf{a}_1\times\mbf{a}_2|} (6|\mbf{a}_1|^2 \cdot ||p(\mbf{a}_1)||^2 ) \geq \frac{2}{3\sqrt{3}\min\{ N_{\mathrm{orb}},3N_{\mathrm{occ}}\}} \max \left\{\delta_{1b,1}^2 - \delta_{1b,1} \delta_{1b,2} + \delta_{1b,2}^2,\delta_{1c,1}^2 - \delta_{1c,1} \delta_{1c,2} + \delta_{1c,2}^2 \right\} 
\eea
where we used $|\mbf{a}_1|^2/|\mbf{a}_1 \times \mbf{a}_2| = 2/\sqrt{3}$. As usual, these bounds show that any RSI off the 1a position will contribute to a lower bound.

We now consider $C_6$ symmetry. In fact, $C_6$ does not introduce any new bounds beyond the $C_3 = C_6^2$ and $C_2 = C_6^3$ bounds because only the $\Gamma$ point has $C_6$ irreps. To obtain our bounds, we needed to take linear combinations of projectors at different high-symmetry points. Therefore with $C_6$, we just reduce the RSIs to the $C_2$ and $C_3$ cases. $C_6$ sets $\delta^{(6)}_{2b,i} = \delta^{(3)}_{1b,i}= \delta^{(3)}_{1c,i}$ and $\delta^{(6)}_{3c,1} = \delta^{(2)}_{1b,1}= \delta^{(2)}_{1c,1}= \delta^{(2)}_{1d,1}$. The constraints in \Eqs{eq:c2LconB}{eq:c3Lcon} become
\bea
\frac{1}{\sqrt{\min \{N_{\mathrm{orb}},3N_{\mathrm{occ}}\}}} |3(e^{\frac{2\pi i}{3}} \delta_{2b,1}+e^{-\frac{2\pi i}{3}} \delta_{2b,2})|  &\leq  \sum_{\mbf{L} = (2\mbf{a}_1+\mbf{a}_2)\mathds{Z}+ (\mbf{a}_1+2\mbf{a}_2)\mathds{Z}+\mbf{a}_1} 3 ||p(\mbf{L})|| = 9 ||p(\mbf{a}_1)||+\dots \\
\frac{1}{\sqrt{\min \{N_{\mathrm{orb}},4N_{\mathrm{occ}}\}}} |4 \delta_{3c}| &\leq \sum_{\mbf{L} = (2\mathds{Z}+1) \mbf{a}_1+ 2\mathds{Z} \mbf{a}_2} 4||p(\mbf{L})|| = 8 ||p(\mbf{a}_1)|| + \dots \ . \\
\eea
Using the concentration lemma, we find that
\bea
G \geq \frac{1}{2|\mbf{a}_1\times \mbf{a}_2|} (6 |\mbf{a}_1|^2 \cdot ||p(\mbf{a}_1)||^2 ) \geq \frac{2}{\sqrt{3}} \max \left\{  \frac{\delta_{2b,1}^2 - \delta_{2b,1} \delta_{2b,2} + \delta_{2b,2}^2}{3\min\{ N_{\mathrm{orb}},3N_{\mathrm{occ}}\}} , \frac{\delta_{3c,1}^2}{4\min\{ N_{\mathrm{orb}},4N_{\mathrm{occ}}\}}  \right\} \ .
\eea
This completes the bounds for the rotational symmetries. We now need to discuss mirror symmetries. If the only symmetry of the model is a mirror then we can use a new approach because mirror is quasi-1D symmetry. Note that the only groups with a mirror and no rotation are $pm$ and $cm$. We only consider $pm$ because $cm$ has only a single high-symmetry Wyckoff position, and thus has no OWC phases.

Mirror symmetry is quasi-1D because it only acts nontrivially on one spatial component. Its Wyckoff positions and high-symmetry lines in the BZ are one-dimensional and lie parallel to the mirror axis. For this reason, the $M$ irreps provide much more data about the projectors than the rotation symmetries, which only provide information at high-symmetry \emph{points}. Without loss of generality, we choose the mirror plane such that $M\hat{x} = - \hat{x}$ and $P(M\mbf{k}) = D[M]P(\mbf{k})D^\dag[M]$. A bound follows immediately:
\bea
G &= \frac{1}{2} \int \frac{dk_xdk_y}{(2\pi)^2} \Tr \lp (\del_{k_x}P)^2 + (\del_{k_y}P)^2 \rp \geq \frac{1}{2} \int \frac{dk_xdk_y}{(2\pi)^2} \Tr (\del_{k_x}P)^2 \\
&= \frac{1}{2\pi}  \int \frac{dk_y}{2\pi} \int_0^\pi dk_x ||\del_{k_x}P||^2 \\
&\geq \frac{1}{2\pi} \int \frac{dk_y}{2\pi} \left|\left| \int_0^\pi dk_x \del_{k_x} P \right| \right|^2 \\
&= \frac{1}{2\pi} \int \frac{dk_y}{2\pi} \left|\left| P(\pi,k_y) - P(0,k_y) \right| \right|^2 \\
&\geq \frac{1}{2\pi \min \{2 N_{\mathrm{occ}},N_{\mathrm{orb}}\}} \int \frac{dk_y}{2\pi} \left| \Tr D[M](P(\pi,k_y) - P(0,k_y)) \right|^2 \\
\eea
where the inequality in the third line is Cauchy-Schwartz, and the inequality in the last line is $||A||^2 \geq |\Tr S A|^2/ \text{Rk}(A)$ with $A=P(\pi,k_y) - P(0,k_y)$ and $S=D[M]$. In a gapped band structure, the mirror eigenvalues cannot change along the high symmetry lines $\bar{\Gamma} = (0,k_y)$ and $\bar{X} = (\pi,k_y)$. Thus $| \Tr D[M](P(\pi,k_y) - P(0,k_y))| =| m(\bar{X}_1)-m(\bar{X}_2) - (m(\bar{\Gamma}_1)-m(\bar{\Gamma}_2))| = 2|m(\bar{\Gamma}_2) - m(\bar{X}_2)| $ is independent of $k_y$. Here $\Gamma_1,X_1$ are the $+1$ mirror irreps and $\Gamma_2, X_2$ are the odd mirror irreps. We also used $m(\bar{X}_1/\bar{\Gamma}_1) = N_{\mathrm{occ}} - m(\bar{X}_2/\bar{\Gamma}_2)$. Plugging in, we find
\bea
G \geq \frac{2|m(\bar{\Gamma}_2) - m(\bar{X}_2)|^2}{\pi \min \{2 N_{\mathrm{occ}},N_{\mathrm{orb}}\}} = \frac{2 \delta_{1b}^2}{\pi \min \{2 N_{\mathrm{occ}},N_{\mathrm{orb}}\}}
\eea
where we used the definition of $\delta_{1b}$, the RSI at the 1b $= \hat{x}/2$ position, from \Ref{2020Sci...367..794S}. Our argument here is similar to the winding number bound from \Ref{PhysRevB.94.245149}, which finds that in 1D, $G \geq \frac{W^2}{N_{\mathrm{occ}}}$ where $W$ is a winding number protected by chiral symmetry.

\subsection{Tables}
\label{app:tables}

 The bounds we have obtained for orbitals at the 1a position depend only on the RSIs (which are equivalent to the symmetry data) and the filling numbers of the model $N_{\mathrm{occ}}$ or $N_{\mathrm{orb}}$. As such, our bounds apply to OWCs and fragile topological phases, greatly expanding upon the Chern number and Euler number bounds. However, we point out that our bounds are not saturated and we expect that they can be improved by overall numerical factors as discussed in the Main Text. We suspect that a more careful use of the constraints on $p(\mbf{R})$ is required, and would be worthy of future work.

We now give tables for the lower bounds in all 2D wallpaper groups with and without time reversal symmetry (TRS, denoted $1'$), with and without spin-orbit coupling (SOC). (In the non-symmorphic groups with a glide symmetry, there are no multiplicity 1 Wyckoff positions, so we exclude these cases.) The tables below are obtained by reducing the RSIs protected by $C_n$ in the presence of mirrors and TRS. For instance in $p3$ which has $C_3$ symmetry, there are two RSIs $\delta_{1b,1}, \delta_{1b,2}$ at the 1b position. Written in terms of Wannier function irreps $A,{}^1E,{}^2E$ at the 1b site, the RSIs are $\delta_{1b,1} = m({}^1E)-m(A),\delta_{1b,2} = m({}^2E)-m(A)$. If mirror symmetry is added at the 1b site, the irreps become $A$ and ${}^1E{}^2E$ which is two dimensional. The two RSIs with $C_3$ are reduced to a single RSI $\delta_{1b,1} = \delta_{1b,2}  \to \delta^{(3m)}_{1b,1} = m({}^1E{}^2E)-m(A)$.  \Ref{2020Sci...367..794S} proves that the RSIs in 2D all groups can be obtained by reduction from the $C_n$ subgroup. Note that without mirrors or time-reversal, the rotation groups with and without SOC are identical (up to an overall factor). However, the reductions are different for groups with and without SOC. Formulas to compute the RSIs from the symmetry data can be found in \Ref{2020Sci...367..794S}.

Briefly, we give an example of how our bounds may be applied to fragile and stable topological insulators, in addition to OWCs. We pick $p2$ as an illustrative example where our bound in \Eq{eq:GboundC2} is
\bea
G \geq \frac{1}{4 \min \{N_{\mathrm{orb}},4N_{\mathrm{occ}}\}} \lp \delta^2_{1b} +\delta_{1c}^2+\delta_{1d}^2 \rp 
\eea
taking a square unit cell $|\mbf{a}_1| = |\mbf{a}_2|$ and $\mbf{a}_1\cdot \mbf{a}_2 = 0$ for convenience. 

First, we study a two-band Chern insulator with symmetry data $\Gamma_1 + X_1 + Y_1 + M_2$. This symmetry data must have an odd Chern number because it has an odd number of $\rho_2$ irreps in the BZ. Using the tables in \Ref{2020Sci...367..794S}, we calculate $\delta_{1a} = \delta_{1b} = \delta_{1c} = -1/2$ and $\delta_{1d} = 1/2$. Thus our bound is
\bea
G \geq \frac{1}{4 \cdot 2} 3 \lp \frac{1}{2} \rp^2 = \frac{3}{32} = .09375 \ .
\eea
We can compare this to the Chern number bound obtained in \Ref{2015NatCo...6.8944P}, which is
\bea
G \geq \frac{|C|}{2\pi} = .159\dots
\eea
for the lowest odd Chern number $C=1$, which is larger by less than a factor of 2. We now consider a four-band fragile insulator with band structure $2\Gamma_1 + 2X_1 + 2Y_1 + 2M_2$, which could be obtained by stacking a $C=1$ Chern insulator with its time-reversed copy. In this case the total Chern number is zero. However, the RSIs are
$\delta_{1a} = \delta_{1b} = \delta_{1c} = -1$ and $\delta_{1d} = 1$, so our bound is nonzero:
\bea
G \geq \frac{3}{4 \cdot 4} = \frac{3}{16} = .1875
\eea
which is larger than the $C=1$ Chern insulator bound but smaller than the $C=2$ bound, which would be obtained if the $C= \pm1$ bands do not mix. In this case, the Wilson loop contains two oppositely winding eigenvalues, which is the same as the Wilson loops of Euler insulators where $C_2\mathcal{T}$ protects the bands from gapping \cite{2018arXiv180710676S}. In this case, we can apply the Euler number bound of \Ref{PhysRevLett.124.167002}, which is equivalent to the $C=2$ bound $G \geq 1/\pi$.

Lastly, we remark that $G$ is the same for the conduction and valence bands of a given model because $\del_\mu(1-P)\del^\mu(1-P) = \del_\mu P \del^\mu P$. Because fragile valence bands can have an OWC complement \cite{2021arXiv210713556S} as their conduction bands, we learn that an OWC can have just as strong a lower bound as a fragile topological insulator.

\begingroup
\renewcommand\arraystretch{1.5}
\begin{center}
	\label{tab:GlbnoSOCnoTRS}
\begin{longtable}{l|l}
		\caption{Lower bounds for $G$ without SOC (spinless)}\\
	\hline
$p1$& \\
\hline
$p2$&$\frac{1}{4 \min \{N_{\mathrm{orb}},4N_{\mathrm{occ}}\}} \lp |\mbf{a}_1|^2 \delta^2_{1b} +|\mbf{a}_2|^2\delta_{1c}^2+\frac{1}{2}\frac{|\mbf{a}_1+\mbf{a}_2|^2|\mbf{a}_1-\mbf{a}_2|^2}{|\mbf{a}_1|^2+|\mbf{a}_2|^2}\delta_{1d}^2 \rp / |\mbf{a}_1\times \mbf{a}_2| $ \\
\hline
$pm$&$\frac{2}{\pi \min \{2 N_{\mathrm{occ}},N_{\mathrm{orb}}\}} \delta_{1b}^2 $ \\
\hline
$pg$& \\
\hline
$cm$& \\
\hline
$p2mm$&$\frac{1}{4 \min \{N_{\mathrm{orb}},4N_{\mathrm{occ}}\}} \lp |\mbf{a}_1|^2 \delta^2_{1b} +|\mbf{a}_2|^2\delta_{1c}^2+\frac{1}{2}\frac{|\mbf{a}_1+\mbf{a}_2|^2|\mbf{a}_1-\mbf{a}_2|^2}{|\mbf{a}_1|^2+|\mbf{a}_2|^2}\delta_{1d}^2 \rp / |\mbf{a}_1\times \mbf{a}_2|$ \\
\hline
$p2mg$& \\
\hline
$p2gg$& \\
\hline
$c2mm$&\\
\hline
$p4$&$\max \left\{ \frac{\delta_{1b,2}^2 + (\delta_{1b,1}-\delta_{1b,3})^2}{8\min\{N_{\mathrm{orb}},2 N_{\mathrm{occ}}\}} ,  \frac{\delta_{2c}^2}{2\min\{N_{\mathrm{orb}},4 N_{\mathrm{occ}}\}} \right\}+ \frac{(\delta_{1b,1}+\delta_{1b,3}-\delta_{1b,2})^2}{4 \min \{N_{\mathrm{orb}},4N_{\mathrm{occ}}\}}$  \\
\hline
$p4mm$&$\max \left\{ \frac{\delta_{1b,2}^2}{8\min\{N_{\mathrm{orb}},2 N_{\mathrm{occ}}\}} ,  \frac{\delta_{2c}^2}{2\min\{N_{\mathrm{orb}},4 N_{\mathrm{occ}}\}} \right\}+ \frac{(2\delta_{1b,1}-\delta_{1b,2})^2}{4 \min \{N_{\mathrm{orb}},4N_{\mathrm{occ}}\}}$  \\
\hline
$p3$&$ \frac{2}{3\sqrt{3}\min\{ N_{\mathrm{orb}},3N_{\mathrm{occ}}\}} \max \left\{\delta_{1b,1}^2 - \delta_{1b,1} \delta_{1b,2} + \delta_{1b,2}^2,\delta_{1c,1}^2 - \delta_{1c,1} \delta_{1c,2} + \delta_{1c,2}^2 \right\} $  \\
\hline
$p3m1$&$ \frac{2}{3\sqrt{3}\min\{ N_{\mathrm{orb}},3N_{\mathrm{occ}}\}} \max \left\{\delta_{1b,1}^2,\delta_{1c,1}^2  \right\} $  \\
\hline
$p31m$&$ \frac{2}{3\sqrt{3}\min\{ N_{\mathrm{orb}},3N_{\mathrm{occ}}\}} (\delta_{2b,1}^2 - \delta_{2b,1} \delta_{2b,2} + \delta_{2b,2}^2 ) $  \\
\hline
$p6$&$ \frac{2}{\sqrt{3}} \max \left\{  \frac{\delta_{2b,1}^2 - \delta_{2b,1} \delta_{2b,2} + \delta_{2b,2}^2}{3\min\{ N_{\mathrm{orb}},3N_{\mathrm{occ}}\}} , \frac{\delta_{3c,1}^2}{4\min\{ N_{\mathrm{orb}},4N_{\mathrm{occ}}\}}  \right\}$  \\
\hline
$p6mm$&$  \frac{2}{\sqrt{3}} \max \left\{  \frac{\delta_{2b,1}^2}{3\min\{ N_{\mathrm{orb}},3N_{\mathrm{occ}}\}} , \frac{\delta_{3c,1}^2}{4\min\{ N_{\mathrm{orb}},4N_{\mathrm{occ}}\}}  \right\}$  \\
\hline
\hline
$p1'$& \\
\hline
$p21'$&$\frac{1}{4 \min \{N_{\mathrm{orb}},4N_{\mathrm{occ}}\}} \lp |\mbf{a}_1|^2 \delta^2_{1b} +|\mbf{a}_2|^2\delta_{1c}^2+\frac{1}{2}\frac{|\mbf{a}_1+\mbf{a}_2|^2|\mbf{a}_1-\mbf{a}_2|^2}{|\mbf{a}_1|^2+|\mbf{a}_2|^2}\delta_{1d}^2 \rp / |\mbf{a}_1\times \mbf{a}_2| $ \\
\hline
$pm1'$&$\frac{2}{\pi \min \{2 N_{\mathrm{occ}},N_{\mathrm{orb}}\}} \delta_{1b}^2 $ \\
\hline
$pg1'$& \\
\hline
$cm1'$& \\
\hline
$p2mm1'$&$\frac{1}{4 \min \{N_{\mathrm{orb}},4N_{\mathrm{occ}}\}} \lp |\mbf{a}_1|^2 \delta^2_{1b} +|\mbf{a}_2|^2\delta_{1c}^2+\frac{1}{2}\frac{|\mbf{a}_1+\mbf{a}_2|^2|\mbf{a}_1-\mbf{a}_2|^2}{|\mbf{a}_1|^2+|\mbf{a}_2|^2}\delta_{1d}^2 \rp / |\mbf{a}_1\times \mbf{a}_2| $ \\
\hline
$p2mg1'$& \\
\hline
$p2gg1'$& \\
\hline
$c2mm1'$&\\
\hline
$p41'$&$\max \left\{ \frac{\delta_{1b,2}^2}{8\min\{N_{\mathrm{orb}},2 N_{\mathrm{occ}}\}} ,  \frac{\delta_{2c}^2}{2\min\{N_{\mathrm{orb}},4 N_{\mathrm{occ}}\}} \right\}+ \frac{(2\delta_{1b,1}-\delta_{1b,2})^2}{4 \min \{N_{\mathrm{orb}},4N_{\mathrm{occ}}\}}$  \\
\hline
$p4mm1'$&$\max \left\{ \frac{\delta_{1b,2}^2}{8\min\{N_{\mathrm{orb}},2 N_{\mathrm{occ}}\}} ,  \frac{\delta_{2c}^2}{2\min\{N_{\mathrm{orb}},4 N_{\mathrm{occ}}\}} \right\}+ \frac{(2\delta_{1b,1}-\delta_{1b,2})^2}{4 \min \{N_{\mathrm{orb}},4N_{\mathrm{occ}}\}}$  \\
\hline
$p31'$&$ \frac{2}{3\sqrt{3}\min\{ N_{\mathrm{orb}},3N_{\mathrm{occ}}\}} \max \left\{\delta_{1b,1}^2,\delta_{1c,1}^2  \right\} $  \\
\hline
$p3m1'$&$ \frac{2}{3\sqrt{3}\min\{ N_{\mathrm{orb}},3N_{\mathrm{occ}}\}} \max \left\{\delta_{1b,1}^2,\delta_{1c,1}^2  \right\} $  \\
\hline
$p31'm$&$ \frac{2}{3\sqrt{3}\min\{ N_{\mathrm{orb}},3N_{\mathrm{occ}}\}} \delta_{2b,1}^2 $  \\
\hline
$p61'$&$\frac{2}{\sqrt{3}} \max \left\{  \frac{\delta_{2b,1}^2}{3\min\{ N_{\mathrm{orb}},3N_{\mathrm{occ}}\}} , \frac{\delta_{3c,1}^2}{4\min\{ N_{\mathrm{orb}},4N_{\mathrm{occ}}\}}  \right\}$  \\
\hline
$p6mm1'$&$\frac{2}{\sqrt{3}}\max \left\{  \frac{\delta_{2b,1}^2}{3\min\{ N_{\mathrm{orb}},3N_{\mathrm{occ}}\}} , \frac{\delta_{3c,1}^2}{4\min\{ N_{\mathrm{orb}},4N_{\mathrm{occ}}\}}  \right\}$  \\
\end{longtable}
\end{center}
\endgroup

\begingroup
\renewcommand\arraystretch{1.5}
\begin{center}
	\label{tab:GlbSOCnoTRS}
\begin{longtable}{l|l}
		\caption{Lower bounds for $G$ with SOC (spinful)}\\
	\hline
$p1$& \\
\hline
$p2$&$\frac{1}{4 \min \{N_{\mathrm{orb}},4N_{\mathrm{occ}}\}} \lp |\mbf{a}_1|^2 \delta^2_{1b} +|\mbf{a}_2|^2\delta_{1c}^2+\frac{1}{2}\frac{|\mbf{a}_1+\mbf{a}_2|^2|\mbf{a}_1-\mbf{a}_2|^2}{|\mbf{a}_1|^2+|\mbf{a}_2|^2}\delta_{1d}^2 \rp / |\mbf{a}_1\times \mbf{a}_2| $ \\
\hline
$pm$&$\frac{2}{\pi \min \{2 N_{\mathrm{occ}},N_{\mathrm{orb}}\}} \delta_{1b}^2 $ \\
\hline
$pg$& \\
\hline
$cm$& \\
\hline
$p2mm$& \\
\hline
$p2mg$& \\
\hline
$p2gg$& \\
\hline
$c2mm$&\\
\hline
$p4$&$ \frac{\delta_{1b,2}^2}{4\min\{N_{\mathrm{orb}},2 N_{\mathrm{occ}}\}} $  \\
\hline
$p4mm$&$ \frac{\delta_{1b,2}^2}{4\min\{N_{\mathrm{orb}},2 N_{\mathrm{occ}}\}} $   \\
\hline
$p3$&$ \frac{2}{3\sqrt{3}\min\{ N_{\mathrm{orb}},3N_{\mathrm{occ}}\}} \max \left\{\delta_{1b,1}^2 - \delta_{1b,1} \delta_{1b,2} + \delta_{1b,2}^2,\delta_{1c,1}^2 - \delta_{1c,1} \delta_{1c,2} + \delta_{1c,2}^2 \right\} $  \\
\hline
$p3m1$&$ \frac{2}{3\sqrt{3}\min\{ N_{\mathrm{orb}},3N_{\mathrm{occ}}\}} \max \left\{\delta_{1b,1}^2, \delta_{1c,1}^2 \right\} $  \\
\hline
$p31m$&$ \frac{2}{3\sqrt{3}\min\{ N_{\mathrm{orb}},3N_{\mathrm{occ}}\}} (\delta_{2b,1}^2 - \delta_{2b,1} \delta_{2b,2} + \delta_{2b,2}^2 ) $  \\
\hline
$p6$&$\frac{2}{\sqrt{3}} \max \left\{  \frac{\delta_{2b,1}^2 - \delta_{2b,1} \delta_{2b,2} + \delta_{2b,2}^2}{3\min\{ N_{\mathrm{orb}},3N_{\mathrm{occ}}\}} , \frac{\delta_{3c,1}^2}{4\min\{ N_{\mathrm{orb}},4N_{\mathrm{occ}}\}}  \right\}$  \\
\hline
$p6mm$&$ \frac{2}{3\sqrt{3}\min\{ N_{\mathrm{orb}},3N_{\mathrm{occ}}\}} \delta_{2b,1}^2 $  \\
\hline
\hline
$p1'$& \\
\hline
$p21'$&  \\
\hline
$pm1'$& \\
\hline
$pg1'$& \\
\hline
$cm1'$& \\
\hline
$p2mm1'$& \\
\hline
$p2mg1'$& \\
\hline
$p2gg1'$& \\
\hline
$c2mm1'$&\\
\hline
$p41'$&$\max \left\{ \frac{2\delta_{1b,2}^2}{8\min\{N_{\mathrm{orb}},2 N_{\mathrm{occ}}\}} ,  \frac{\delta_{2c}^2}{2\min\{N_{\mathrm{orb}},4 N_{\mathrm{occ}}\}} \right\}+ \frac{(\delta_{1b,1}+\delta_{1b,3}-\delta_{1b,2})^2}{4 \min \{N_{\mathrm{orb}},4N_{\mathrm{occ}}\}}$  \\
\hline
$p4mm1'$&$ \frac{\delta_{1b,2}^2}{4\min\{N_{\mathrm{orb}},2 N_{\mathrm{occ}}\}} $   \\
\hline
$p31'$&$ \frac{2}{3\sqrt{3}\min\{ N_{\mathrm{orb}},3N_{\mathrm{occ}}\}} \max \left\{\delta_{1b,1}^2,\delta_{1c,1}^2 \right\} $  \\
\hline
$p3m1'$&$ \frac{2}{3\sqrt{3}\min\{ N_{\mathrm{orb}},3N_{\mathrm{occ}}\}} \max \left\{\delta_{1b,1}^2, \delta_{1c,1}^2 \right\} $  \\
\hline
$p31m1'$&$ \frac{2}{3\sqrt{3}\min\{ N_{\mathrm{orb}},3N_{\mathrm{occ}}\}} \delta_{2b,1}^2 $  \\
\hline
$p61'$&$ \frac{2\delta_{2b,1}^2}{3\sqrt{3}\min\{ N_{\mathrm{orb}},3N_{\mathrm{occ}}\}} $  \\
\hline
$p6mm1'$&$ \frac{2}{3\sqrt{3}\min\{ N_{\mathrm{orb}},3N_{\mathrm{occ}}\}} \delta_{2b,1}^2 $  \\
\end{longtable}
\end{center}
\endgroup

\subsection{Two Lemmas}
\label{app:lemmas}

In this section, we prove two simple lemmas used in obtaining the lower bounds. First we prove a matrix norm inequality
\bea
\label{eq:mnineq}
||A||^2 \geq \frac{1}{\text{Rk } A} | \Tr S A|, \qquad
\eea
where $||A||^2 = \Tr A^\dag A$ is the Frobenius norm and \Eq{eq:mnineq} holds for any square matrix $A$ and any unitary matrix $S$. Our result is a simple extension of \Ref{wolkowicz1980bounds} which proves \Eq{eq:mnineq} for $S=\mathbb{1}$. To be self-contained, we prove our result from scratch. First, let $\sigma_i > 0, i = 1, \dots, r$ denote the singular values of $A$ and $r = \text{Rk } A$, i.e. $A = U^\dag \Sigma V$ where $U,V$ are unitary and $\Sigma = \text{diag}(\sigma_1,\dots, \sigma_r,0,\dots,0)$. Then we have
\bea
||A||^2 &= \sum_{i=1}^r \sigma_i^2 = \lp \sum_{i=1}^r \sigma_i^2  \rp \lp \sum_{i=1}^r \frac{1}{\sqrt{r}^2}  \rp \geq \lp \sum_{i=1}^r  \frac{\sigma_i }{\sqrt{r}} \rp^2 = \frac{1}{r} \lp \sum_{i=1}^r \sigma_i\rp^2
\eea
using Cauchy-Schwartz for the inequality. We now use
\bea
|\Tr A| = |\Tr \Sigma V U^\dag | = \left| \sum_{i,j=1}^r \Sigma_{ij} [V U^\dag]_{ij} \right| = \left| \sum_{i=1}^r \sigma_{i} [V U^\dag]_{ii} \right| \leq  \sum_{i=1}^r \sigma_{i}  | [V U^\dag]_{ii}| \leq  \sum_{i=1}^r \sigma_{i}
\eea
because $V U^\dag$ is a unitary matrix, so $|[V U^\dag]_{ii}| \leq 1$. With these results, we obtain
\bea
||A||^2 \geq \frac{1}{r} \lp \sum_{i=1}^r \sigma_i\rp^2 \geq \frac{1}{\text{Rk }A} |\Tr A|^2 \ .
\eea
To introduce the free unitary matrix $S$, we use $\text{Rk } SA = \text{Rk } A$ because $S$ is invertible, so
\bea
||A||^2  = ||SA||^2 \geq \frac{1}{\text{Rk }SA} |\Tr SA|^2 = \frac{1}{\text{Rk }A} |\Tr SA|^2
\eea
which proves \Eq{eq:mnineq}. In the Main Text, our application of this inequality is tight. \\

We now prove the ``concentration lemma" which allows us to perform an extremization subject to some constraints. Specifically, we want to solve the following problem:
 \bea
 \label{eqoptapp}
R_{min}^2 = \min_{\psi_\mbf{R}} \sum_\mbf{R} \frac{1}{2} |\mbf{R}|^2 |\psi_\mbf{R}|^2
 \eea
for $\mbf{R}$ in the Bravais lattice and where the minimization is taken over the space of $\{|\psi_\mbf{R}| \geq 0\}$ satisfying the following constraints
 \bea
 \label{eq:constraints}
(1): & \quad \sum_{\mbf{R}} |\psi_\mbf{R}|^2 = N_{\mathrm{occ}}, \\
(2): &  \quad  |\psi_\mbf{R}| = |\psi_{g\mbf{R}}| \\
(3): &  \quad |B_a| \leq \sum_{\mbf{R} \in L_a} |\psi_\mbf{R}|
 \eea
 where $L_a = g L_a$ is a subset of the Bravais lattice symmetric under $g$, $ |B_a|$ are known, and $a = 1, \dots, N_a$. (We assume that the $|B_a|$ in (3) are small enough that (1) is not contradictory.) Constraint (1) is normalization, so we can think of $\psi_\mbf{R}$ as a wavefunction, and $R_{min}^2$ as the minimum expectation value of a quadratic potential. Constraint (2) requires that $|\psi_\mbf{R}|$ is symmetric under $g$. Constraint (3) is a linear constraint which forces some $|\psi_\mbf{L}|$ to be nonzero if $|B_a| \neq 0$. Without (3), we could take $|\psi_0|^2 = N_{\mathrm{occ}}$ and $|\psi_{\mbf{R}\neq0}| = 0$, in which case $R_{min}^2 = 0$ would be trivial. Thus $(3)$ is required for a nontrivial lower bound. In our problem, $(3)$ represents the constraints of symmetry eigenvalues on the projector. Intuitively, we expect the minimum to occur when $\psi_\mbf{R}$ saturates constraint (3) by taking $|\psi_\mbf{R}|$ nonzero only on the small $\mbf{R} \in L_a$. Indeed, this is the case.

Consider two normalized wavefunctions $|\psi_{\mbf{R}}|$ and $|\psi'_{\mbf{R}}| = \psi_{\mbf{R}} +  t_1 \delta_{\mbf{R},\mbf{R}_1} - t_2 \delta_{\mbf{R}, \mbf{R}_2}$ where $t_2 \in (0, |\psi_{\mbf{R}_2}|)$. Because both wavefunctions are normalized, $t_1$ and $t_2$ must satisfy
\bea
\label{eq:t1t2}
\sum_{\mbf{R}} |\psi_\mbf{R}|^2 &= \sum_{\mbf{R}} |\psi'_\mbf{R}|^2 = t_1^2 + 2 t_1 |\psi_{\mbf{R}_1}|   + t_2^2 - 2 t_2 |\psi_{\mbf{R}_2}| +  \sum_{\mbf{R}} |\psi_\mbf{R}|^2, \\
 \implies \quad t_1 &= - |\psi_{\mbf{R}_1}| + \sqrt{|\psi_{\mbf{R}_1}|^2 + |\psi_{\mbf{R}_2}|^2 -(t_2 - |\psi_{\mbf{R}_2}|)^2} \\
\eea
and we observe that for $t_2 > 0$, $t_1 >0$. Assume that $|\psi_\mbf{R}|$ is nonzero at $\mbf{R}_2$. Then increasing $t_2$ \emph{strictly} lowers the amplitude of $|\psi'_\mbf{R}|$ at $\mbf{R}_2$ and \emph{strictly} increases it at $\mbf{R}_1$. We will use this property to find the minimum of \Eq{eqoptapp}. Let us take $|\mbf{R}_2| > |\mbf{R}_1|$. We now prove that
\bea
\sum_\mbf{R} \frac{1}{2} |\mbf{R}|^2 |\psi_\mbf{R}|^2 > \sum_\mbf{R} \frac{1}{2} |\mbf{R}|^2 |\psi'_\mbf{R}|^2, \quad t> 0 \ . \\
\eea
This is a direct calculation
\bea
\sum_\mbf{R} \frac{1}{2} |\mbf{R}|^2 |\psi'_\mbf{R}|^2 &=\sum_\mbf{R} \frac{1}{2} |\mbf{R}|^2 |\psi_\mbf{R}|^2+ |\mbf{R}_1|^2 (t_1^2 + 2 t_1 |\psi_{\mbf{R}_1}| )  + |\mbf{R}_2|^2 (t_2^2 - 2 t_2|\psi_{\mbf{R}_2}| ) \\
\eea
Note that $(t_1+ \psi_{\mbf{R}_1})^2 - \psi_{\mbf{R}_1}^2 = \psi_{\mbf{R}_2}^2 -(t_2 - \psi_{\mbf{R}_2})^2$ by \Eq{eq:t1t2}, and hence
\bea
\sum_\mbf{R} \frac{1}{2} |\mbf{R}|^2 |\psi'_\mbf{R}|^2 &=\sum_\mbf{R} \frac{1}{2} |\mbf{R}|^2 |\psi_\mbf{R}|^2 + |\mbf{R}_1|^2[(t_1+ \psi_{\mbf{R}_1})^2- \psi_{\mbf{R}_1}^2] - |\mbf{R}_2|^2[\psi_{\mbf{R}_2}^2 -(t_2 - \psi_{\mbf{R}_2})^2] \\
\sum_\mbf{R} \frac{1}{2} |\mbf{R}|^2 |\psi_\mbf{R}|^2 - \sum_\mbf{R} \frac{1}{2} |\mbf{R}|^2 |\psi'_\mbf{R}|^2&= |\mbf{R}_2|^2 [\psi_{\mbf{R}_2}^2 -(t_2 - \psi_{\mbf{R}_2})]^2 - | \mbf{R}_1|^2[(t_1+ \psi_{\mbf{R}_1})^2- \psi_{\mbf{R}_1}^2]  \\
 &= (|\mbf{R}_2|^2 - |\mbf{R}_1|^2 )[ \psi_{\mbf{R}_2}^2 -(t_2 - \psi_{\mbf{R}_2})^2]  \geq 0 \\
\eea
where equality only holds at $t_2 = 0$ because $ (|\mbf{R}_2|^2 -  |\mbf{R}_1|^2 ) >0$. We observe that shifting the weight of $|\psi_{\mbf{R}}|$ from higher to lower harmonics strictly decreases $\sum_\mbf{R} \frac{1}{2} |\mbf{R}|^2 |\psi_\mbf{R}|^2$. Thus the minimum $R_{min}^2$ is obtained when $|\psi_\mbf{R}|$ is nonzero only on the symmetry-related $\mbf{R} \in L_a$ with smallest $|\mbf{R}|^2$ such that (3) is saturated, i.e. $|\psi_\mbf{R}|$ is as small as possible subject to $|B_a|$, and the normalization constraint (1) can be satisfied by tuning $|\psi_{\mbf{R}=0}|$.

We can apply the concentration lemma to lower bound $G$ because
\bea
\label{eq:Goptboundapp}
G = \frac{1}{2\Omega_c} \sum_\mbf{R} |\mbf{R}|^2 ||p(\mbf{R})||^2 \geq \min_{|\psi_\mbf{R}|} \frac{1}{2\Omega_c} \sum_\mbf{R} |\mbf{R}|^2 |\psi_\mbf{R}|^2
\eea
where the minimization can be performed over any space of $\{|\psi_\mbf{R}| > 0\}$ such that $|\psi_\mbf{R}| = ||p(\mbf{R})|| $ is admissible. We can impose the constraints in \Eq{eq:constraints} on $|\psi_\mbf{R}|$ such that $|\psi_\mbf{R}| = ||p(\mbf{R})||$ is still admissible. Then using the concentration lemma, we can solve the $|\psi_\mbf{R}|$ minimization yielding a lower bound on the superfluid weight from \Eq{eq:Goptboundapp}. The results are shown in \App{app:realspaceproj}.

\section{Lower Bounds for General Orbital positions}
\label{app:genorb}

We now discuss how to generalize the lower bounds to general orbital locations. Our result will be a lower bound depend on the RSIs of the occupied bands, with prefactors dependent on the locations of the orbitals in the unit cell.

The important feature of having orbitals off the 1a position is that the embedding matrix $V[\mbf{G}]$ is nontrivial. The first effect of a nontrivial embedding matrix appears when Fourier transforming $G$ to real space as we now show. Recall that from \Eq{eq:Vperiod}, we know that $P(\mbf{k})$ is not strictly periodic for general orbitals. Instead:
\bea
P(\mbf{k} + 2\pi \mbf{b}_i) = V[2\pi \mbf{b}_i] P(\mbf{k}) V^\dag[2\pi \mbf{b}_i] ,
\eea
recalling that the embedding matrix is $V_{\al \be}[\mbf{G}] = \exp \lp - i \mbf{G} \cdot \mbf{r}_\al \rp \delta_{\al \be}$ (unsummed) where $\mbf{r}_\al$ are the orbital locations. (Although $P(\mbf{k})$ is not periodic, $G$ is well-defined because $\Tr \del_i P \del^i P$ \emph{is} periodic on the BZ.)
Because $P(\mbf{k})$ is not periodic on the BZ, it does not have a Fourier representation. To circumvent this difficulty, we consider a transformed operator $\tilde{P}(\mbf{k}) \equiv V^\dag_L(\mbf{k}) P(\mbf{k}) V_R(\mbf{k})$ which is periodic. The auxiliary matrices $V_{L,R}(\mbf{k})$ are so far undetermined, but we require them to be unitary and to obey $V_{L,R}(\mbf{k}+\mbf{G}) = V[\mbf{G}] V_{L,R}(\mbf{k})$. As such,
\bea
\label{eq:Vperiodiity}
\tilde{P}(\mbf{k}+\mbf{G}) = V^\dag_L(\mbf{k}+\mbf{G}) P(\mbf{k}+\mbf{G}) V_R(\mbf{k}+\mbf{G}) = V^\dag_L(\mbf{k}) P(\mbf{k}) V_R(\mbf{k})  = \tilde{P}(\mbf{k})
\eea
so $\tilde{P}(\mbf{k})$ is periodic on the BZ. Hence $\tilde{P}(\mbf{k})$ has a Fourier representation:
\bea
\label{eq:PFourier}
V^\dag_L(\mbf{k})  P(\mbf{k})  V_R(\mbf{k})&= \sum_{\mbf{R}} \tilde{p}(\mbf{R}) e^{-i \mbf{R} \cdot \mbf{k}} , \qquad \tilde{p}(\mbf{R}) = \int \frac{dk^1dk^2}{(2\pi)^2} e^{i \mbf{R} \cdot \mbf{k}} V_L^\dag(\mbf{k})P(\mbf{k}) V_R(\mbf{k})
\eea
where $\mbf{R} \in \mathds{Z} \mbf{a}_1+\mathds{Z} \mbf{a}_2$ are the lattice vectors and $k^i \in(0,2\pi)$ are the dimensionless crystal momenta. We refer to $\tilde{p}(\mbf{R})$ as the harmonics of $P(\mbf{k})$. There are many choice of $V_{L,R}(\mbf{k})$. The simplest choice is taking $V_L(\mbf{k}) = V_R(\mbf{k}) = V(\mbf{k})$ where
\bea
\label{eq:Vkdef}
V_{\al \be}(\mbf{k}) &= \exp \lp - i \mbf{k} \cdot \mbf{r}_\al \rp \delta_{\al \be} \qquad \text{(unsummed)}
\eea
which is just the interpolation of $V[\mbf{G}]$ to all $\mbf{k}$. Physically, $V(\mbf{k})$ is the representation of the unitary ``position" operator $e^{-i \mbf{k} \cdot \mbf{X}}$ on the orbitals. However, we will now show it is necessary to make a different choice of $V_{L,R}$ which depends on the orbital representations $D[C_n]$. Explicitly, we saw in \App{app:realspaceproj} that lower bounds on the Fourier harmonics were obtained by bounding $||P(\mbf{K}_1) \pm \dots P(\mbf{K}_n)||$ with the quantity $|\Tr D[C_n](P(\mbf{K}_1) \pm \dots P(\mbf{K}_n))|$ which could be evaluated in terms of the symmetry data because $\Tr D[C_n] P(\mbf{K}_1)$ depended only on the irreps $\chi$ at $\mbf{K}$. However, when there is a nontrivial embedding matrix, the trace formula (\Eq{eq:trPtoirrep}) generalizes to
\bea
 \Tr V[C_n \mbf{K}-\mbf{K}]^\dag D[C_n]  P(\mbf{K})  &= \sum_\chi m(\chi) \chi[C_n].
\eea
where the embedding matrix $V[C_n \mbf{K}-\mbf{K}]^\dag D[C_n]$ appears which depends on $\mbf{K}$. We now show that there is a choice of $V_{L,R}(\mbf{k})$ which allows us to obtain the requisite $V[C_n \mbf{K}-\mbf{K}]^\dag$ in the projector bound. To determine the superfluid weight lower bound protected by a fixed symmetry $\tilde{g}=C_n$, we set
\bea
\label{eq:VLR}
\boxed{
P_{\tilde{g}}(\mbf{k}) = V^\dag_{\tilde{g}}(\mbf{k}) P(\mbf{k}) V(\mbf{k}), \quad V_{\tilde{g}}(\mbf{k})  =  D^\dag[\tilde{g}] V(\tilde{g} \mbf{k}) D[\tilde{g}]
}
\eea
from which it is clear that $V(\mbf{k})$ and $V_{\tilde{g}}(\mbf{k})$ are unitary. It remains to show that they obey the periodicity condition \Eq{eq:Vperiodiity}. $V(\mbf{k})=V(\mbf{k}+\mbf{G})$ follows from  \Eq{eq:Vkdef}, and $V_{\tilde{g}}(\mbf{k})$ follows from a brief calculation:
\bea
V_{\tilde{g}}(\mbf{k}+\mbf{G}) &= D^\dag[\tilde{g}] V(\tilde{g} \mbf{k} + \tilde{g} \mbf{G}) D[\tilde{g}]\\
 &=D^\dag[\tilde{g}] V[\tilde{g} \mbf{G}]  V(\tilde{g} \mbf{k}) D[\tilde{g}]\\
&=D^\dag[\tilde{g}]V[\tilde{g} \mbf{G}] D[g] D^\dag[\tilde{g}] V(\tilde{g} \mbf{k}) D[\tilde{g}]\\
&=V[\mbf{G}] D^\dag[\tilde{g}] V(\tilde{g} \mbf{k})D[\tilde{g}] \\
&=V[\mbf{G}] V_{\tilde{g}}(\mbf{k}) \\
\eea
where we used \Eq{eq:DgVgproof} in the second to last line. We now show that that \Eq{eq:VLR} allows us to obtain the symmetry data bounds. We compute
\bea
\label{eq:DgcancelP}
\Tr D[\tilde{g}] P_{\tilde{g}}(\mbf{K}) &= \Tr D[\tilde{g}] \lp D^\dag[\tilde{g}] V^\dag(\tilde{g} \mbf{K}) D[\tilde{g}]  P(\mbf{K}) V(\mbf{K}) \rp  \\
&= \Tr V(\mbf{K}) V^\dag(\tilde{g} \mbf{K}) D[\tilde{g}]  P(\mbf{K})  \\
&= \Tr V^\dag(\tilde{g} \mbf{K}-\mbf{K}) D[\tilde{g}]  P(\mbf{K})  \\
&= \sum_\chi m(\tilde{g}) \chi[\tilde{g}] \ .
\eea
It is important to note that $P_{\tilde{g}}$ depends on the choice of $\tilde{g}$. We would not be able to obtain the simple result of \Eq{eq:DgcancelP} if we tried to evaluate $\Tr D[\tilde{h}] P_{\tilde{g}}(\mbf{K})$ for $\tilde{h} \neq \tilde{g}$. As such, we are only able to calculate a lower bound from a single symmetry $\tilde{g}$, even when the space group contains multiple symmetry operators, e.g. $C_4$ and $C_2$. In cases like these, we obtain a bound for each symmetry individually. This result may be improved by future work. 

We now need to evaluate the real space expression for $G$. To do so, we need to take $\mbf{k}$-derivatives of $P_{\tilde{g}}(\mbf{k}) = V^\dag_{\tilde{g}}(\mbf{k}) P(\mbf{k}) V(\mbf{k})$, and it will be useful to use the explicit formula:
\bea
\label{eq:VLexplicit}
\null [V_{\tilde{g}}(\mbf{k})]_{\al \be} &= \exp \lp -i \mbf{k} \cdot (\mbf{r}_{\al} + \tilde{\mbf{A}}_\al) \rp \delta_{\al \be}\qquad \text{(unsummed)}
\eea
where $\tilde{\mbf{A}}_\al \mod \mbf{a}_i = 0$ is a lattice vector determined by the orbital locations as will will now show. It is convenient to define $\pmb{\Lambda}_{\al \be} = \mbf{r}_{\al} \delta_{\al \be}$ (unsummed) as the matrix of orbital positions within the 0th unit cell, so $V(\mbf{k}) = e^{i \mbf{k} \cdot \pmb{\Lambda}}$. We now calculate $D[\tilde{g}] \pmb{\Lambda} D^\dag[\tilde{g}]$. It is convenient to use a Wigner-Seitz unit cell which respects $\tilde{g}$ and where $| (\mbf{r}_\al - \mbf{r}_\be) \cdot \mbf{b}_i| < 1$ because $\mbf{r}_\al$ and $\mbf{r}_\be$ are in the same unit cell by definition. Now note that if $\tilde{\mbf{r}}_\al$ is an orbital of the model, then $\tilde{g} \tilde{\mbf{r}}_\al$ is also an orbital of the model. If $\mbf{r}_\al$ is not at a high-symmetry Wyckoff position, then $\mbf{r}_{\al'} = \tilde{g} \mbf{r}_\al \neq \mbf{r}_{\al}$ is a distinct orbital within the same unit cell (recall that the center of the $\tilde{g} = C_n$ symmetry is at the center of the Wigner-Seitz unit cell). If $\mbf{r}_\al$ is a high-symmetry Wyckoff position, then $\tilde{g} \mbf{r}_\al$ may \emph{not} be in the same unit cell. For instance, the 1d position $\mbf{a}_1/2+\mbf{a}_2/2$ of the unit cell at the origin is taken to $-(\mbf{a}_1/2+\mbf{a}_2/2)$ by $C_2$, but $-(\mbf{a}_1/2+\mbf{a}_2/2)$ is an orbital in the unit cell $-\mbf{a}_1-\mbf{a}_2$. In general, for a Wyckoff position of multiplicity $m$ under $\tilde{g}$, there are $m$ distinct orbital positions $\mbf{r}_{\al_1}, \dots, \mbf{r}_{\al_m}$  in the unit cell with $\mbf{r}_{\al_{n+1}} = \tilde{g} \mbf{r}_{\al_n} - \tilde{\mbf{A}}_{\al_n}$ where $ \mbf{r}_{\al_{m+1}} =  \mbf{r}_{\al_1}$ by convention. This defines $\tilde{\mbf{A}}_{\al_1},\dots,\tilde{\mbf{A}}_{\al_m}$ for every Wyckoff position, which we enumerate in \Tab{tab:Aal}.

\begin{table}[h]
    \centering
\begin{tabular}{c|cc}
$\tilde{g} = M$ & 1a & 1b \\
\hline
$\mbf{r}_{\al} $ & 0 & $\mbf{a}_1/2$  \\
$\tilde{\mbf{A}}_{\al} $ & 0 & $-\mbf{a}_1$ \\
\end{tabular} \quad
\begin{tabular}{c|cccc}
$\tilde{g} = C_2$ & 1a & 1b & 1c & 1d \\
\hline
$\mbf{r}_{\al} $ & 0 & $\mbf{a}_1/2$ &  $\mbf{a}_2/2$ &  $\mbf{a}_1/2+\mbf{a}_2/2$\\
$\tilde{\mbf{A}}_{\al_1} $ & 0 & $-\mbf{a}_1$ &  $-\mbf{a}_2$ &  $-\mbf{a}_1-\mbf{a}_2$\\
\end{tabular} \quad
\begin{tabular}{c|ccc}
$\tilde{g} = C_4$ & 1a & 1b & 2c\\
\hline
$\mbf{r}_{\al} $ & 0 & $\mbf{a}_1/2+\mbf{a}_2/2$ &  $\{\mbf{a}_1/2,\mbf{a}_2/2\}$ \\
$\tilde{\mbf{A}}_{\al_1} $ & 0 & $-\mbf{a}_1$ &  0 \\
$\tilde{\mbf{A}}_{\al_2} $ &  &  &  $-\mbf{a}_1$ \\
\end{tabular} \\
\begin{tabular}{c|ccc}
$\tilde{g} = C_3$ & 1a & 1b & 1c\\
\hline
$\mbf{r}_{\al} $ & 0 & $\mbf{a}_1/3+2\mbf{a}_2/3$ &  $2\mbf{a}_1/3+\mbf{a}_2/3$ \\
$\tilde{\mbf{A}}_{\al_1} $ & 0 & $-\mbf{a}_1-\mbf{a}_2$ & $-\mbf{a}_1$ \\
\end{tabular} \qquad
\begin{tabular}{c|ccc}
$\tilde{g} = C_6$ & 1a & 2b & 3c\\
\hline
$\mbf{r}_{\al} $ & 0 & $\{\mbf{a}_1/3+\mbf{a}_2/3,2\mbf{a}_1/3-\mbf{a}_2/3\}$ &  $\{\mbf{a}_1/2,\mbf{a}_2/2,-\mbf{a}_1/2+\mbf{a}_2/2\}$ \\
$\tilde{\mbf{A}}_{\al_1} $ & 0  & $0$ & $0$ \\
$\tilde{\mbf{A}}_{\al_2} $ &  & $-\mbf{a}_1$ & $0$ \\
$\tilde{\mbf{A}}_{\al_3} $ & &  & $-\mbf{a}_1$ \\
\end{tabular}

\caption{Lattice vectors. We compute $\tilde{\mbf{A}}_\al$ for each case of $\tilde{g}$. We take $\mbf{a}_2 = \tilde{g} \mbf{a}_1$ for $\tilde{g} \neq C_2$. For a Wyckoff position of multiplicity $m$ with positions $\mbf{r}_{\al_{n+1}} = \tilde{g} \mbf{r}_{\al_n}$ such that $\mbf{r}_{\al_{m+1}} = \mbf{r}_{\al_{1}}$, we compute $\tilde{\mbf{A}}_{\al_n} = \tilde{g} \mbf{r}_{\al_n} - \mbf{r}_{\al_{n+1}}$. The value of  $\tilde{\mbf{A}}_\al$ does depend on the convention of the Wyckoff position, but is always a lattice vector.
 \label{tab:Aal}} 
\end{table}

Explicitly, we can write the representation of $\tilde{g}$ on the orbitals from \Eq{eq:defineDg} as
\bea
D_{\al\be}[\tilde{g}] = \la_{\al} \delta_{\mbf{r}_\al, \tilde{g}\mbf{r}_\be - \tilde{\mbf{A}}_\al}
\eea
where $|\la_\al| = 1$ is a phase that encodes the rotation/mirror irreps of the orbital.  We can choose the order of the $\al$ indices to block diagonalize $\Lambda$ and $D[\tilde{g}]$ so that each block (of dimension $m$) contains orbitals within the same multiplicity-m Wyckoff position. Without loss of generality, we focus on a single block $\pmb{\Lambda}'$. Plugging in, we find
 \bea
\label{eq:DLD1}
\left[ D[\tilde{g}] \pmb{\Lambda}' D^\dag[\tilde{g}] \right]_{\al \be} &= \sum_{\al' \be'} D_{\al \al'}[\tilde{g}] \mbf{r}_{\al'} \delta_{\al' \be'} D_{\be \be'}^*[\tilde{g}] = \sum_{\al'} D_{\al \al'}[\tilde{g}] \mbf{r}_{\al'} D_{\be \al'}^*[\tilde{g}] \\
&= \sum_{\al'} \mbf{r}_{\al'}  \la_{\al}\la^*_{\be}  \delta_{\mbf{r}_\al, \tilde{g}\mbf{r}_{\al'} - \tilde{\mbf{A}}_\al}  \delta_{\mbf{r}_\be, \tilde{g}\mbf{r}_{\al'} - \tilde{\mbf{A}}_\be} \\
&= \la_{\al}\la^*_{\be}  \delta_{\mbf{r}_\al+\tilde{\mbf{A}}_\al, \mbf{r}_{\be} + \tilde{\mbf{A}}_\be}  \sum_{\al'} \mbf{r}_{\al'}    \delta_{\mbf{r}_\be, \tilde{g}\mbf{r}_{\al'} - \tilde{\mbf{A}}_\be} \\
&= \la_{\al}\la^*_{\be}  \delta_{\mbf{r}_\al+\tilde{\mbf{A}}_\al, \mbf{r}_{\be} + \tilde{\mbf{A}}_\be}  \tilde{g}^{-1}(\mbf{r}_{\be} +\tilde{\mbf{A}}_\be) \sum_{\al'}   \delta_{\mbf{r}_\be, \tilde{g}\mbf{r}_{\al'} - \tilde{\mbf{A}}_\be} \ . \\
\eea
The sum over $\al'$ is equal to 1 because there is always a single position $\mbf{r}_{\al'}$ in the Wyckoff position satisfying $\mbf{r}_\be + \tilde{\mbf{A}}_\be =  \tilde{g}\mbf{r}_{\al'}$ (see \Tab{tab:Aal}). Similarly within a single Wyckoff position, $\delta_{\mbf{r}_\al+\tilde{\mbf{A}}_\al, \mbf{r}_{\be} + \tilde{\mbf{A}}_\be}  = \delta_{\al,\be}$ as can be checked exhaustively from \Tab{tab:Aal}. Thus we obtain
\bea
\left[ D[\tilde{g}] \pmb{\Lambda}' D^\dag[\tilde{g}] \right]_{\al \be} &=  \la_{\al}\la^*_{\be} \delta_{\al \be} \tilde{g}^{-1}(\mbf{r}_{\be} +\tilde{\mbf{A}}_\be) =\delta_{\al \be}\tilde{g}^{-1} (\mbf{r}_\al + \tilde{\mbf{A}}_\al)  \\
\eea
which holds for all blocks $\pmb{\Lambda}'$ and for the full matrix $\pmb{\Lambda}$. A simple case is when all orbitals $\mbf{r}_\al$ correspond to multiplicity-1 Wyckoff positions (which may be different from each other). Such positions are only mapped to themselves under $\tilde{g}$ because $\tilde{\mbf{A}}_\al  = \tilde{g}\mbf{r}_\al - \mbf{r}_\al$, so $D[\tilde{g}]$ is diagonal (block diagonal with $1\times1$ blocks), and $D[\tilde{g}] \pmb{\Lambda}D^\dag[\tilde{g}]  = \pmb{\Lambda}$. This matches \Eq{eq:DLD1} using $\tilde{\mbf{A}}_\al  = \tilde{g}\mbf{r}_\al - \mbf{r}_\al $. Exponentiating, we find from \Eq{eq:VLR} that
\bea
V_{\tilde{g}}(\mbf{k}) &=  D^\dag[\tilde{g}] \exp \lp - i \tilde{g}\mbf{k} \cdot \pmb{\Lambda} \rp D[\tilde{g}] = \exp \lp - i \tilde{g}\mbf{k} \cdot D[\tilde{g}^{-1}] \pmb{\Lambda} D^\dag[\tilde{g}^{-1}] \rp \\
&= \exp \lp - i \tilde{g}\mbf{k} \cdot \tilde{g} (\pmb{\Lambda} + \tilde{\mbf{A}}) \rp \\
&= \exp \lp - i \mbf{k} \cdot (\pmb{\Lambda} + \tilde{\mbf{A}}) \rp \\
\eea
where $[\tilde{\mbf{A}}]_{\al \be} = \tilde{\mbf{A}}_{\al} \delta_{\al \be}$ (unsummed).

Returning to \Eq{eq:PFourier}, we find that in components
\bea
\label{eq:Palbe}
P_{\al \be}(\mbf{k}) &= \sum_{\mbf{R},\al'\be'} [V_{\tilde{g}}(\mbf{k})]_{\al \al'} \tilde{p}_{\al' \be'}(\mbf{R}) [V^\dag(\mbf{k})]_{\be' \be} e^{-i \mbf{R} \cdot \mbf{k}} = \sum_{\mbf{R}}  \tilde{p}_{\al \be}(\mbf{R}) e^{-i (\mbf{R} + \mbf{r}_\al + \tilde{\mbf{A}}_\al - \mbf{r}_\be) \cdot \mbf{k}} \\
\pmb{\nabla} P_{\al \be}(\mbf{k}) &= \sum_{\mbf{R}} i (\mbf{R} + \mbf{r}_\al + \tilde{\mbf{A}}_\al - \mbf{r}_\be) \tilde{p}_{\al \be}(\mbf{R}) e^{-i (\mbf{R} + \mbf{r}_\al + \tilde{\mbf{A}}_\al - \mbf{r}_\be) \cdot \mbf{k}} \ . \\
\eea
Because all the $\mbf{k}$-dependence is in the exponent, we find that
\bea
\label{eq:Ginrealspace}
G = \int \frac{d^2k}{(2\pi)^2} \frac{1}{2} \Tr \pmb{\nabla} P \cdot  \pmb{\nabla} P  = \sum_{\mbf{R},\al \be} \frac{1}{2\Omega_c} |\mbf{R}_{\al \be}|^2 |\tilde{p}_{\al \be}(\mbf{R})|^2
\eea
where we use the shorthand $\mbf{R}_{\al \be} = \mbf{R} + \mbf{r}_\al +  \tilde{\mbf{A}}_\al - \mbf{r}_\be$ and $\al,\be$ are unsummed. We emphasize that $p_{\al\be}(\mbf{R})$ and $\tilde{\mbf{A}}_\al$ depend on the choice of $\tilde{g}$.

The last result we need is a normalization identity for the harmonics. Using \Eq{eq:PFourier}, we prove
\bea
\sum_{\mbf{R}} ||\tilde{p}(\mbf{R})||^2 &= \int \frac{d^2k}{(2\pi)^2}  \frac{d^2k'}{(2\pi)^2} \sum_\mbf{R} e^{i \mbf{R} \cdot (\mbf{k}-\mbf{k}')} \Tr \lp V_{\tilde{g}}^\dag(\mbf{k}')P(\mbf{k}') V(\mbf{k}')  \rp^\dag V_{\tilde{g}}^\dag(\mbf{k})P(\mbf{k}) V(\mbf{k})   \\
&= \int \frac{d^2k}{(2\pi)^2}  \Tr \lp V_{\tilde{g}}^\dag(\mbf{k})P(\mbf{k}) V(\mbf{k})  \rp^\dag V_{\tilde{g}}^\dag(\mbf{k})P(\mbf{k}) V(\mbf{k})  \\
&= \int \frac{d^2k}{(2\pi)^2}  \Tr P(\mbf{k}) P(\mbf{k}) \\
&= N_{\mathrm{occ}}
\eea
because $\Tr P(\mbf{k})^2 = \Tr P(\mbf{k}) = N_{\mathrm{occ}}$. Thus, the normalization property of $||\tilde{p}(\mbf{R})||$ is the same for $||p(\mbf{R})||$ (see \App{app:realspaceproj}). This is guaranteed because the Frobenius norm is unitarily invariant. 

We now want to be able to write $G$ in terms of $||\tilde{p}(\mbf{R})||$ to apply the bounds on $||\tilde{p}(\mbf{R})||$ obtained from the symmetry data. We use \Eq{eq:Ginrealspace} to establish
\bea
G = \sum_{\mbf{R},\al \be} \frac{1}{2\Omega_c} |\mbf{R}_{\al \be}|^2 |\tilde{p}_{\al \be}(\mbf{R})|^2 \geq \sum_{\mbf{R}} \frac{1}{2\Omega_c} ||\tilde{p}(\mbf{R})||^2 \min_{\al \be}  |\mbf{R}_{\al \be}|^2
\eea
where $||\tilde{p}(\mbf{R})||^2 = \sum_{\al \be} |\tilde{p}_{\al \be}(\mbf{R})|^2$ is the Frobenius norm. From here, the procedure for determining lower bounds on $G$ is very similar to \App{app:realspaceproj}, but the bounds will depend on $\min_{\al \be}  |\mbf{R}_{\al \be}|^2$, which must be computed for the orbital positions of each model under consideration. This is possible by enumeration because $|\mbf{R}_{\al \be}|^2$ depends only on the orbital positions. For convenience, we define
\bea
M^2(\mbf{R}) = \min_{\al \be} |\mbf{R} _{\al\be}|^2 \ . \\
\eea
We now determine the bounds for $\tilde{g} = C_2 ,C_4, C_3$. Because $P_{\tilde{g}}(\mbf{k})$ is only compatible with a specific $\tilde{g}$, one computes bounds in a given space group by obtaining a list of bounds computed from each possible $\tilde{g}$. Thus we need only treat the abelian groups $p2,p3,p4$. We do not need to consider $p6$ because only a single high-symmetry momentum has $C_6$ irreps, and our bounds require linear combinations of projectors at \emph{different} high symmetry points to show that $\del P$ is nonzero. Hence bounds follows from $p2$ and $p3$ separately. We also do not need to consider $pm$ because there are only two Wyckoff positions with $M$. Hence in an OWC phase we can assume only one of the Wyckoff positions has orbitals, but that case is equivalent to only occupying 1a and so has already been covered in \App{app:tables}.

\subsection{$\tilde{g} = C_2$}
\label{eq:tildegC2}

We follow the arguments of \App{app:realspaceproj} closely. By taking linear combinations of projectors,
\bea
\label{eq:P2}
P_{\tilde{g}}(\Gamma) + P_{\tilde{g}}(X) + P_{\tilde{g}}(Y) + P_{\tilde{g}}(M) &=  \sum_{\mbf{L} \in L_a} 4\tilde{p}(\mbf{L}), \qquad L_a = 2\mathds{Z} \mbf{a}_1+ 2\mathds{Z} \mbf{a}_2 \\
P_{\tilde{g}}(\Gamma) - P_{\tilde{g}}(X) + P_{\tilde{g}}(Y) - P_{\tilde{g}}(M) &=  \sum_{\mbf{L} \in L_b} 4\tilde{p}(\mbf{L}) , \qquad L_b = (2\mathds{Z}+1) \mbf{a}_1+ 2\mathds{Z} \mbf{a}_2 \\
P_{\tilde{g}}(\Gamma) + P_{\tilde{g}}(X) - P_{\tilde{g}}(Y) - P_{\tilde{g}}(M) &=  \sum_{\mbf{L} \in L_c} 4\tilde{p}(\mbf{L}), \qquad L_c = 2\mathds{Z} \mbf{a}_1+ (2\mathds{Z}+1) \mbf{a}_2 \\
P_{\tilde{g}}(\Gamma) - P_{\tilde{g}}(X) - P_{\tilde{g}}(Y) + P_{\tilde{g}}(M) &=  \sum_{\mbf{L} \in L_{d}} 4\tilde{p}(\mbf{L}) , \qquad L_{d} = (2\mathds{Z}+1) \mbf{a}_1+ (2\mathds{Z}+1) \mbf{a}_2
\eea
we will find constraints on $||\tilde{p}(\mbf{R})||$ on certain sublattices. Taking Frobenius norms, applying the triangle inequality, and using \Eq{eq:DgcancelP} to evaluate traces, we find
\bea
\label{eq:c2constraintsgen}
\frac{1}{\sqrt{\min \{N_{\mathrm{orb}},4N_{\mathrm{occ}}\}}} |\delta_{1a}| &\leq \sum_{\mbf{L} \in L_a} ||\tilde{p}(\mbf{L})|| \\
\frac{1}{\sqrt{\min \{N_{\mathrm{orb}},4N_{\mathrm{occ}}\}}} |\delta_{1b}| &\leq \sum_{\mbf{L} \in L_b} ||\tilde{p}(\mbf{L})|| \\
\frac{1}{\sqrt{\min \{N_{\mathrm{orb}},4N_{\mathrm{occ}}\}}} |\delta_{1c}| &\leq \sum_{\mbf{L} \in L_c} ||\tilde{p}(\mbf{L})|| \\
\frac{1}{\sqrt{\min \{N_{\mathrm{orb}},4N_{\mathrm{occ}}\}}} |\delta_{1d}| &\leq \sum_{\mbf{L} \in L_d} ||\tilde{p}(\mbf{L})|| \\
\eea
which is essentially the same result derived in \App{app:realspaceproj}. Note that all the sublattices $L_w$ are disjoint.  Here we also evaluated the symmetry data bound on the $L_a$ lattice, which yields the 1a RSI. With these constraints, we want to use the concentration lemma to bound
\bea
G \geq \sum_{\mbf{R}} \frac{1}{2\Omega_c} ||\tilde{p}(\mbf{R})||^2 M^2(\mbf{R}) \geq  \min_{|\psi_\mbf{R}|} \sum_{\mbf{R}} \frac{1}{2\Omega_c} |\psi_\mbf{R}|^2 M^2(\mbf{R})
\eea
where $|\psi_\mbf{R}|$ also obeys \Eq{eq:c2constraintsgen} with $||p(\mbf{R})|| \to |\psi_\mbf{R}|$ and the normalization constraint $\sum_\mbf{R} |\psi_\mbf{R}|^2 = N_{\mathrm{occ}}$. Define
\bea
\label{eq:minweight}
M^2_w = \min_{\mbf{L} \in L_w} M^2(\mbf{L}) \geq 0
\eea
which depends \emph{only} on the orbitals of the model, and so can be explicitly calculated. The concentration lemma says that the minimum over $\psi_\mbf{R}$ is obtained by taking $|\psi_\mbf{R}|$ nonzero only on $\mbf{R} \in L_a$ such that $\min_{\mbf{L} \in L_a} M^2(\mbf{L})= M^2(\mbf{R})$. This is because $\sum_{\mbf{R}} \frac{1}{2} |\psi_\mbf{R}|^2 M^2(\mbf{R})$ is strictly decreasing when $|\psi_\mbf{R}|$ is decreased and $|\psi_{\mbf{R}'}|$ is increased (preserving the normalization) for $M^2(\mbf{R}) > M^2(\mbf{R}')$. We would like to take $|\psi_\mbf{R}|$ nonzero only on the positions with minimum $M^2(\mbf{L}_w)$ in each sublattice such that the value of $|\psi_\mbf{R}|$ saturates the bounds in \Eq{eq:c2constraintsgen}. However, in this case $\sum_\mbf{R} |\psi_\mbf{R}|^2$ may be \emph{less} than $N_{occ}$. Nevertheless, $|\psi_\mbf{R}|$ would still give a (non-optimal) lower bound because $G$ is strictly decreasing under decreasing $|\psi_\mbf{R}|$. For simplicity, we report these non-optimal lower bounds, but we emphasize that a more detailed analysis beyond the scope of this work can improve them. Saturating the bounds in \Eq{eq:c2constraintsgen} on the minimum weights determined in \Eq{eq:minweight}, we find a lower bound given by
\bea
G \geq \frac{1}{\min \{N_{\mathrm{orb}},4N_{\mathrm{occ}}\}} \sum_{w=1a,1b,1c,1d} \frac{1}{2\Omega_c} \delta_w^2 M_w^2
\eea
which reduces to \Eq{eq:GboundC2}when all orbitals are at the 1a position since $M^2_{1a} = 0$ because $\mbf{R}= 0 \in L_a$, so only the 1b,1c, and 1d RSIs appear. 

\subsection{$\tilde{g} = C_4$}

With $C_4$, we can take the linear combinations
\bea
P(\Gamma) + P(M) &=  \sum_{\mbf{L} \in L_a} 2p(\mbf{L}) , \qquad L_a = \mathds{Z} (\mbf{a}_1 + \mbf{a}_2)+ \mathds{Z} (\mbf{a}_1 -\mbf{a}_2) \\
P(\Gamma) - P(M) &=  \sum_{\mbf{L} \in L_b} 2p(\mbf{L}) , \qquad L_b = (\mathds{Z}+\frac{1}{2}) (\mbf{a}_1 + \mbf{a}_2)+ (\mathds{Z}+\frac{1}{2}) (\mbf{a}_1 -\mbf{a}_2) \ . \\
\eea
Taking Frobenius norms, applying the triangle inequality, and using \Eq{eq:DgcancelP} to evaluate traces, we find
\bea
\label{eq:c4constraintsgen}
\frac{|\delta_{1a,2} + i(\delta_{1a,1} - \delta_{1a,3})|}{\sqrt{\min \{N_{\mathrm{orb}},2N_{\mathrm{occ}}\}}}  &\leq  \sum_{\mbf{L} \in L_a} ||p(\mbf{L})||  \\
\frac{|\delta_{1b,2} + i (\delta_{1b,1} - \delta_{1b,3})|}{\sqrt{\min \{N_{\mathrm{orb}},2N_{\mathrm{occ}}\}}}   &\leq  \sum_{\mbf{L} \in L_b} ||p(\mbf{L})|| \ . \\
\eea
With these constraints, we use the concentration lemma to bound
\bea
G \geq \sum_{\mbf{R}} \frac{1}{2\Omega_c} ||\tilde{p}(\mbf{R})||^2 M^2(\mbf{R}) \geq  \min_{|\psi_\mbf{R}|} \sum_{\mbf{R}} \frac{1}{2\Omega_c} |\psi_\mbf{R}|^2 M^2(\mbf{R})
\eea
where $|\psi_\mbf{R}|$ also obeys \Eq{eq:c4constraintsgen} with $||p(\mbf{R})|| \to |\psi_\mbf{R}|$ and the normalization constraint $\sum_\mbf{R} |\psi_\mbf{R}|^2 = N_{\mathrm{occ}}$. Define
\bea
M^2_w = \min_{\mbf{L} \in L_w} M^2(\mbf{L}) \geq 0
\eea
which depend \emph{only} on the orbitals of the model, and so can be explicitly calculated. Applying the concentration lemma and allowing the normalization lowered as discussed \App{eq:tildegC2}, we find
\bea
G \geq \frac{1}{\min \{N_{\mathrm{orb}},2N_{\mathrm{occ}}\}} \sum_{w=1a,1b} \frac{1}{2\Omega_c} (\delta_{w,2}^2 + (\delta_{w,1}-\delta_{w,3})^2) M_w^2  \ .
\eea
Again we emphasize this is a non-optimal bound and can be improved. 

\subsection{$\tilde{g} = C_3$}

With $C_3$, we can take the linear combinations
\bea
P(\Gamma) + P(K) + P(K') &= \sum_{\mbf{L} \in L_a} 3 p(\mbf{L}), \qquad L_a = (2\mbf{a}_1+\mbf{a}_2)\mathds{Z}+ (\mbf{a}_1+2\mbf{a}_2)\mathds{Z} \\
P(\Gamma) + e^{- \frac{2\pi i}{3}} P(K) + e^{\frac{2\pi i}{3}} P(K') &= \sum_{\mbf{L} \in L_b} 3 p(\mbf{L}), \qquad  L_b = (2\mbf{a}_1+\mbf{a}_2)(\mathds{Z} - \frac{1}{3})+ (\mbf{a}_1+2\mbf{a}_2)(\mathds{Z} + \frac{2}{3}) \\
P(\Gamma) + e^{\frac{2\pi i}{3}} P(K) + e^{-\frac{2\pi i}{3}} P(K') &= \sum_{\mbf{L} \in L_c} 3 p(\mbf{L}), \qquad  L_c = (2\mbf{a}_1+\mbf{a}_2)(\mathds{Z} - \frac{2}{3})+ (\mbf{a}_1+2\mbf{a}_2)(\mathds{Z} + \frac{1}{3}) \\
\eea
Taking Frobenius norms, applying the triangle inequality, and using \Eq{eq:DgcancelP} to evaluate traces, we find
\bea
\label{eq:c3constraintsgen}
\frac{|e^{\frac{2\pi i}{3}} \delta_{1a,1} + e^{-\frac{2\pi i}{3}}\delta_{1a,2} |}{\sqrt{\min \{N_{\mathrm{orb}},3N_{\mathrm{occ}}\}}}  &\leq  \sum_{\mbf{L} \in L_a} ||p(\mbf{L})||  \\
\frac{|e^{\frac{2\pi i}{3}} \delta_{1b,1} + e^{-\frac{2\pi i}{3}}\delta_{1b,2} |}{\sqrt{\min \{N_{\mathrm{orb}},3N_{\mathrm{occ}}\}}} &\leq  \sum_{\mbf{L} \in L_b} ||p(\mbf{L})||  \\
\frac{|e^{\frac{2\pi i}{3}} \delta_{1c,1} + e^{-\frac{2\pi i}{3}}\delta_{1c,2} |}{\sqrt{\min \{N_{\mathrm{orb}},3N_{\mathrm{occ}}\}}} &\leq  \sum_{\mbf{L} \in L_c} ||p(\mbf{L})||  \\
\eea
With these constraints, we use the concentration lemma to bound
\bea
G \geq \sum_{\mbf{R}} \frac{1}{2\Omega_c} ||\tilde{p}(\mbf{R})||^2 M^2(\mbf{R}) \geq  \min_{|\psi_\mbf{R}|} \sum_{\mbf{R}} \frac{1}{2\Omega_c} |\psi_\mbf{R}|^2 M^2(\mbf{R})
\eea
where $|\psi_\mbf{R}|$ also obeys \Eq{eq:c3constraintsgen} with $||p(\mbf{R})|| \to |\psi_\mbf{R}|$ and the normalization constraint $\sum_\mbf{R} |\psi_\mbf{R}|^2 = N_{\mathrm{occ}}$. Define
\bea
M^2_w = \min_{\mbf{L} \in L_w} M^2(\mbf{L}) \geq 0
\eea
which depend \emph{only} on the orbitals of the model, and so can be explicitly calculated. Applying the concentration lemma as in \App{eq:tildegC2}, we find
\bea
G \geq \frac{1}{\min \{N_{\mathrm{orb}},3N_{\mathrm{occ}}\}} \sum_{w=1a,1b,1c} \frac{1}{2\Omega_c} (\delta^2_{w,1} - \delta^2_{w,1} \delta^2_{w,2} +\delta^2_{w,2}) M_w^2  \ .
\eea
The results of this Appendix show that nonzero lower bounds can be obtained in the case of generic orbital positions, but are suboptimal and can be improved with future work. The study of optimal bounds in the generic case is enriched by the existence of non-compact obstructed atomic insulators \cite{2021arXiv210713556S} and may be a fruitful area of study.

\end{document}